\documentclass[twocolumn]{pasj01}
\usepackage{natbib} 
\usepackage{lscape}
\usepackage{color}

\newcommand{\lya}{${\rm Ly}\alpha \ $}

\newcommand{\wobs}{$\omega_{\rm obs}(\theta)$}
\newcommand{\Ao}{A_{\omega}}
\newcommand{\Aocor}{A_{\rm \omega,\, corr}}

\Accepted{$\langle$February, 2019$\rangle$}

\begin{document}
\title{The dominant origin of diffuse Ly$\alpha$ halos around LAEs explored by SED fitting and clustering analysis} 

\author{Haruka \textsc{Kusakabe}\altaffilmark{1}%
}
\altaffiltext{1}{Department of Astronomy, Graduate School of Science, The University of Tokyo, 7-3-1 Hongo, Bunkyo-ku, Tokyo 113-0033, Japan}
\email{kusakabe@astron.s.u-tokyo.ac.jp}

\author{Kazuhiro \textsc{Shimasaku}\altaffilmark{1,2}}
\altaffiltext{2}{Research Center for the Early Universe, The University of Tokyo, 7-3-1 Hongo, Bunkyo-ku, Tokyo 113-0033, Japan}

\author{Rieko \textsc{Momose}\altaffilmark{1,3}}
\altaffiltext{3}{Department of Physics, National Tsing Hua University, 101 Section 2 Kuang Fu Road, Hsinchu 30013, Taiwan}

\author{Masami \textsc{Ouchi}\altaffilmark{4,5}}
\altaffiltext{4}{Institute for Cosmic Ray Research, The University of Tokyo, 5-1-5 Kashiwanoha, Kashiwa, Chiba 277-8582, Japan}
\altaffiltext{5}{Kavli Institute for the Physics and Mathematics of the Universe (Kavli IPMU, WPI), The University of Tokyo, 5-1-5 Kashiwanoha, Kashiwa, Chiba 277-8583, Japan}

\author{Kimihiko \textsc{Nakajima}\altaffilmark{6,7,8}}
\altaffiltext{6}{Niels Bohr Institute, University of Copenhagen, Blegdamsvej 17, 2100 Copenhagen, Denmark}
\altaffiltext{7}{Cosmic Dawn Center (DAWN), Niels Bohr Institute, University of Copenhagen, Lyngbyvej 2, 2100 Copenhagen \O, Denmark}
\altaffiltext{8}{DTU-Space, Technical University of Denmark, Elektrovej 327, 2800 Kgs.\ Lyngby,Denmark}

\author{Takuya \textsc{Hashimoto}\altaffilmark{ 9, 10}}
\altaffiltext{9}{College of General Education, Osaka Sangyo University, 3-1-1 Nakagaito, Daito, Osaka 574-8530, Japan}
\altaffiltext{10}{National Astronomical Observatory of Japan, 2-21-1 Osawa, Mitaka, Tokyo 181-8588, Japan}

\author{Yuichi \textsc{Harikane}\altaffilmark{5,11}}
\altaffiltext{11}{Department of Physics, Graduate School of Science, The University of Tokyo, 7-3-1 Hongo, Bunkyo, Tokyo 113-0033, Japan}

\author{John \textsc{D. Silverman}\altaffilmark{4}}

\author{Peter \textsc{L. Capak}\altaffilmark{12, 13}}
\altaffiltext{12}{California Institute of Technology, MC 105-24, 1200 East California Blvd., Pasadena, CA 91125, USA}
\altaffiltext{13}{Infrared Processing and Analysis Center, California Institute of Technology, MC 100-22, 770 South Wilson Ave., Pasadena, CA 91125, USA}

\maketitle

\begin{abstract}
The physical origin of diffuse Ly$\alpha$ halos (LAHs) around star-forming galaxies is still a matter of debate. We present the dependence of LAH luminosity ($L({\rm Ly}\alpha)_H$) on the stellar mass ($M_\star$), $SFR$, color excess ($E(B-V)_\star$), and dark matter halo mass ($M_{\rm h}$) of the parent galaxy for $\sim 900$ Ly$\alpha$ emitters (LAEs) at $z\sim2$ divided into ten subsamples. We calculate $L({\rm Ly}\alpha)_H$ using the stacked observational relation between $L({\rm Ly}\alpha)_H$ and central Ly$\alpha$ luminosity of Momose et al. (2016, MNRAS, 457, 2318), which we find agrees with the average trend of VLT/MUSE-detected individual LAEs. We find that our LAEs have relatively high $L({\rm Ly}\alpha)_H$ despite low $M_\star$ and $M_{\rm h}$, and that $L({\rm Ly}\alpha)_H$ remains almost unchanged with $M_\star$ and perhaps with $M_{\rm h}$. These results are incompatible with the cold stream (cooling radiation) scenario and the satellite-galaxy star-formation scenario, because the former predicts fainter $L({\rm Ly}\alpha)_H$ and both predict steeper $L({\rm Ly}\alpha)_H$ vs. $M_\star$ slopes. We argue that LAHs are mainly caused by Ly$\alpha$ photons escaping from the main body and then scattered in the circum-galactic medium. This argument is supported by LAH observations of H$\alpha$ emitters (HAEs). When LAHs are taken into account, the Ly$\alpha$ escape fractions of our LAEs are about ten times higher than those of HAEs with similar $M_\star$ or $E(B-V)_\star$, which may partly arise from lower HI gas masses implied from lower $M_{\rm h}$ at fixed $M_\star$, or from another Ly$\alpha$ source in the central part.
\end{abstract}

\section{Introduction}\label{sec:intro}
A Ly$\alpha$ halo (LAH) is a diffuse, spatially extended structure of Ly$\alpha$ emission seen around star-forming galaxies. LAHs around local galaxies, as well as around active galactic nuclei (AGNs) and quasi-stellar objects (QSOs), can be detected individually because they are relatively bright \citep[e.g.,][and reference therein]{Keel1999, Kunth2003, Hayes2005,  Goto2009, Ostlin2009, Hayes2013, Matsuda2011}. LAHs around high-redshift ($z$) galaxies are much fainter, but they have been detected in stacked narrow-band images (tuned to redshifted Ly$\alpha$ emission) of $100$ -- $4000$ star-forming galaxies at $z\sim2$--$6$ \citep[e.g.,][see also a stacking study of spectra of $\sim80$ LAEs at $z\sim2$--$4$ by \citet{Guaita2017}]{Hayashino2004, Steidel2011, Matsuda2012,Feldmeier2013, Momose2014, Momose2016, Xue2017}. Very recently, LAHs around $\sim 170$ star forming galaxies at $z\sim3$--$6$ have been detected individually by deep integral field spectroscopy with VLT/MUSE \citep{Wisotzki2016,Leclercq2017,Wisotzki2018}. Since the existence of LAHs has now been established, the next question is what is their physical origin(s).

Theoretical studies have proposed several physical origins of LAHs: resonant scattering in the CGM, cold streams (gravitational cooling radiation), star formation in satellite galaxies (one-halo term), fluorescence (photo-ionization), shock heating by gas outflows, and major mergers \citep[e.g.,][]{Haiman2000, Taniguchi2000, Cantalupo2005, Mori2006, Laursen2007, Zheng2011, Rosdahl2012, Yajima2013, Lake2015, Mas-Ribas2016}. The former three are generally considered for high-$z$ star-forming galaxies \citep[e.g.,][]{Lake2015}, while the latter three are preferred for giant Ly$\alpha$ nebulae (Ly$\alpha$ blobs; LABs) and/or bright QSOs \citep[e.g.,][]{Mori2006, Kollmeier2010,  Yajima2013,  Momose2018arXiv}. 

Understanding the origin of LAHs provides crucial information on the circum-galactic medium (CGM), which is closely linked to galaxy formation and evolution. It also enables us to estimate the escape fraction of Ly$\alpha$ emission from central galaxies correctly. If resonant scattering mainly drives LAHs, the Ly$\alpha$ luminosity of LAHs should be included in the calculation of the Ly$\alpha$ escape fraction. LAHs are also important for studies of cosmic reionization because their spatial extent can be used as a probe of the intergalactic medium (IGM) ionization fraction.

Lyman $\alpha$ emitters (LAEs) are suitable objects for studying the nature of LAHs because a large sample of LAEs at a fixed redshift as needed for a stacking analysis can be constructed relatively easily from a narrow-band imaging survey \citep[][]{Matsuda2012,Feldmeier2013, Momose2014, Momose2016, Xue2017}. LAEs are typically low-stellar-mass young galaxies with low metallicities and low-dust contents hosted in low-mass dark matter halos \citep[e.g.,][ and reference therein]{Pirzkal2007,  Lai2008, Ono2010b, Nakajima2014, Kusakabe2015,  Kojima2017, Ouchi2018}. They are detected owing to efficient Ly$\alpha$ escapes, which are suggested to stem partly from these physical properties such as low-dust attenuation \citep[e.g.,][]{Finkelstein2009b}. 

\citet{Matsuda2012} have found that LAEs in a large-scale overdense region at $z=3.1$ have large ($\sim 100$--$200$ \AA) EWs if LAH components are included. They suggest that those LAHs may partly originate from shock heating due to gas outflows or cold streams, although they have not ruled out other possibilities. On the other hand, \citet{Momose2016} have stacked $\sim3600$ LAEs in field regions at $z\sim2$ to find that some subsamples have relatively small Ly$\alpha$ EWs fully consistent with pop I\hspace{-.1em}I star formation, suggesting that the cold stream scenario is not preferred. Finding no correlation between  the spatial extent (the scale length, $r_s$) and the surface number density for LAEs at $z\sim 3$--$4$, \citet{Xue2017} have suggested that star formation in satellite galaxies is not the dominant contributor to LAHs \citep[see however, ][]{Matsuda2012}. They have also found that the radial profile of LAHs is very close to that predicted by models of resonant scattering in \citet{Dijkstra2012}, leaving only little room for the contribution from satellites galaxies and cold streams modeled by \citet{Lake2015}. Note, however, that \citet{Lake2015}'s model reproduces the radial profile of LAHs seen in LAEs at $z\sim3$ in \citet{Momose2014}. More recently, \citet{Leclercq2017} have measured LAH properties of $\sim150$ individual LAEs at $z\sim3$--$6$ using VLT/MUSE. They argue that a significant contribution from star formation in satellite galaxies is somewhat unlikely since the UV component of LAEs is compact and not spatially offset from the center of their LAHs, while having not given a firm conclusion on other origins. 

To summarize, although there are a number of observational studies on the origin of LAHs, their results are not very conclusive, nor consistent with each other \citep[][see also \citet{Steidel2011} ]{Matsuda2012, Feldmeier2013, Momose2016, Wisotzki2016, Xue2017, Leclercq2017}. This is partly because correlations of LAH properties with properties of central galaxies have not been fully studied. Especially important may be correlations with the dark matter halo mass and stellar mass of central galaxies, because they can be directly compared with theoretical predictions \citep[e.g.,][]{Rosdahl2012}. Although \citet{Leclercq2017} have discussed a correlation between the Ly$\alpha$ luminosity of LAHs and the UV luminosity of central galaxies, they have not estimated those masses. $SFRs$ and dust attenuation are also important quantities to discuss the scattering origin of LAHs. 

Another problem is that $r_s$, the scale-length of LAHs that is often used to discuss the origin of LAHs in previous studies, is not robust against measurement errors. Indeed, the dependence of $r_s$ on Ly$\alpha$ luminosity found in individually detected MUSE LAEs is not consistent with the average dependence obtained by \citet{Momose2016} from stacked images. In contrast, as we will see later, relations between the Ly$\alpha$ luminosity of central galaxies and that of LAHs found in \citet{Momose2016} is in good agreement with those seen in individual MUSE-LAEs in \citet{Leclercq2017}. This suggests that Ly$\alpha$ luminosity is more robust against systematic errors from stacking. 

In this paper, we study the dependence of LAH luminosity on stellar properties and dark matter halo mass using $\sim900$ star-forming LAEs at $z\sim2$ to identify the dominant origin of LAHs around LAEs. Section $2$ summarizes the data and sample used in this study. In section $3$, we construct subdivided samples based on UV, Ly$\alpha$, and $K$-band properties. We present methods to derive the Ly$\alpha$ luminosities of LAHs as well as the stellar properties and dark matter halo masses of subdivided LAEs in section $4$. After showing results in section $5$, we discuss the origin of LAHs and high Ly$\alpha$ escape fractions in section $6$. Conclusions are given in Section 7.

Throughout this paper, we adopt a flat cosmological model with the matter density $\Omega_{\rm m} = 0.3$, the cosmological constant $\Omega_{\Lambda} = 0.7$, the baryon density $\Omega_{b} = 0.045$, the Hubble constant $H_{0} = 70~{\rm km\,s^{-1}Mpc^{-1}}\,(h_{100}=0.7)$, the power-law index of the primordial power spectrum $n_{\rm s} = 1$, and the linear amplitude of mass fluctuations $\sigma_{8}=0.8$, which are consistent with the latest Planck results \citep{Planckcollaboration2016}. We assume a Salpeter initial mass function \citep[IMF: ][]{Salpeter1955} with a mass range of $0.1$--$100$ $M_\odot$\footnote{\label{ft:imf}To rescale stellar masses in previous studies assuming a Chabrier or Kroupa IMF \citep{Kroupa2001, Chabrier2003}, we divide them by a constant factor of 0.61 or 0.66, respectively. Similarly, to convert SFRs in the literature with a Chabrier or Kroupa IMF, we divide them by a constant factor of 0.63 or 0.67, respectively.}. Magnitudes are given in the AB system \citep{Oke1983} and coordinates are given in J2000. Distances are expressed in comoving units. We use ``log'' to denote a logarithm with a base $10$ (${\rm log}_{10}$).

%
%
%
\section{Data and sample}
\subsection{Sample selection}\label{subsec:selection}
\citet{Kusakabe2018a} have constructed large samples of $z=2.2$ LAEs in four deep fields: the Subaru/XMM-Newton Deep Survey (SXDS) field \citep{Furusawa2008}, the Cosmic Evolution Survey (COSMOS) field \citep{Scoville2007}, the Hubble Deep Field North \citep[HDFN:][]{ Capak2004}, and the Chandra Deep Field South \citep[CDFS:][]{Giacconi2001}.In this study, we only use their SXDS and COSMOS samples. We do not use the HDFN sample because the $R$-band image of this field is not deep enough to derive the UV slope for faint LAEs. We also do not use the CDFS sample because the $i$, $z$, and $H$ data are too shallow to perform reliable SED fitting as has been pointed out by \citet{Kusakabe2018a}.

We summarize the sample selection and the estimation of the contamination fraction detailed in \citet{Kusakabe2018a}. LAEs at $z=2.14$--$2.22$ are selected using the narrow band $NB387$ \citep{Nakajima2012} as described in selection papers \citep[][]{Nakajima2012, Nakajima2013, Konno2016, Kusakabe2018a}. The threshold of the rest-frame equivalent width, $EW_0$, of \lya emission is $EW_0({\rm Ly}\alpha) \geq 20$--$30$\AA$\,$ \citep[see figure 1 in ][]{Konno2016}. The $NB387$ limiting magnitude is $25.7$ mag for the SXDS sample and $26.1$ mag for the COSMOS sample ($2''$ diameter aperture, 5$\,\sigma$). We only use LAEs with $NB387$ total (i.e., aperture-corrected; see table \ref{tbl:data}) magnitude brighter than $25.5$ mag. All sources detected in either X-ray, UV, or radio have been removed since they are regarded as AGNs. Our entire sample consists of $897$ LAEs from $\simeq 1980$ square arcminutes. The survey area of each field is shown in table \ref{tbl:data}. 

\citet{Kusakabe2018a} have conservatively estimated the fraction of possible interlopers in their LAE samples to be $10\, \pm\, 10\%$, where interlopers are categorized into spurious sources,  AGNs without an X-ray, UV, or radio counterpart, foreground/background galaxies, and $z=2.2$ LAEs with low $EW_0({\rm Ly}\alpha)$ which happen to meet the color selection due to photometric errors. See sections 2.2 and 3.2 of \citet{Kusakabe2018a} for details. We use this contamination fraction to obtain true clustering amplitudes from observed ones in section \ref{subsec:acf}.

\subsection{Imaging data for SED fitting}\label{subsec:data}
Most of the data used in this work are the same as those used in \citet{Kusakabe2018a}, except that the NIR imaging data are replaced to new ones in this work. We overview the data used in SED fitting in the two fields below. 

We use ten broadband images for SED fitting: five optical bands -- $B, V, R$ (or $r$), $i$ (or $i'$), and $z$ (or $z'$); three NIR bands -- $J$, $H$, and $K$ (or $Ks$); and two mid-infrared (MIR) bands -- IRAC ch1 and ch2. The PSFs of the images are matched in each field. The aperture corrections to convert $3''$ MIR aperture magnitudes to total magnitudes are taken from \citet[][see table\ref{tbl:data}]{Ono2010b}. For each field, a K-band or NIR detected catalog is used to obtain secure IRAC photometry in section \ref{subsec:stack_photo} and to divide the LAEs into subsamples in section \ref{subsec:criteria}. 

\begin{description}
\item[SXDS field]
The images used for SED fitting are as follows: $B, V, R, i'$, and $z'$ images with Subaru/Suprime-Cam from the Subaru/XMM-Newton Deep Survey project \citep[SXDS]{Furusawa2008}; $J, H$, and $K$ images from the data release $11$ of the UKIRT/WFCAM UKIDSS/UDS project \citep[][Almaini et al. in prep.]{Lawrence2007}; Spitzer/IRAC $3.6$~$\mu$m (ch1) and $4.5$~$\mu$m (ch2) images from the Spitzer Large Area Survey with Hyper-Suprime-Cam (SPLASH) project \citep[SPLASH; PI: P. Capak; Capak et al. in prep.; ][]{Mehta2018}. All images are publicly available except the SPLASH data. The aperture corrections for optical and NIR images are given in \citet{Nakajima2013}. The catalog used to clean IRAC photometry and to obtain $K$-band counterparts is constructed from the $K$-band image.

\item[COSMOS field]
We use the publicly available $B, V, r', i'$, and $z'$ images with Subaru/Suprime-Cam by the Cosmic Evolution Survey \citep[COSMOS:][]{Capak2007, Taniguchi2007} and $J, H$, and $Ks$ images with the VISTA/VIRCAM from the third data release of the UltraVISTA survey \citep{McCracken2012}. We also use Spitzer/IRAC ch1 and ch2 images from the SPLASH project \citep{Laigle2016}. The aperture corrections for the optical images are taken from \citet{Nakajima2013} and those for the NIR images follow \citet{McCracken2012}. The catalog used to clean IRAC photometry and to obtain $K$-band counterparts is the one given by \citet{Laigle2016}, for which sources have been detected in a combined z'YJHKs image.
\end{description}

\begin{table*}
\tbl{Details of the data. }{
\begin{tabular}{l|cccc|cccc}
\hline
&\multicolumn{4}{|c|}{SXDS ($\sim1240\,{\rm arcmin^2}$, $600^{(a)}$ LAEs)} &  \multicolumn{4}{|c}{COSMOS ($\sim740\,{\rm arcmin^2}$, $297^{(a)}$ LAEs)}\\ 
band & PSF & aperture  & aperture  & $5\sigma$ limit & PSF & aperture  & aperture&   $5\sigma$ limit \\ 
 & ($''$)& diameter ($''$)&correction (mag)& (mag) &($''$)& diameter ($''$)&correction (mag)&(mag)\\ 
 & (1) & (2)  & (3)  & (4) & (1) & (2)  & (3)  & (4)\\  \hline
$NB387$ & 0.88 & 2.0 & 0.17 & 25.7 & 0.95 & 2.0 & 0.25 & 26.1\\ 
$B$ & 0.84 & 2.0 & 0.17 & 27.5--27.8 & 0.95 & 2.0 & 0.12 & 27.5\\ 
$V$ & 0.8 & 2.0 & 0.15 & 27.1--27.2 & 1.32 & 2.0 & 0.33 & 26.8\\ 
$R$ ($r'$) & 0.82 & 2.0 & 0.16 & 27.0--27.2 & 1.04 & 2.0 & 0.19 & 26.8\\ 
$i'$ ($I$) & 0.8 & 2.0 & 0.16 & 26.9--27.1 & 0.95 & 2.0 & 0.12& 26.3\\ 
$z'$ & 0.81 & 2.0 & 0.16 & 25.8 -- 26.1 & 1.14 & 2.0 & 0.25 & 25.4\\ 
$J$ & 0.85 & 2.0 & 0.15 & 25.6 & 0.79 & 2.0 & 0.3 & 24.6--24.8\\ 
$H$ & 0.85 & 2.0 & 0.15 & 25.1  & 0.76 & 2.0 & 0.2 & 24.3--24.4\\ 
$K$ ($Ks$) & 0.85 & 2.0 & 0.16 & 25.3 & 0.75 & 2.0 & 0.2 & 23.9--24.6\\ 
IRAC ch1 & 1.7 & 3.0 & 0.52  & 24.9$^{(b)}$ & 1.7 & 3.0 & 0.52 & 25.4$^{(b)}$\\ 
IRAC ch2 & 1.7 & 3.0 & 0.55 & 24.9$^{(b)}$ & 1.7 & 3.0 & 0.55 & 25.1$^{(b)}$\\ \hline 
\end{tabular}
}
\tabnote{Note. (1) The FWHM of the PSF, (2) aperture diameter in photometry, (3) aperture correction, and (4) $5\sigma$ limiting magnitude with a $2''$ diameter aperture are shown for each band. Values in parentheses show the area used in clustering analysis. (a) The number of LAEs in the SXDS field is slightly different from that in \citet{Kusakabe2018a} since we use $NB387$ images before PSF matching to other selection-band images for photometry. (b) The limiting magnitude measured in areas with no sources \citep[see][]{Laigle2016, Mehta2018}. 
 }\label{tbl:data}
\end{table*}

\section{Subsamples}\label{sub:subsample}
\begin{table*}
\tbl{Subsample definition.}{
\begin{tabular}{ll|ccc}
\hline
subsample & criteria & COSMOS & SXDS & total \\ \hline
bright UV (MuvB) & $M_{UV}\leq-19.2\,{\rm mag}$ & 123 (123, 9) & 293 (257, 52) &416 (380, 61)\\
faint UV (MuvF) & $M_{UV}>-19.2\,{\rm mag}$ &173 (173, 13) & 302 (257, 47) & 475 (430, 60)\\
blue $\beta$ (betaB) & $\beta\leq-1.6$ & 80 (80, 5)	& 389 (334, 74)	 & 469 (414, 79)\\
red $\beta$ (betaR) & $\beta>-1.6$ & 216	(216, 17)	& 206 (180, 25) & 422 (396, 42)\\
bright Ly$\alpha$ (lyaB) & $ L({\rm Ly\alpha})_{ps}\geq1.2	\times10^{42}\,{\rm erg\,s^{-1}}$ &	211	(211, 14)	& 236 (218, 41) & 447 (429, 55)\\
faint Ly$\alpha$ (lyaF) & $ L({\rm Ly\alpha})_{ps}<1.2\times10^{42}\,{\rm erg\,s^{-1}}$ &	85 (85, 8) & 359 (296, 58) & 444 (381, 66)\\
large EW	(ewL) & $EW_{0,\,ps}({\rm Ly\alpha}) \geq34\,$\AA & 222 (222, 16) & 228 (205, 35) & 450 (427, 51)\\
small EW	(ewS) & $EW_{0,\,ps}({\rm Ly\alpha}) <34\,$\AA & 74 (74, 6) & 367 (309, 64) & 441 (383, 70) \\
bright $K$ (KB) & $ m_{\rm K}\leq25\,{\rm mag}$ & 112 (112, 11) & 178 (177, 35) & 290 (144, 46)\\
faint $K$ (KF) & $ m_{\rm K}>25\,{\rm mag}$ & 184 (184, 11) & 417	(337, 64)	& 601 (236, 75)\\\hline
\end{tabular}
}
\tabnote{Note. The selection criterion and the numbers of objects for each subsample. The number outside the bracket indicates the number of objects for clustering analysis, while the numbers in the bracket are for SED fitting: the left one corresponds to objects with UV to NIR photometry and the right one to those with clean ch1 and ch2 photometry.
} \label{tbl:subsample}
\end{table*}

A vast majority of our LAEs are too faint to estimate stellar masses on individual basis.
To study how LAH luminosity depends on stellar and dark matter halo masses, we therefore divide the entire sample into subsamples in accordance with the following five quantities which are expected to correlate with stellar mass, and perform a stacking analysis on each subsample. 
(i) $K$-band apparent magnitude, $m_{\rm K}$, known as a good tracer of stellar mass \citep[e.g.,][]{Daddi2004}. 
(ii) Rest-frame UV absolute magnitude, $M_{\rm UV}$, which is related to $SFR$ and hence expected to trace stellar mass through the star formation main sequence \citep[e.g.,][]{Speagle2014a}. 
(iii) UV spectral slope $\beta$ ($f_{\lambda} \propto \lambda^\beta$), an indicator of dust attenuation and may correlate with stellar mass \citep[e.g., ][]{Reddy2010}. 
(iv) Ly$\alpha$ luminosity $L({\rm Ly\alpha})$ and (v) rest-frame Ly$\alpha$ equivalent width $EW_0({\rm Ly}\alpha)$, both of which possibly anti-correlate with stellar mass according to Ando relation \citep[][see also \citet{Shimakawa2017a} ]{Ando2006, Ando2007}.

While only $30$--$40$\% of our LAEs are detected in the $K$ band with $m_{\rm K}\lesssim25.0$ (see section \ref{subsec:criteria}), the other four quantities can be measured for almost all objects because they need only optical imaging data, which are deep enough as shown in table \ref{tbl:data}. We divide the whole sample of each field into two subsamples in accordance with each of $m_{\rm K}$, $M_{\rm UV}$, $\beta$, $L({\rm Ly\alpha})$, and $EW_0({\rm Ly}\alpha)$; further division makes stacked SEDs too noisy to do reliable SED fitting. Among the five quantities, $m_K$ and $M_{\rm UV}$ are expected to correlate with $M_\star$ most tightly. The subsamples by $\beta$, $L({\rm Ly\alpha})$, and $EW_0({\rm Ly}\alpha)$ are useful to check the results obtained for the $m_K$ and $M_{\rm UV}$ subsamples, because these three quantities are affected by the NB selection bias differently from $m_K$ and $M_{\rm UV}$ as discussed in appendix \ref{sec: appendix_NBbias} (see figure \ref{fig:param}). As shown later, all five subsample pairs give similar results.

\subsection{UV and Ly$\alpha$ properties}\label{subsec:ulya}
For each object, we measure $M_{\rm UV}, \beta, L({{\rm Ly}\alpha})$, and $EW_0({{\rm Ly}\alpha})$ from $NB387$, $B$, $V$, and $R$ magnitudes in the following manner. First, we approximate the UV SED of the object by a simple SED composed of a power-law continuum and a Ly$\alpha$ line centered at rest-frame $1216$ \AA:
\begin{eqnarray}
\label{eq:uvlya1}
f_{\nu} {\rm (erg\,s^{-1}\, cm^{-2}\, Hz^{-1})} &=& A\,  10^{-0.4(m_{\rm UV(1+z)}+48.60)}    \left ( \frac{\nu_{\lambda=UV(1+z)}}{\nu}\right) ^{\beta+2}\\\nonumber
&+& F_{\rm Ly\alpha}\, \delta(\nu-\nu_{\lambda=1216(1+z)}),
\end{eqnarray}
where $A$, $m_{\rm UV(1+z)}$, and $F_{\rm Ly\alpha}$ are the IGM attenuation factor from \citet{Madau1995}, the apparent UV magnitude (corresponding to $M_{\rm UV}$), and the Ly$\alpha$ flux ${\rm(erg\,s^{-1}\, cm^{-2})}$, respectively. 
The apparent magnitude of the model SED in a given band $i$ is calculated from its transfer function $T_i(\lambda)$ as below: 
\begin{equation}
\label{eq:uvlya2}
m_{i, model}= -2.5\log_{10}\left(  \frac{ \int f_{\nu} c/\lambda^2 T_i(\lambda) {d\lambda}}{\int c/\lambda^2 T_i(\lambda) {d\lambda}}   \right)-48.6,  
\end{equation}
where $c$ is the speed of light. 

We fit this model SED to the apparent magnitudes of the object with $M_{\rm UV}$, $\beta$, and $F_{\rm Ly\alpha}$ as free parameters. We search for the best-fit parameter values that minimize
\begin{equation}
\chi^2 = \Sigma_{i=NB, B, V, R} \left (\frac{m_{i} - m_{i, model}}{\sigma m_{i}} \right)^2, 
\end{equation}
where $m_{i}$ and $\sigma m_{i}$ are the $i$-th band apparent magnitude and its 1 $\sigma$ error, respectively. We calculate apparent magnitudes from $2''$ diameter aperture magnitudes \citep[see][]{Kusakabe2018a} assuming that our LAEs are point sources in all four bands including $NB387$ which detects Ly$\alpha$ emission. We also assume that their Ly$\alpha$ lines are located at the peak of the response function of $NB387$ and do not correct for flux loss. The best-fit $F_{\rm Ly\alpha}$ is obtained by solving $\frac{\partial \chi^2}{\partial F_{\rm Ly\alpha}}=0$. Hereafter, we refer to the $L({{\rm Ly}\alpha})$ and $EW_0({{\rm Ly}\alpha})$ obtained with the assumption of point sources as $L({{\rm Ly}\alpha})_{ps}$ and $EW_{0, ps}({{\rm Ly}\alpha})$. Since the best-fit $EW_0({{\rm Ly}\alpha})$ is derived from the other three parameters, the degree of freedom is one.

Among the $897$ LAEs, six sources are undetected in at least one of the three broad bands. We do not use these objects in the following analyses because the four quantities derived from the SED fitting are highly uncertain. 

\subsection{Subsample construction}\label{subsec:criteria}
\begin{figure*}[ht]
 \includegraphics[width=0.95\linewidth]{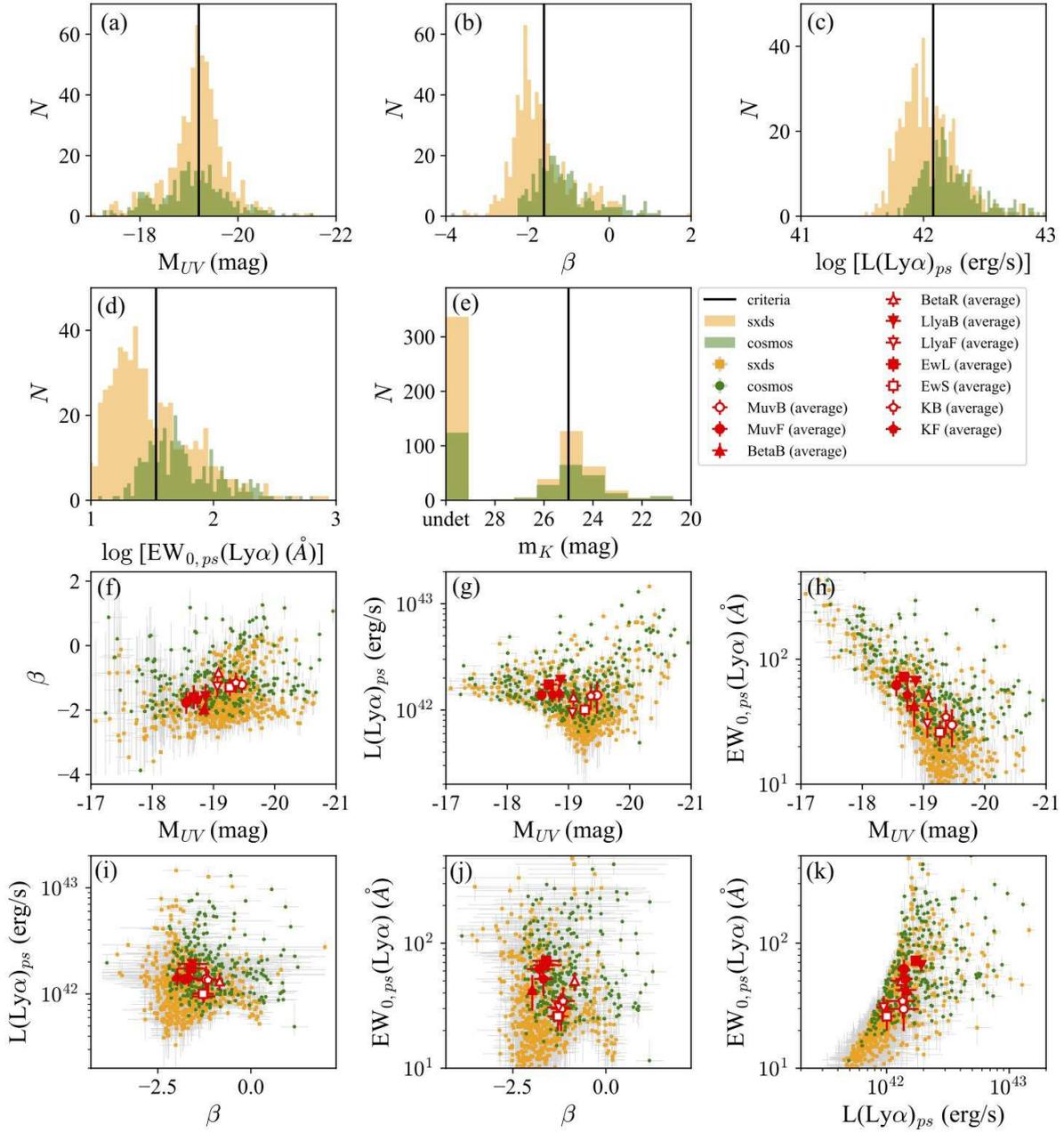}
  \caption{
The distribution of the five quantities used to divide our LAEs into subsamples. Panels (a) -- (e) show histograms: (a) $M_{\rm UV}$, (b) $\beta$, (c) $L({\rm Ly\alpha})_{ps}$, (d) $EW_{0, ps}({\rm Ly}\alpha)$, and (e) $m_{\rm K}$, with orange and green colors corresponding to the SXDS and COSMOS fields, respectively. Black lines indicate the boundaries of the two subsamples. Panels (f) -- (k) are scatter plots: (f) $\beta$ vs. $M_{\rm UV}$, (g) $L({\rm Ly\alpha})_{ps}$ vs. $M_{\rm UV}$, (h) $EW_{0, ps}({\rm Ly}\alpha)$ vs.  $M_{\rm UV}$,  (i) $L({\rm Ly\alpha})_{ps}$ vs. $\beta$, (j) $EW_{0, ps}({\rm Ly}\alpha)$ vs. $\beta$, and (k) $EW_{0, ps}({\rm Ly}\alpha)$ vs. $L({\rm Ly\alpha})_{ps}$, with the same color coding as panels (a)--(e). Red symbols represent averages over the two fields, where different symbols correspond to different classifications: open (filled) circles for bright (faint) $M_{\rm UV}$, open (filled) triangles for red (blue) $\beta$, open (filled) inverted triangles for faint (bright) $L({\rm Ly\alpha})_{ps}$, open (filled) squares for large (small) EW, and  open (filled) pentagons for bright (faint) $m_{\rm K}$.
}
 \label{fig:param}
\end{figure*}

Since we divide LAEs into two subsamples in accordance with each of the five quantities, we have a total of ten subsamples for each field. The boundaries of the subsamples are defined from the distribution of the five quantities, which is shown in figure \ref{fig:param}.

Our LAEs are widely distributed over the four UV and Ly$\alpha$ properties as shown in figures \ref{fig:param} (a) -- (d). The distribution of $M_{\rm UV}$, $\beta$, $L({\rm Ly\alpha})_{ps}$, and $EW_{0, ps}({\rm Ly}\alpha)$ is different between the two fields. This is possibly because of systematic offsets of the zero-point magnitudes (ZPs) of the optical images adopted in the original papers\footnote{ZP offsets of optical broad bands can shift the relation between $M_{\rm UV}$ and $\beta$ (figure \ref{fig:param} [f]). They have a larger effect on smaller-$EW_{0, ps}({\rm Ly}\alpha)$ objects in the $EW_{0, ps}({\rm Ly}\alpha)$ vs. $M_{\rm UV}$ plot (figure \ref{fig:param} [e]), since the contribution of the UV continuum flux in $NB387$ is larger for such objects (see Appendix \ref{sec:appendix_zp} for more details). Because of the NB-selection bias (see also Appendix \ref{sec: appendix_NBbias}), small-$EW_{0, ps}({\rm Ly}\alpha)$ objects tend to have bright $M_{\rm UV}$.} as has been discussed in both the COSMOS \citep{Capak2007,Ilbert2009,Skelton2014} and SXDS \citep{Yagi2013,Skelton2014} fields. However, these papers often claim opposite error directions (see Appendix \ref{sec:appendix_zp} for more details). Another possible reason for the different distribution is field-to-field variance from large scale structure (cosmic variance). In this paper, we use the original ZPs following \citet{Kusakabe2018a} and include ZP uncertainties in the flux-density errors in the calculations given in sections \ref{subsec:ulya} and \ref{sec:sed}.  Although the causes of the different distributions and the correct ZPs remain to be unclear, a pair of subsamples (with the same definition) from the two fields give consistent 
SED fitting results and Ly$\alpha$ luminosities in most cases (see figure \ref{fig:IRX}(b), figure \ref{fig:MS}(b), and Table \ref{tbl:Lya} in Appendix \ref{sec:appendix_Llya}). 

We define the boundary for the four UV and Ly$\alpha$ quantities so that the two subsamples have roughly comparable sizes:
\begin{equation}
M_{\rm UV}=-19.2\,{\rm mag},
\end{equation}
\begin{equation}
\beta=-1.6,
\end{equation}
\begin{equation}
L({\rm Ly\alpha})_{ps}=1.2\times10^{42}\,{\rm erg\,s^{-1}},
\end{equation}
and 
\begin{equation}
EW_{0, ps}({\rm Ly\alpha})=34\,\mathrm{\mathring{A}}
\end{equation}
as indicated by black lines in figure \ref{fig:param} (a) -- (d). The numbers of the LAEs in the eight subsamples are shown in table \ref{tbl:subsample}.

For each field, we also construct two subsamples divided by $m_{\rm K}$.  The $K$-band catalog mentioned in section \ref{subsec:data} effectively include sources with $m_{\rm K}\lesssim25$ mag. Indeed, the $5\sigma$ limiting magnitude of the SXDS $K$-band image is $25.3$ mag and the detection image for the COSMOS catalog, a combined $z'YJHKs$ image, reaches deeper than $25.3$ mag ($5\sigma$). As a result, about $30$--$40\%$ of the LAEs in each field have a $K$-band counterpart with $m_{\rm K}< 25.0$ as shown in figure \ref{fig:param}(e). Therefore, we define the $K$-magnitude boundary as:
\begin{equation}
m_{\rm K}=25.0\,{\rm mag}.
\end{equation} 
Note that the COSMOS $K$ image is composed of Deep and Ultradeep stripes. Since this could add an artificial pattern in the sky distribution of $K$-divided subsamples, we do not use the $K$-divided subsamples for clustering analysis.

We derive the four UV and Ly$\alpha$ quantities for each subsample from a median-stacked SED (see section \ref{sec:sed}) in the same manner as in section \ref{subsec:ulya}. We then calculate average values over the two field, e.g., the average $M_{\rm UV}$ of the two faint-$M_{\rm UV}$ subsamples, as shown by red symbols in figures \ref{fig:param} (f) -- (k). They are located in the middle of the distribution of individual sources (orange and green points), implying that the average SEDs of the subsamples represent well individual LAEs. We find that the subsamples with red $\beta$, faint $L({\rm Ly\alpha})_{ps}$, small $EW_{0, ps}({\rm Ly}\alpha)$, and bright $m_{\rm K}$ as well as bright $M_{\rm UV}$ have bright $M_{\rm UV}$ as shown by red open symbols. Note that the lower left part in figures \ref{fig:param} (g) and (h) and the upper left part in figure \ref{fig:param}(k) show a selection bias: LAEs with faint $M_{\rm UV}$ can be detected only if they have bright $L({\rm Ly\alpha})_{ps}$.

%
\section{Methods}\label{sec:method}
The Ly$\alpha$ luminosities of LAHs are estimated from a stacked observational relation obtained by \citet{Momose2016}.  We do not perform a stacking analysis of LAHs on our own subsamples since their sample sizes, which are one ninth to one half of the subsample sizes ($\sim700$ each) in \citet{Momose2016}, are not large enough to obtain reliable results. Parameters that characterize stellar populations and the mass of dark matter halos are derived from SED fitting and clustering analysis, respectively, in the same manner as in \citet{Kusakabe2018a}.

\subsection{LAH luminosities}\label{subsec:lah}
The LAHs of LAEs have been studied either by a stacking analysis of large samples or using individually detected objects.
\citet{Momose2016} have used stacked images of $\sim 700$ LAEs in each subsample (in total $\sim3600$) at $z \sim 2$ to compare Ly$\alpha$ luminosities within $r=40$ kpc ($\sim5''$) to those within $r=1''$ ($\sim8$ kpc). They have estimated an empirical relation between the two Ly$\alpha$ luminosities from $\sim3000$ LAEs that are the parent sample of our $\sim 900$ LAEs.  On the other hand, \citet{Leclercq2017} have measured Ly$\alpha$ luminosities for $3\leq{z}\leq6$ LAEs with an individually detected LAH by fitting a two component model consisting of halo and continuum-like components. We define three kinds of Ly$\alpha$ luminosities as below.

\begin{description}
\item[$L({\rm Ly}\alpha)_{\rm C}$]
Ly$\alpha$ luminosity at the central part, i.e., the main body of the object where stars are being formed. In \citet{Leclercq2017}, it corresponds to the continuum-like component of Ly$\alpha$ luminosities. We assume that the Ly$\alpha$ luminosities within $r=1''$ in 2D images in \citet{Momose2016} are approximately equal to $L({\rm Ly}\alpha)_{\rm C}$. The aperture size $r=1''$ ($\sim8$ kpc) is often used in photometry with ground-based telescopes for point sources, since it is comparable to their typical PSF size and hence $r=1''$ fluxes are nearly equal to total fluxes. \citet{Leclercq2017} show that the scale length ($r_s$) of the continuum-like component of LAEs is typically smaller than $1$ kpc, ensuring our assumption that LAEs are point sources.

\item[$L({\rm Ly}\alpha)_{\rm H}$]
Ly$\alpha$ luminosity of the LAH. In \citet{Leclercq2017}, it approximately corresponds to the halo component of Ly$\alpha$ luminosity. We assume that the Ly$\alpha$ luminosities falling in the annulus of $8 \le r \le 40$ kpc in \citet{Momose2016} approximately equal to $L({\rm Ly}\alpha)_{\rm H}$.  In \citet{Momose2016}, the typical $r_s$ of the stacked Ly$\alpha$ emission including the LAH component is $\sim10$ kpc, and LAHs are found to extend up to $r\sim40$ kpc. 

\item[$L({\rm Ly}\alpha)_{\rm tot}$]
Total Ly$\alpha$ luminosity. In \citet{Leclercq2017}, it corresponds to a sum of $L({\rm Ly}\alpha)_{\rm C}$ and $L({\rm Ly}\alpha)_{\rm H}$. we assume that the Ly$\alpha$ luminosities within $40$ kpc in \citet{Momose2016} approximately equal to $L({\rm Ly}\alpha)_{\rm tot}$. 
\end{description}

\begin{figure*}[ht] 
\begin{flushleft}
      \includegraphics[width=1.0\linewidth]{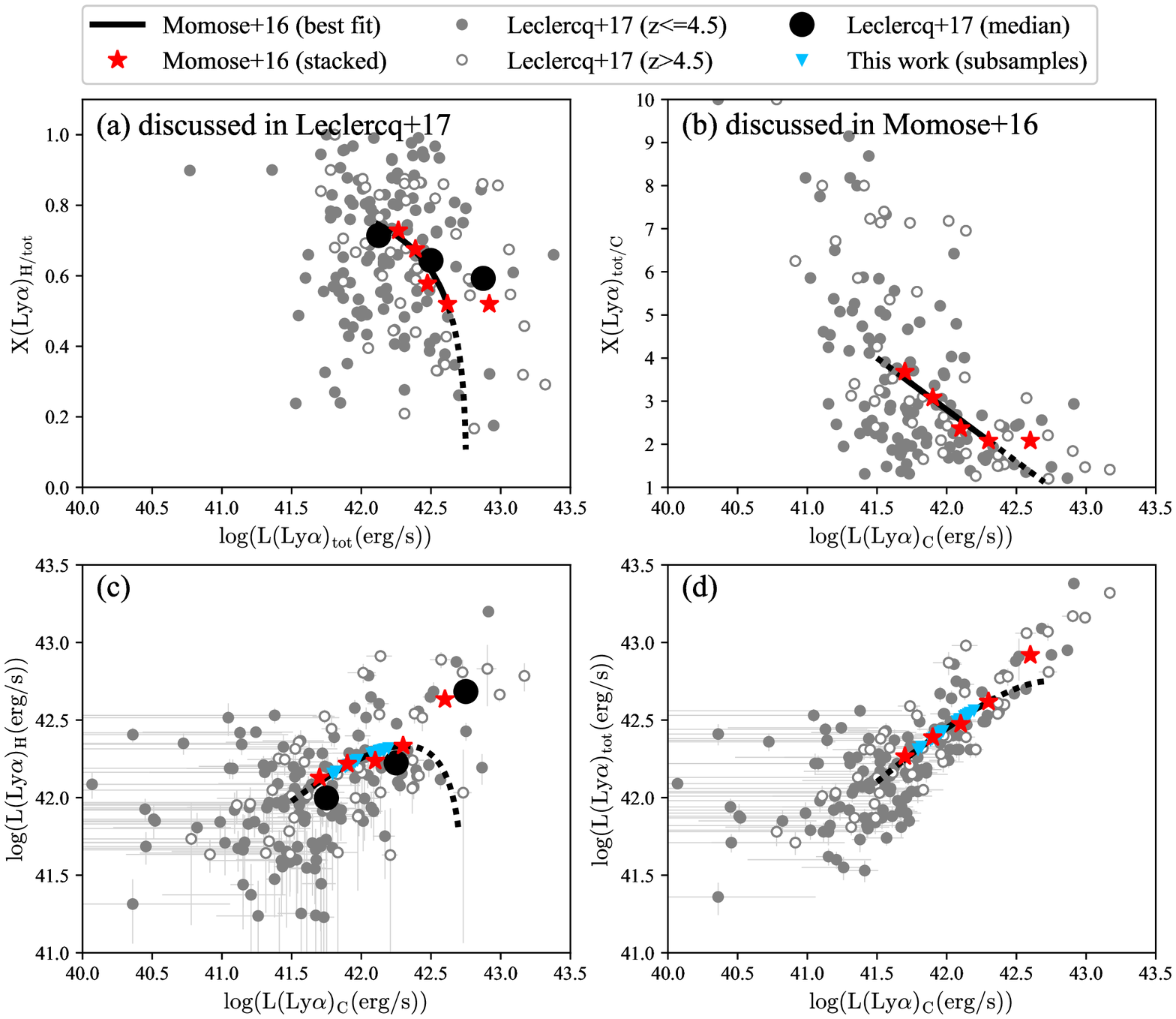}  
\end{flushleft}      
  \caption{
  Relation between $L({\rm Ly}\alpha)_{\rm H}$ and $L({\rm Ly}\alpha)_{\rm C}$ in four different presentations. 
(a) $L({\rm Ly}\alpha)_{\rm H}$/$L({\rm Ly}\alpha)_{\rm tot}$ vs. $L({\rm Ly}\alpha)_{\rm tot}$; 
(b) $L({\rm Ly}\alpha)_{\rm tot}$/$L({\rm Ly}\alpha)_{\rm C}$ vs. $L({\rm Ly}\alpha)_{\rm C}$;
(c) $L({\rm Ly}\alpha)_{\rm H}$ vs. $L({\rm Ly}\alpha)_{\rm C}$;
and (d) $L({\rm Ly}\alpha)_{\rm tot}$ vs $L({\rm Ly}\alpha)_{\rm C}$. 
Red stars and black lines indicate, respectively, the stacked results and their best-fit relation given by \citet{Momose2016}. Black solid lines and dotted lines represent the relation in a range of validity in this work and out side the range, respectively (see footnote \ref{ft:M16details} for more details). The best-fit linear relation is determined in panel (b) and is shown in equation \ref{eq:m16}. The grey filled and open circles represent MUSE--LAEs at $z\sim3-4.5$ and $z\sim4.5-6$ in \citet{Leclercq2017}, where  errors are only shown in panels (c) and (d). The black filled circles in panels (a) and (c) show the median of the MUSE--LAEs over a range of $\log_{10}[ L({\rm Ly}\alpha)_{\rm tot} {\rm (erg/s)}]\simeq42.0$--$43.0$ and $\log_{10}[ L({\rm Ly}\alpha)_{\rm C} {\rm (erg/s)}]\simeq41.5$--$43.0$, respectively.  Skyblue inverted triangles in panels (c) and (d) show the $L({\rm Ly}\alpha)_{\rm H}$ and $L({\rm Ly}\alpha)_{\rm tot}$ of our subsamples calculated from $L({\rm Ly}\alpha)_{\rm C}$ using the  stacked observational relation in \citet{Momose2016}. Note that the y-axis depends on the x-axis by construction in panels (a), (b) and (d), while panel (c) has independent axis  as described in footnote \ref{ft:lah_dependence}. (Color online)
}
  \label{fig:lah}
\end{figure*}

\citet{Momose2016} have found that LAEs with fainter $L({\rm Ly}\alpha)_{\rm C}$ have a higher $L({\rm Ly}\alpha)_{\rm tot}$ to $L({\rm Ly}\alpha)_{\rm C}$ ratio, X$(L_{\rm Ly\alpha})_{\rm tot/C}$, as shown in their figure 14. This means that the relative contribution of the halo component to the total Ly$\alpha$ luminosity increases with decreasing $L({\rm Ly}\alpha)_{\rm C}$. The best-fitting linear function between X$(L_{\rm Ly\alpha})_{\rm tot/C}$ and $L({\rm Ly}\alpha)_{\rm C}$, shown as their equation 2 is: 
\begin{equation}
{\rm X}(L_{\rm Ly\alpha})_{\rm tot/C}=103.6-2.4\,\log_{10}[{L({\rm Ly}\alpha)_{\rm C}\, {\rm(erg/s)}}]. 
\label{eq:m16}
\end{equation} 
This equation is calculated over $41.5 < \log_{10}(L({\rm Ly}\alpha)_{\rm C}) < 42.7$\footnote{They use images with the PSF matched to $1''.32$ in FWHM. Here we have corrected a typo in their equation 2 and revised the range of $\log_{10}(L({\rm Ly}\alpha)_{\rm C})$. We conclude that this equation is valid over $41.7 < \log_{10}(L({\rm Ly}\alpha)_{\rm C}) < 42.3$ from the discussions below.\label{ft:M16details} } and is shown in figure \ref{fig:lah}(b). 

\citet{Leclercq2017} have used the MUSE Hubble Ultra Deep Field survey data to detect LAHs for $145$ star forming galaxies (essentially all are LAEs) at $3\leq{z}\leq6$ individually. They have measured the size and $L({\rm Ly}\alpha)_{\rm H}$ of Ly$\alpha$ halos as well as $L({\rm Ly}\alpha)_{\rm C}$. They do not find a significant evolution of the LAH size with redshift. This result is consistent with that obtained by \citet{Momose2014} with stacked LAEs at $z\simeq2.2$--$6.6$, implying that the difference in redshift can be ignored in a comparison of the two studies. Indeed, there is no clear redshift evolution in the relations of MUSE LAEs shown by gray filled circles ($z<=4.5$) and gray open circles ($z>4.5$) in figure \ref{fig:lah} described below.

In figure \ref{fig:lah}, we compare the stacked observational relation of LAEs at $z=2.2$ in \citet{Momose2016} (black lines and red stars) with the individual results by \citet{Leclercq2017} (gray and black circles), where X$({\rm Ly}\alpha)_{x/y}$ indicates the Ly$\alpha$ luminosity ratio of the component $x$ to the component $y$. Figure \ref{fig:lah}(a) is originally discussed in \citet{Leclercq2017}, while figure \ref{fig:lah}(b) is used to determine the best-fit linear relation (equation\ref{eq:m16}) in \citet{Momose2016}. Black lines in figures \ref{fig:lah} (a), (c), and (d) are converted from one in figure \ref{fig:lah}(b). It is notable that the y-axis depends on the x-axis by construction \footnote{We regard $L({\rm Ly}\alpha)_{\rm C}$ and $L({\rm Ly}\alpha)_{\rm H}$ as two independent parameters in the measurements in \citet{Leclercq2017} and \citet{Momose2016}. Even if objects are randomly distributed in the $L({\rm Ly}\alpha)_{\rm C}$ and $L({\rm Ly}\alpha)_{\rm H}$ plane (panel [c]), we will see a 'correlation' in the other three panels because the y axis of these panels is a combination of $L({\rm Ly}\alpha)_{\rm C}$ and $L({\rm Ly}\alpha)_{\rm H}$. Instead, the y-axis of panels (a) and (b) are not affected by a flux-limited detection bias for a sample with a wide range of redshift.  \label{ft:lah_dependence}} in figures \ref{fig:lah} (a), (b) and (d). 
We find that all five stacked data points (red stars) lie in the middle of the distribution of individual MUSE-LAEs (grey circles) over a range of $\log_{10}[ L({\rm Ly}\alpha)_{\rm C} {\rm (erg/s)}]\simeq41.7$--$42.6$ or $\log_{10}[ L({\rm Ly}\alpha)_{\rm tot} {\rm (erg/s)}]\simeq42.3$--$42.8$. It is also found that the median values of individual MUSE-LAEs (black filled circles in figures \ref{fig:lah} [a] and [c]) are located near the stacked values. 
This means that the stacked results represent the average halo luminosities of LAEs despite the fact that there is a great variation in halo luminosity among objects. The best-fit relation shown by a black line traces well the stacked points except for the brightest one. This is because the brightest point already deviates from the best-fit linear relation determined in figure \ref{fig:lah}(b) while the other four are on the relation. Based on figure \ref{fig:lah}(a), \citet{Leclercq2017} have concluded that there is no significant correlation between $L({\rm Ly}\alpha)_{\rm tot}$ and X$({\rm Ly}\alpha)_{\rm H/tot}$ on the basis of a Spearman rank correlation coefficient of $-0.05$ (see their figure 7 and their section 5.3.1). Although the existence of a correlation is not clear, and a further test is needed, figures \ref{fig:lah} (a) and (c) indicate that the stacked results (red stars) also trace the median trend of individual MUSE LAEs (black filled circles).

In this work, we estimate average $L({\rm Ly}\alpha)_{\rm H}$ and $L({\rm Ly}\alpha)_{\rm tot}$ for each subsample from the stacked relation (equation\ref{eq:m16}) as well as average $L({\rm Ly}\alpha)_{\rm C}$ by multiplying average $L({\rm Ly}\alpha)_{ps}$ (in section \ref{subsec:ulya}) by $0.77$ as an inverse aperture correction of $1''.32$ PSF (see table \ref{tbl:Lya} in appendix). The $L({\rm Ly}\alpha)_{\rm C}$ values of our subsamples are found to be within the range shown by skyblue inverted triangles in figures \ref{fig:lah} (c) and (d) where the stacked relation traces well the stacked points.  The typical $1\sigma$ uncertainties in the individual data points in \citet{Momose2016} are propagated to uncertainties in $L({\rm Ly\alpha})_{\rm H}$ and $L({\rm Ly\alpha})_{\rm tot}$ of $\sim22\%$ and $\sim16\%$, respectively.  \citet{Momose2016} also present a stacked relation (anti-correlation) between ${\rm X}(L_{\rm Ly\alpha})_{\rm tot/C}$ and $EW_{0,ps}(L_{\rm Ly\alpha})$. Using this relation instead of equation \ref{eq:m16} gives nearly the same $L({\rm Ly\alpha})_{\rm H}$ and $L({\rm Ly\alpha})_{\rm tot}$ values (see appendices \ref{sec:appendix_LAH} and \ref{sec:appendix_Llya}).

\subsection{SED fitting}\label{sec:sed}
We derive parameters that characterize the stellar populations of our subsamples in the two fields by fitting SEDs based on stacked multiband images. We use $810$ LAEs ($\sim91$\% of the entire sample, $891$) that have data in all ten broadband filters ($B, V, R, i, z, J, H, K, {\rm ch1}$, and ${\rm ch2}$). To obtain secure IRAC photometry, some prescriptions are adopted in previous studies \citep[e.g.,][]{Vargas2014, Kusakabe2018a, Malkan2017}. In this paper, we follow \citet{Kusakabe2018a} and only use LAEs that are not contaminated by other objects in the ch1 and ch2 images. To do so, we exclude LAEs that have either one or more neighbors or a high sky background through a two-step cleaning process. We are thus left with $121$ LAEs for stacking of ch1 and ch2 images \citep[see section 4.1 in ][for more detail]{Kusakabe2018a}. We briefly describe stacking analysis, photometry, and SED models below. A detailed description can be found in \citet{Kusakabe2018a}. 

\subsubsection{Stacking Analysis and Photometry}\label{subsec:stack_photo}
For each band, we use the task IRAF/imcombine to create a median-stacked image at the NB387 source positions from images of size 50$''$ $\times$ 50$''$ that are cut out with IRAF/imcopy task. While a stacked SED is not necessarily a good representation of individual objects \citep{Vargas2014}, stacking is still useful for our faint objects to obtain a rest-frame UV to NIR SED. 

An aperture flux is measured for each stacked image using the task PyRAF/phot with the same parameters in \citet{Kusakabe2018a}. We use an aperture diameter of $2''$ for the $NB387$, optical, and NIR band images and $3''$ for the MIR (IRAC) images following \citet{Ono2010b}. For each of the ch1 and ch2 images, we obtain the net $3''$-aperture flux density by subtracting an offset of the sky background as described in section 4.2 of \citet{Kusakabe2018a}. All aperture magnitudes are corrected for Galactic extinction, $\rm{E(B-V)_b}$, of $0.020$ and $0.018$ for the SXDS and COSMOS fields, respectively \citep{Schlegel1998}. 

The aperture magnitudes are converted into total magnitudes using the aperture correction values summarized in table \ref{tbl:data}. The 1$\sigma$ uncertainty in the total magnitudes is the sum of the errors in photometry, aperture correction, and the ZP. For the ch1 and ch2 data, the errors in sky subtraction are also included. The photometric errors are determined in the same procedure as \citet{Kusakabe2015}. The aperture correction errors in the $NB387$, optical, and NIR bands are estimated to be $0.03$ mag, and those in the ch1 and ch2 bands are set to $0.05$ mag. The ZP errors for all bands are set to be $0.1$ mag. The stacked SEDs thus obtained for individual subsamples are shown in figures \ref{fig:sed_sxds} and \ref{fig:sed_cosmos} in appendix.

\subsubsection{SED models}\label{subsec:sed_model}
We perform SED fitting on the stacked SEDs with model SEDs in the same manner as in \citet{Kusakabe2018a}. The model SEDs are constructed by adding nebular emission (lines and continuum) to the stellar population synthesis model GALAXEV \citep{Bruzual2003, Ono2010b}. We assume constant star formation history, 0.2$Z_{\odot}$ stellar metallicity, and $\rm{E(B-V)}_{\rm gas}=\rm{E(B-V)}_{\star}$ \citep{Erb2006b} following previous SED studies of LAEs \citep[e.g.,][]{Ono2010b, Vargas2014}. We also assume an SMC-like dust extinction model for the attenuation curve \citep[hereafter an SMC-like attenuation curve;][]{Gordon2003} since it is suggested to be more appropriate for LAEs at $z\sim2$ and low-mass star forming galaxies at $z\geq2$ than the Calzetti curve \citep{Calzetti2000} in \citet{Kusakabe2015} and \citet[][]{Reddy2018}, respectively. The Lyman continuum escape fraction, $f^{\rm ion}_{\rm esc}$, is fixed to $0.2$ \citep{Nestor2013}. We also examine the case of the Calzetti attenuation curve for comparison with previous studies and conservative discussion. The case without nebular emission ($f^{\rm ion}_{\rm esc}=1$) has been examined and discussed in \citet{Kusakabe2018a}.

We search for the best-fitting model SED to the stacked SED of each subsample that minimizes $\chi^2$ and derive the following stellar parameters: stellar mass ($M_{\star}$), color excess ($\rm{E(B-V)}_{\star}$), age, and $SFR$. The stellar mass is calculated by solving $\frac{\partial \chi^2}{\partial M_{\star}} = 0$, while the $SFR$ is determined from {$M_{\star}$} and age. Thus, the degree of freedom is $7$. The 1$\sigma$ confidence interval in each stellar parameter is obtained from the range of the values giving  $\chi^2\leq\chi^2_{\rm min} +1$, where $\chi^2_{\rm min}$ is the minimum $\chi^2$ value. Figures \ref{fig:sed_sxds} and \ref{fig:sed_cosmos} in appendix shows the best-fit SEDs and table \ref{tbl:sed_field} in appendix summarize the results of the best-fit parameters in the two fields. The field-average values are shown in table \ref{tbl:sed_average}.

\begin{table*}
\tbl{ The field-average values of stellar parameters, $f_{\rm esc}(\rm Ly\alpha)_{\rm tot}$, and the $q$-parameter.
}{
\begin{tabular}{l|cccc|cc}
\hline
subsample &$M_{\star}$ & $E(B-V)_{\star}$ & Age & $SFR$ & $f_{\rm esc}(\rm Ly\alpha)_{\rm tot}$ & $q$-parameter \\ 
 & ($10^8 M_{\odot}$) & (mag) & (Myr) & ($M_{\odot}$yr$^{-1}$) & & \\
 & (1) & (2) & (3) & (4) & (5) & (6) \\\hline
 \multicolumn{7}{c}{SMC-like attenuation curve } \\\hline
bright UV & $14.1\pm2.1$ & $0.08\pm0.01$ & $ 240\pm  14$ &  $ 6.8\pm 1.3$ & $0.37\pm0.00$ & $0.80^{+0.11}_{-0.09}$ \\
faint UV & $4.1\pm0.1$ & $0.03\pm0.01$ & $ 280\pm  46$ &  $ 1.7\pm 0.3$ &   $1.43\pm0.17$ & $-0.69^{+0.30}_{-0.62}$ \\  
blue $\beta$ & $4.8\pm2.4$ & $0.02\pm0.00$ & $ 246\pm 145$ &  $ 2.1\pm 0.0$ & $1.21\pm0.10$ & $-0.52^{+0.25}_{-0.27}$ \\ 
red $\beta$ & $14.0\pm0.9$ & $0.10\pm0.01$ & $ 286\pm   0$ &  $ 5.8\pm 0.4$ & $0.43\pm0.08$ & $0.57^{+0.14}_{-0.11}$ \\  
bright Ly$\alpha$ & $7.4\pm0.8$ & $0.04\pm0.02$ & $ 346\pm  80$ &  $ 2.2\pm 0.7$ &$1.20\pm0.35$ & $-0.28^{+0.57}_{-0.57}$ \\ 
faint Ly$\alpha$ & $12.3\pm1.0$ & $0.07\pm0.01$ & $ 360\pm   0$ &  $ 4.2\pm 0.3$ & $0.49\pm0.00$ & $0.64^{+0.10}_{-0.08}$ \\  
large EW & $5.4\pm1.6$ & $0.04\pm0.02$ & $ 338\pm  19$ &  $ 1.8\pm 0.6$ & $1.34\pm0.42$ & $-0.46^{+0.60}_{-0.71}$ \\ 
small EW & $13.7\pm3.4$ & $0.07\pm0.01$ & $ 353\pm  40$ &  $ 5.0\pm 0.7$ & $0.42\pm0.02$ & $0.79^{+0.14}_{-0.11}$ \\  
bright $K$ & $18.3\pm2.2$ & $0.09\pm0.01$ & $ 265\pm  84$ &  $ 6.5\pm 1.2$ & $0.36\pm0.01$ & $0.72^{+0.09}_{-0.08}$ \\ 
faint $K$ &  $3.6\pm0.4$ & $0.04\pm0.01$ & $ 160\pm  44$ &  $ 2.3\pm 0.2$ & $1.03\pm0.04$ & $-0.04^{+0.06}_{-0.07}$ \\\hline
 \multicolumn{7}{c}{the Calzetti attenuation curve } \\\hline
bright UV & $12.9\pm1.6$ & $0.15\pm0.02$ & $ 118\pm  21$ &  $11.7\pm 3.4$ & $0.20\pm0.02$ & $0.96^{+0.16}_{-0.12}$ \\ 
faint UV & $2.9\pm0.3$ & $0.10\pm0.03$ & $  73\pm  37$ &  $ 3.3\pm 1.3$ & $0.74\pm0.25$ & $0.27^{+0.46}_{-0.26}$ \\  
blue $\beta$ & $3.4\pm2.4$ & $0.06\pm0.02$ & $ 106\pm 112$ &  $ 2.9\pm 0.6$ & $0.87\pm0.08$ & $0.21^{+0.21}_{-0.14}$ \\  
red $\beta$ & $13.7\pm2.6$ & $0.18\pm0.00$ & $ 133\pm  30$ &  $11.8\pm 0.3$ & $0.21\pm0.02$ & $0.78^{+0.05}_{-0.05}$ \\ 
bright Ly$\alpha$ & $4.2\pm0.6$ & $0.14\pm0.05$ & $  39\pm  24$ &  $ 6.1\pm 4.2$ & $0.43\pm0.29$ & $0.55^{+1.06}_{-0.34}$ \\ 
faint Ly$\alpha$ & $12.0\pm1.2$ & $0.14\pm0.02$ & $ 189\pm  11$ &  $ 7.1\pm 1.1$ & $0.27\pm0.02$ & $0.84^{+0.15}_{-0.11}$ \\ 
large EW & $3.7\pm0.8$ & $0.14\pm0.03$ & $  60\pm  11$ &  $ 4.9\pm 2.6$ & $0.50\pm0.24$ & $0.46^{+0.51}_{-0.26}$ \\
small EW & $13.2\pm3.6$ & $0.14\pm0.02$ & $ 191\pm  11$ &  $ 8.8\pm 1.9$ & $0.24\pm0.03$ & $0.92^{+0.18}_{-0.13}$ \\ 
bright $K$ & $11.2\pm2.7$ & $0.20\pm0.02$ & $  46\pm  24$ &  $17.9\pm 6.4$ & $0.13\pm0.02$ & $0.93^{+0.14}_{-0.11}$ \\ 
faint $K$ &  $2.3\pm0.9$ & $0.11\pm0.03$ & $  32\pm  25$ &  $ 4.1\pm 1.8$ & $0.56\pm0.18$ & $0.49^{+0.46}_{-0.25}$ \\\hline
\end{tabular}
}\label{tbl:sed_average}
\tabnote{Note. (1) Stellar mass, (2) color excess, (3) age, (4) $SFR$, (5) $f_{\rm esc}(\rm Ly\alpha)_{\rm tot}$ calculated from $SFR$ and $L({\rm Ly\alpha})_{\rm tot}$, and (6) $q$ calculated from $f_{\rm esc}(\rm Ly\alpha)_{\rm tot}$ and $E(B-V)_{\star}$.
} 
\end{table*}

%

\subsection{Clustering analysis}\label{sec:cl}
We derive the angular two-point correlation functions (ACFs) of our subsamples from clustering analysis and convert the correlation lengths into bias factors and then into dark matter halo masses in the same manner as in \citet{Kusakabe2018a}. We briefly describe our methods below. 

\subsubsection{Angular correlation function}\label{subsec:acf}
The ACF, {\wobs}, for a given subsample is measured by the calculator given in \citet{Landy1993}: 
\begin{equation}
\omega_{\rm obs}(\theta)=\frac{DD(\theta)-2DR(\theta)+RR(\theta)}{RR(\theta)},
\end{equation}
where DD($\theta$), RR($\theta$), and DR($\theta$) are the normalized numbers of galaxy-galaxy, galaxy-random, and random-random pairs, respectively. We use a random sample composed of $100,000$ sources with the same geometrical constraints as the data sample. The sky distributions of the LAEs and the random sources in the two fields are shown in figure 2 of \citet{Kusakabe2018a}. Following \citet{Guaita2010}, the $1\,\sigma$ uncertainties in ACF measurements are estimated as: 
\begin{equation}
\Delta\omega_{\rm obs}(\theta)=\frac{1+\omega(\theta)}{\sqrt{DD_0(\theta)}}, 
\label{eq:e_LS1993}
\end{equation}
where $DD_0(\theta)$ is the raw number of galaxy-galaxy pairs. 

We approximate the spatial correlation function of LAEs by a power law: 
\begin{equation}
\xi(r)=\left(\frac{r}{r_0}\right)^{-\gamma},
\end{equation}
where $r$, $r_0$, and $\gamma$ are the spatial separation between two objects in comoving scale, the correlation length, and the slope of the power law, respectively \citep{Totsuji1969, Zehavi2004}. We then convert $\xi(r)$ into the ACF, $\omega_{\rm model}(\theta)$, following \citet{Simon2007}, and describe it as:
\begin{equation}
\omega_{\rm model}(\theta) 
= \left(\frac{r_0 \ h^{-1}_{100}{\rm Mpc}}{1\ h^{-1}_{100}{\rm Mpc}}\right)^{\gamma}\, \omega_{\rm model,\, 0}(\theta),
\end{equation}
where $\omega_{\rm model,\, 0}(\theta)$ is the ACF in the case of $r_0= 1 \ h^{-1}_{100}{\rm Mpc}$. The correlation amplitude of the ACF at $\theta=1''$, $A_{\omega}$, is $\omega_{\rm model}(\theta=1'') $.

An observationally obtained ACF, $\omega_{\rm obs}(\theta)$, includes an offset due to the fact that the measurements are made over a limited area.  
This offset is given by the integral constraint (IC), 
\begin{equation}
\omega(\theta)=\omega_{\rm obs}(\theta)+IC,
\end{equation}
\begin{equation}
IC= \frac{\Sigma_{\theta} RR(\theta)\,\omega_{\rm model,\, 0}(\theta)}{\Sigma_{\theta} RR(\theta)}\left(\frac{r_0 \ h^{-1}_{100}{\rm Mpc}}{1\ h^{-1}_{100}{\rm Mpc}}\right)^{\gamma}\,,
\end{equation}
where $\omega(\theta)$ is the true ACF. We fix $\gamma$ to the fiducial value $1.8$ following previous studies \citep[e.g., ][]{Ouchi2003} and fit the $\omega_{\rm model}(\theta)$ to this $\omega(\theta)$ by minimizing $\chi^2$ over $\sim40''$ $-1000''$, where we avoid the one-halo term at small scales and large sampling noise at large scales. The best-fit field-average correlation amplitude, $\Ao$, is calculated analytically by minimizing the summation of $\chi^2$ over the two fields in the same manner as in \citet{Kusakabe2018a}. The $1\,\sigma$ fitting error in $A_{\omega}$, $\Delta \Ao$, is estimated from $\chi^2_{\rm min} +1$, where $\chi^2_{\rm min}$ is the minimum $\chi^2$ value. 

The correlation amplitude corrected for randomly distributed foreground and background interlopers, $\Aocor$, is given by
\begin{equation}
\Aocor = \frac{A_{\omega}}{(1-f_{\rm c})^2},
\end{equation}
where  $f_{\rm c}$ is the contamination fraction. The contamination fraction of our LAEs is estimated to be $10\,\pm\,10$\% ($0$--$20$\%) conservatively (see section \ref{subsec:selection}) and the error range in $\Aocor$ includes both the no correction case and the maximum correction case \citep[e.g.,][]{Khostovan2018}. The $1\,\sigma$ error in the contamination-corrected correlation amplitude, $\Delta \Aocor$, is derived from error propagation of $\Ao$ and $f_{\rm c}$:
\begin{equation}
\frac{\Delta\Aocor }{\Aocor} \simeq \sqrt{\left(\frac{\Delta\Ao}{\Ao}\right)^2 + \left(\frac{2\Delta f_{\rm c}}{f_{\rm c}}\right) ^2  },
\label{eq:delta_a}
\end{equation}
where $\Delta f_{\rm c}(= 0.1)$ is the uncertainty in the contamination estimate. The value of the contamination-corrected correlation length, $r_{\rm 0,\,corr}$ and its $1\,\sigma$ error are calculated from $\Aocor$ and $\Delta \Aocor$. Figure \ref{fig:ACF} in appendix shows the best-fit ACFs and table \ref{tbl:acf} summarizes the results of the clustering analysis.

\subsubsection{Bias factor and dark matter halo mass}\label{subsec:bias}
The galaxy-matter bias, $b_{\rm g}$, is defined as 
 \begin{equation}
 b_{\rm g}(r)=\sqrt{\frac{\xi(r)}{\xi_{\rm DM}(r,z)}}, 
 \end{equation}
 where $\xi_{\rm DM}(r,z)$ is the spatial correlation function of underlying dark matter calculated with the linear dark matter power spectrum \citep{Eisenstein1998, Eisenstein1999}. We estimate the effective galaxy-matter bias, $b_{\rm g,\, eff}$, at $r=8\, h^{-1}_{100}{\rm Mpc}$ following previous clustering analyses \citep[e.g., ][]{Ouchi2003} using a suite of cosmological codes called Colossus \citep{Diemer2015}. The obtained $b_{\rm g,\, eff}$ is converted into the peak height in the linear density field, $\nu$, by the formula given in \citet{Tinker2010}. The effective dark matter halo mass is derived from $\nu$ with the top-hat window function and the linear dark matter power spectrum \citep{Eisenstein1998, Eisenstein1999} using a cosmological package for Python called CosmoloPy\footnote{http://roban.github.com/CosmoloPy/}. The effective bias and the effective halo mass of each subsample is listed in table \ref{tbl:acf}. 

\begin{table*}
\tbl{Clustering Measurements for the eight subsamples.
}{
\begin{tabular}{lcccccc}
\hline
subsamples & $A_{\omega}$ &   $A_{\rm \omega,\, corr}$  &  $r_{\rm 0,\, corr}$ &  $b_{\rm g,\,eff}$  &$M_{\rm h}$ & reduced $\chi^2$\\
 &  &     &  $(\mathrm{h^{-1}_{100}Mpc})$ &   &$(\mathrm{\times10^{10}\ {\rm M_{\odot}} })$ & \\
& (1) & (2) & (3) & (4) & (5) & (6) \\
\hline 
bright UV & 1.03 $\pm$  0.82 & 1.28 $\pm$  1.05 &$ 1.20^{+ 0.48}_{- 0.74}$& $ 0.66^{+ 0.23}_{- 0.38}$& $ <0.2^{(7)}$& 1.46\\
faint UV & 3.65 $\pm$  1.25 & 4.51 $\pm$  1.84 &$ 2.42^{+ 0.51}_{- 0.61}$& $ 1.23^{+ 0.23}_{- 0.29}$& $ 4.4^{+ 8.8}_{- 4.0}$& 1.34\\ 
blue $\beta$ & 1.12 $\pm$  0.74 & 1.38 $\pm$  0.97 &$ 1.25^{+ 0.43}_{- 0.61}$& $ 0.68^{+ 0.21}_{- 0.31}$& $<0.2^{(7)}$& 0.91\\ 
red $\beta$ & 4.29 $\pm$  1.37 & 5.29 $\pm$  2.06 &$ 2.65^{+ 0.53}_{- 0.63}$& $ 1.34^{+ 0.24}_{- 0.29}$& $ 7.6^{+12.4}_{- 6.5}$& 0.52\\ 
bright Ly$\alpha$ &  3.96 $\pm$  1.29 & 4.89 $\pm$  1.93 &$ 2.53^{+ 0.51}_{- 0.62}$& $ 1.29^{+ 0.23}_{- 0.29}$& $ 5.9^{+10.4}_{- 5.1}$& 0.85\\ 
faint Ly$\alpha$ &  5.39 $\pm$  1.27 & 6.65 $\pm$  2.16 &$ 3.00^{+ 0.51}_{- 0.59}$& $ 1.50^{+ 0.23}_{- 0.27}$& $15.2^{+16.8}_{-10.8}$& 1.81\\ 
large EW &  3.27 $\pm$  1.27 & 4.04 $\pm$  1.81 &$ 2.28^{+ 0.52}_{- 0.64}$& $ 1.17^{+ 0.24}_{- 0.30}$& $ 3.0^{+ 7.4}_{- 2.8}$& 0.64\\
small EW &  4.90 $\pm$  1.26 & 6.05 $\pm$  2.05 &$ 2.85^{+ 0.50}_{- 0.59}$& $ 1.43^{+ 0.23}_{- 0.27}$& $11.5^{+14.3}_{- 8.7}$& 1.75\\ 
\hline
\end{tabular}
}
\label{tbl:acf}
\tabnote{Note. (1) Correlation amplitude without contamination correction; (2) contamination-corrected correlation amplitude used to derive (3)--(5); (3) correlation length; (4) effective bias factor, (5) dark matter halo mass; and (6) reduced $\chi^2$ value. (7) $1\,\sigma$ upper limit of $M_{\rm h}$ (see appendix \ref{sec:appendix_acf}). The field-average best fit values are calculated from equation 13 in \citet{Kusakabe2018a}. 
}
\end{table*}

\section{Results}\label{sec:results}

\begin{figure*}[ht]
\begin{flushleft}    
      \includegraphics[width=0.95\linewidth]{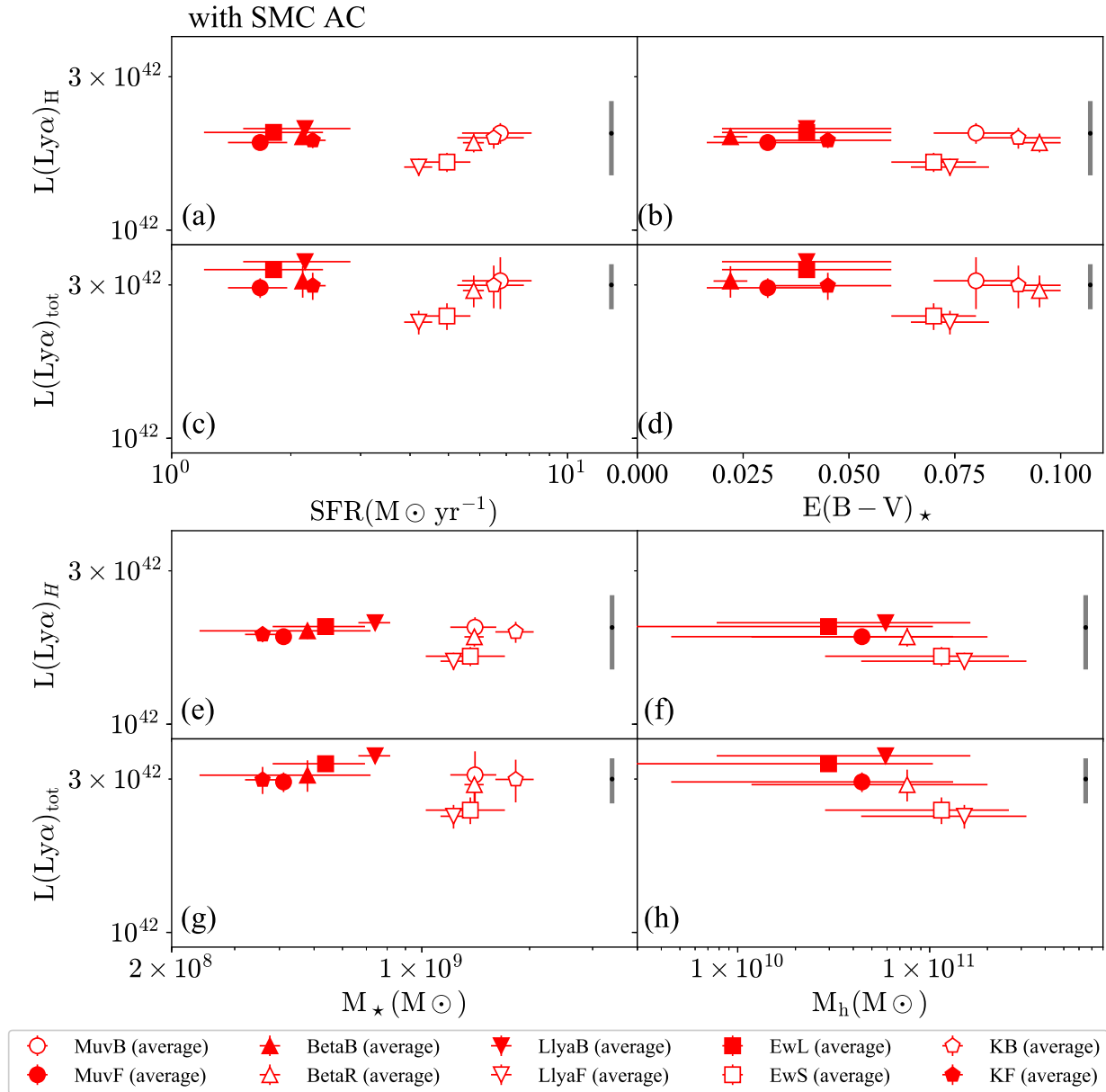}
\end{flushleft}      
  \caption{
$L({\rm Ly\alpha})_{\rm H}$ and $L({\rm Ly\alpha})_{\rm tot}$ as a functions of stellar parameters and dark matter halo mass for an SMC curve: (a) $L({\rm Ly\alpha})_{\rm H}$ vs. $SFR$, (b) $L({\rm Ly\alpha})_{\rm H}$ vs. $E(B-V)_\star$, (c) $L({\rm Ly\alpha})_{\rm tot}$ vs. $SFR$, (d) $L({\rm Ly\alpha})_{\rm tot}$ vs. $E(B-V)_\star$, (e) $L({\rm Ly\alpha})_{\rm H}$ vs. $M_\star$, (f) $L({\rm Ly\alpha})_{\rm H}$ vs. $M_{\rm h}$, (g) $L({\rm Ly\alpha})_{\rm tot}$ vs. $M_\star$, and (h) $L({\rm Ly\alpha})_{\rm tot}$ vs. $M_{\rm h}$. 
All values are field average values. Different symbols indicate different subsamples: 
open (filled) circles for bright (faint) $M_{\rm UV}$, 
open (filled) triangles for red (blue) $\beta$, 
open (filled) inverted triangles for faint (bright) $L({\rm Ly\alpha})_{ps}$, 
open (filled) squares for small (large) $EW_{0,ps}({{\rm Ly}\alpha})$, 
and open (filled) pentagons for bright (faint) $m_{\rm K}$. 
The typical $1\sigma$ uncertainties in the individual data in \citet{Momose2016} are propagated to uncertainties in $L({\rm Ly\alpha})_{\rm H}$ and $L({\rm Ly\alpha})_{\rm tot}$ of $\sim22\%$ and $\sim16\%$, respectively. Gray error bars of black dots in panels (a)--(d) and (e)--(f) show those uncertainties at $L({\rm Ly\alpha})_{\rm H}=2\times10^{42}\,{\rm erg\,s^{-1}}$ and 
$L({\rm Ly\alpha})_{\rm tot}=3\times10^{42}\,{\rm erg\,s^{-1}}$, respectively. The vertical error bars of the red symbols are derived from the fitting errors in $L({\rm Ly}\alpha)_{ps}$. 
$M_{\rm h}$ are not calculated for the $m_K$-divided subsamples. The $M_{\rm h}$ values for the bright $M_{\rm UV}$ and blue $\beta$ subsamples are not shown because they are not constrained well owing to too weak clustering signals. (Color online)
}
  \label{fig:Lh}
\end{figure*}

\begin{figure*}[ht]
      \includegraphics[width=0.75\linewidth]{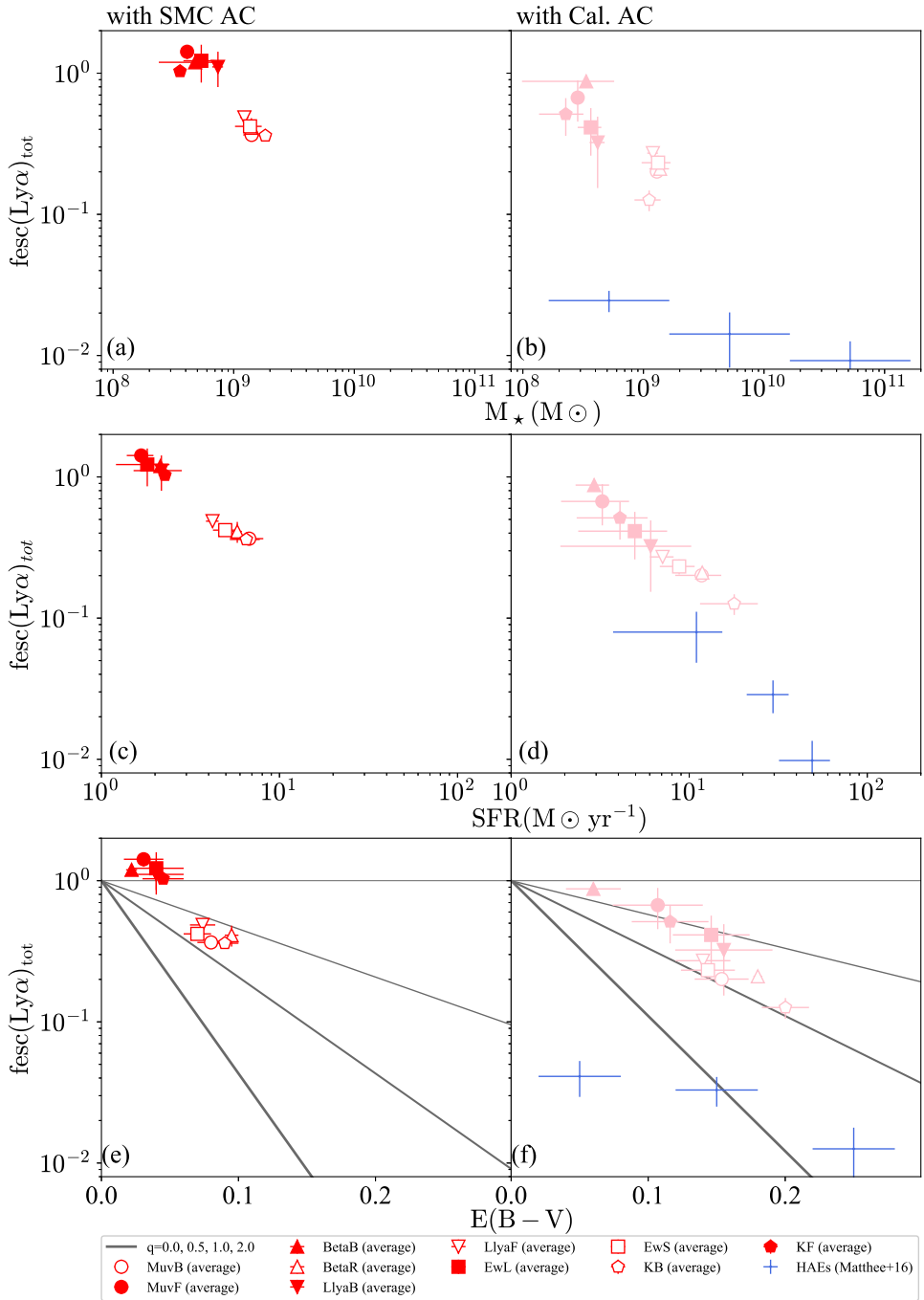}
  \caption{
    $fesc({\rm Ly\alpha})_{\rm tot}$ as a functions of $M_\star$ (panels [a] and [b]), $SFR$ ([c] and [d]), and $E(B-V)$ ([e] and [f]) for the two attenuation curves. All values are field average values. Different symbols indicate different subsamples: 
open (filled) circles for bright (faint) $M_{\rm UV}$, 
open (filled) triangles for red (blue) $\beta$, 
open (filled) inverted triangles for faint (bright) $L({\rm Ly\alpha})_{ps}$, 
open (filled) squares for small (large) $EW_{0,ps}({{\rm Ly}\alpha})$, 
and open (filled) pentagons for bright (faint) $m_{\rm K}$. 
Blue crosses indicate HAEs in \citet{Matthee2016}, whose Ly$\alpha$ luminosities are derived from $6''$ aperture photometry. Dark gray solid lines show models for four different $q$ values, $q=0.0, 0.5, 1.0$, and $2.0$ with increasing thickness. Stellar parameters are derived with the assumption of $E(B-V)_\star=E(B-V)_{g}$. (Color online)
}
  \label{fig:fesc}
\end{figure*}

\begin{figure*}[ht]
\begin{flushleft}
      \includegraphics[width=0.95\linewidth]{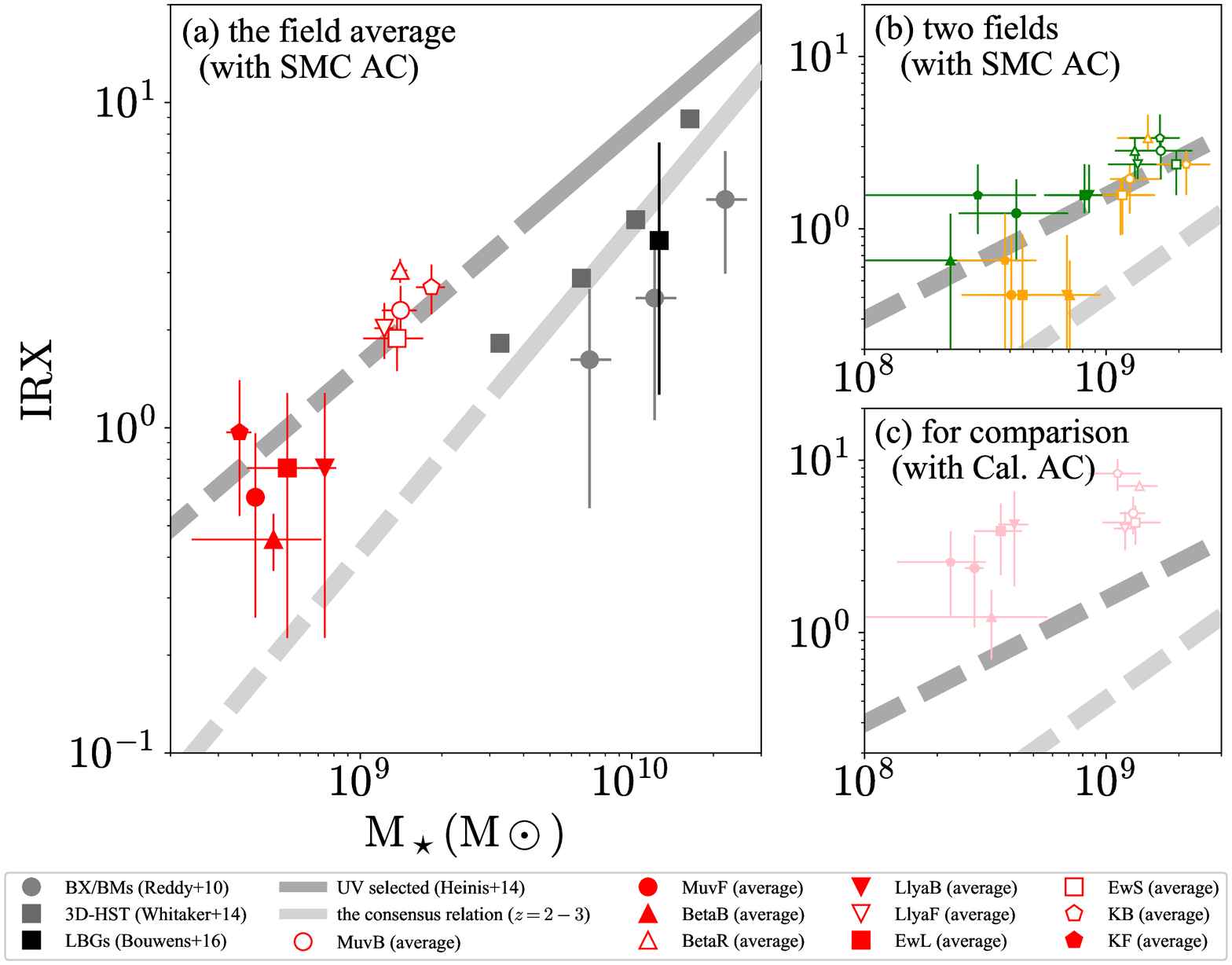}
\end{flushleft}      
  \caption{
$IRX$ vs.$M_{\star}$. (a) Field average values of our ten subsamples with an assumption of an SMC-like attenuation curve (red symbols),
(b) results before averaging (green and orange symbols), and (c) field average values with an assumption of a Calzetti curve (pink symbols), plotted with some literature results. In panels (a) and (c), different subsamples are shown by different symbols: 
open (filled) circles for bright (faint) $M_{\rm UV}$, 
open (filled) triangles for red (blue) $\beta$, 
open (filled) inverted triangles for faint (bright) $L({\rm Ly\alpha})_{ps}$, 
open (filled) squares for small (large) $EW_{0,ps}({{\rm Ly}\alpha})$, 
and open (filled) pentagons for bright (faint) $m_{\rm K}$.
Dark gray squares, dark gray circles, a black square, a dark gray solid line and a light gray solid line represent, respectively, 3D-HST galaxies at $z\sim2$ in \citet{Whitaker2014}, UV selected galaxies at $z\sim2$ in \citet{Reddy2010}, LBGs at $z\sim2-3$ in \citet{Bouwens2016}, UV-selected galaxies at $z\sim1.5$ in \citet{Heinis2014} and the consensus relation of them determined by \citet{Bouwens2016}. Dark and light gray dashed lines indicate extrapolations of gray solid lines. In panel (b), orange and green symbols indicate, respectively, the SXDS and COSMOS subsamples with an SMC-like attenuation curve (with SMC AC). All data are rescaled to a Salpeter IMF according to footnote \ref{ft:imf}. (Color online)
}
  \label{fig:IRX}
\end{figure*}

\begin{figure*}[ht]
\begin{flushleft}
      \includegraphics[width=0.95\linewidth]{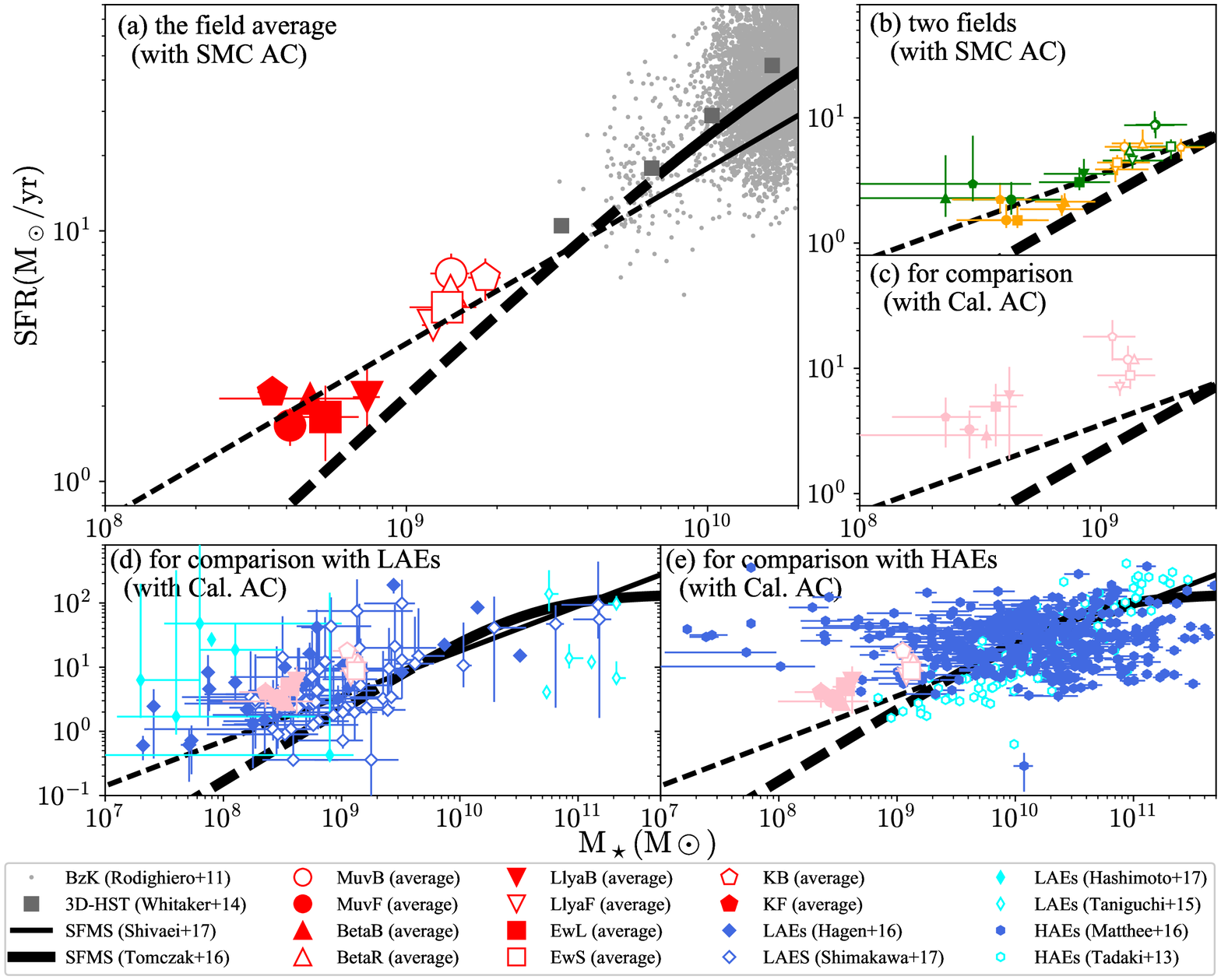}
\end{flushleft}      
  \caption{
$SFR$ vs. $M_{\star}$. 
(a) Field average values of our ten subsamples with an SMC-like attenuation curve (red symbols),
(b) results before averaging (green and orange symbols), and (c)-(e) field average values with a Calzetti curve (pink symbols), plotted with some literature results. In panels (a) and (c)--(e), different subsamples are shown by different symbols: 
open (filled) circles for bright (faint) $M_{\rm UV}$, 
open (filled) triangles for red (blue) $\beta$, 
open (filled) inverted triangles for faint (bright) $L({\rm Ly\alpha})_{ps}$, 
open (filled) squares for small (large) $EW_{0,ps}({{\rm Ly}\alpha})$, 
and open (filled) pentagons for bright (faint) $m_{\rm K}$.
In panel (b), orange and green symbols indicate, respectively, the SXDS and COSMOS subsamples with an SMC-like attenuation curve (with SMC AC). In panels (c)--(e), pink symbols show the average values of the subsamples over the two fields with a Calzetti attenuation curve (with Cal. AC). 
Dark gray squares, light gray dots, thick black solid lines, and thin black solid lines represent, respectively, 3D-HST galaxies at $z\sim2$ in \citet{Whitaker2014}, BzK galaxies at $z\sim2$ in \citet{Rodighiero2011}, the SFMS at $z\sim2$ in \citet{Tomczak2016}, and the SFMS at $z\sim2$ in \citet{Shivaei2017}. Thick and thin black dashed lines indicate extrapolations of the black solid lines. 
In panel (d), filled blue diamonds, open blue diamonds, filled cyan thin diamonds, and open cyan thin diamonds indicate LAEs at $z\sim2$--$3$ in \citet{Hagen2016, Shimakawa2017a,Hashimoto2017a} and \citet{Taniguchi2015}, respectively. In panel (e), filled blue hexagons and open cyan hexagons show HAEs at $z\sim2$--$3$ in \citet{Matthee2016} and \citet{Tadaki2013}. 
All data are rescaled to a Salpeter IMF according to footnote \ref{ft:imf}. (Color online)
}
  \label{fig:MS}
\end{figure*}

The field-average results of the SED fitting and clustering analysis are shown in tables \ref{tbl:sed_average} and \ref{tbl:acf}, respectively. In sections \ref{subsec:r_lah} and \ref{subsec:r_fesc}, we focus on their LAH luminosities and Ly$\alpha$ escape fractions, respectively. In section \ref{subsec:r_sfirx}, we compare the infrared excess ($IRX$) and star formation mode of our subsamples with the average relations of star forming galaxies and examine whether they are normal galaxies in terms of these two properties, which will be employed in the discussion of the origin of LAHs in section \ref{sec:d_lah}.

\subsection{Halo and total Ly$\alpha$ luminosities}\label{subsec:r_lah}
Figure \ref{fig:Lh} plots $L({\rm Ly}\alpha)_{\rm H}$ and $L({\rm Ly}\alpha)_{\rm tot}$ against $SFR$, $E(B-V)_\star$, $M_\star$, and $M_{\rm h}$. The ten subsamples have similar $L({\rm Ly}\alpha)_{\rm H}$ of $\sim 2\times10^{42}\,{\rm erg\,s^{-1}}$, and similar $L({\rm Ly}\alpha)_{\rm tot}$ of $\sim 2\times10^{42}$--$4\times10^{42}\,{\rm erg\,s^{-1}}$ within a factor of 1.5 (see also table \ref{tbl:Lya} in appendix). This is expected from the small difference in $L({\rm Ly}\alpha)_{ps}$ between the subsamples as described in the next paragraph. What we newly find is that $L({\rm Ly}\alpha)_{\rm H}$ and $L({\rm Ly}\alpha)_{\rm tot}$ remain almost unchanged when $M_\star$ increases by factor $2$--$5$. This has not been confirmed with SED fitting (including nebular emission in models). 

The nearly constant (or even slightly decreasing) $L({\rm Ly}\alpha)_{\rm H}$ against $M_\star$ is a result of two competing trends. One is that $L({\rm Ly}\alpha)_{\rm C}$ is constant or decreases with $M_\star$ as expected from the $L({\rm Ly}\alpha)_{ps}$ vs. $M_{\rm UV}$ plot (figure \ref{fig:param} [g]), and the other is that $L({\rm Ly}\alpha)_{\rm H}/L({\rm Ly}\alpha)_{\rm C}$ decreases with $L({\rm Ly}\alpha)_{\rm C}$ as found from equation (\ref{eq:m16}). Let us take the $L({\rm Ly}\alpha)$--divided and $K$--divided subsamples as two examples. For the former subsamples, the $L({\rm Ly}\alpha)_{\rm C}$ of the massive subsample is factor 2.5 {\it lower} than that of the less massive one, but the difference is reduced to factor 1.5 in $L({\rm Ly}\alpha)_{\rm H}$ because objects with lower $L({\rm Ly}\alpha)_{\rm C}$ have higher $L({\rm Ly}\alpha)_{\rm H}/L({\rm Ly}\alpha)_{\rm C}$. For the latter, the two subsamples have almost the same $L({\rm Ly}\alpha)_{\rm C}$ and hence almost the same $L({\rm Ly}\alpha)_{\rm H}$. The slightly decreasing trend of $L({\rm Ly}\alpha)_{\rm tot}$ with mass is due to the fact that $L({\rm Ly}\alpha)_{\rm tot}/L({\rm Ly}\alpha)_{\rm C}$ decreases with $L({\rm Ly}\alpha)_{\rm C}$ more mildly than $L({\rm Ly}\alpha)_{\rm H}/L({\rm Ly}\alpha)_{\rm C}$ does.

Figure \ref{fig:Lh} shows that $L({\rm Ly}\alpha)_{\rm H}$ and $L({\rm Ly}\alpha)_{\rm tot}$ are also nearly independent of $SFR$, $E(B-V)_\star$, and $M_{\rm h}$, although the uncertainties in $M_{\rm h}$ are relatively large. The fact that differently defined subsamples follow a common trend in each panel indicates that the nearly constant $L({\rm Ly}\alpha)_{\rm H}$ and $L({\rm Ly}\alpha)_{\rm tot}$ against $M_\star$ and the other three parameters are real; it is unlikely that grouping the LAEs into two by the five quantities has erased strong mass dependence which otherwise would be visible. We discuss the physical origins of diffuse Ly$\alpha$ halos from these results in section \ref{sec:d_lah}. 

\subsection{Escape fraction of Ly$\alpha$ photons}\label{subsec:r_fesc}
Following previous studies, we define the escape fraction of Ly$\alpha$ photons, $f_{\rm esc}({\rm Ly}\alpha)$, as the ratio of observed Ly$\alpha$ luminosity, $L({\rm Ly}\alpha)_{obs}$, to intrinsic Ly$\alpha$ luminosity, $L({\rm Ly}\alpha)_{int}$, produced in the galaxy due to star formation \citep[e.g.,][]{Atek2008,Kornei2010}:
\begin{equation}
f_{\rm esc}({\rm Ly}\alpha)=\frac{L({\rm Ly}\alpha)_{{obs}}}{L({\rm Ly}\alpha)_{int}}= \frac{SFR_{Ly\alpha}}{SFR_{\rm tot}}, 
\end{equation}
where $SFR_{\rm tot}$ is the total (i.e., dust-corrected) star formation rate and $SFR_{Ly\alpha}$ is the star formation rate converted from $L({\rm Ly}\alpha)_{obs}$ as below: 
\begin{equation}
SFR_{Ly\alpha} \,({\rm M_\odot\,yr^{-1}})=9.1\times10^{-43}\,L({\rm Ly}\alpha)_{obs} \,{\rm(erg\,s^{-1})} 
\label{eq:sfr_lya}
\end{equation}
\citep{Brocklehurst1971,Kennicutt1998}. 
In this work, we derive $f_{\rm esc}({\rm Ly}\alpha)$ from $L({\rm Ly}\alpha)_{\rm tot}$ (total Ly$\alpha$ escape fraction,  $f_{\rm esc}({\rm Ly}\alpha)_{\rm tot}$; see table \ref{tbl:sed_average}) unlike previous studies which have ignored the contribution from the LAH \citep[e.g.,][]{Blanc2011, Kusakabe2015, Oteo2015}. For $SFR_{\rm tot}$ we use the one obtained from the SED fitting.  This definition of $f_{\rm esc}({\rm Ly}\alpha)$ thus assumes that all Ly$\alpha$ photons including those of the LAH are produced from star formation in the central galaxy. We discuss the possibility of the existence of additional Ly$\alpha$ sources later. 

Figure \ref{fig:fesc} shows $f_{\rm esc}({\rm Ly}\alpha)_{\rm tot}$ as a functions of $M_\star$, $SFR$, and $E(B-V)$ for the ten subsamples. All values are field-average values. For a thorough discussion, results with a Calzetti curve are also shown (figures \ref{fig:fesc} [b], [d], and [f]) as well as those with an SMC curve (the other panels). Two interesting features are seen in these figures. 

First, $f_{\rm esc}({\rm Ly}\alpha)_{\rm tot}$ anti-correlates with $M_\star$, $SFR$, and $E(B-V)$ regardless of the assumed curve. Similar anti-correlations have been found for HAEs by \citet{Matthee2016} who have measured total Ly$\alpha$ luminosities on a $6''$ diameter aperture, corresponding to $24$ kpc in radius (blue crosses in the Calzetti-curve panels; see also footnote \ref{ft:M2016}). Any galaxy population may have such anti-correlations. Indeed, an anti-correlation between $f_{\rm esc}({\rm Ly}\alpha)$ and $E(B-V)$ is found for star forming galaxies at $z\sim0$--$3$ \citep[e.g.,][]{Hayes2011,Blanc2011,Atek2014a,Hayes2014}. 
Although Ly$\alpha$ halos are not included in their calculations, these results imply an anti-correlation between $f_{\rm esc}({\rm Ly}\alpha)_{\rm tot}$ and $E(B-V)$ since $L({\rm Ly}\alpha)_{\rm tot}$ increases with $L({\rm Ly}\alpha)_{\rm C}$ as seen in figure \ref{fig:lah}(d).  

Second, our LAEs have very high $f_{\rm esc}({\rm Ly}\alpha)_{\rm tot}$ values. For an SMC-like curve, they are higher than $\sim 30\%$, with some exceeding $100\%$. Using a Calzetti curve makes $f_{\rm esc}({\rm Ly}\alpha)_{\rm tot}$ lower but still in a range of $\sim 10$--$100\%$. The typical $f_{\rm esc}({\rm Ly}\alpha)_{\rm tot}$ of the LAE sample is $\sim1$ dex higher than that of the HAE sample, which is similar to the result obtained in \citet{Sobral2017}. More importantly, a large $f_{\rm esc}({\rm Ly}\alpha)_{\rm tot}$ difference is found even in comparison at a fixed $M_\star$, $SFR$, and $E(B-V)_\star$. We discuss mechanisms by which LAEs can achieve such high escape fractions in section \ref{sec:d_fesc}. 

\subsection{$IRX$ and star formation mode}\label{subsec:r_sfirx}
Star-forming galaxies have a positive correlation that more massive ones have higher $IRXs$. The $IRX \equiv L_{\rm IR}/L_{\rm UV}$ is an indicator of dustiness, where $L_{\rm IR}$ and $L_{\rm UV}$ are IR ($8$--$1000\mu{m}$) and UV ($1530$\AA) luminosities, respectively \citep[e.g., ][]{Reddy2010, Whitaker2014, Alvarez-Marquez2016, Fudamoto2017, McLure2018, Koprowski2018}. Average $M_\star$-$IRX$ relations have been obtained by several studies at $z\sim2$ \citep{Heinis2014, Bouwens2016}. Another important correlation seen in star-forming galaxies is that more massive galaxies have higher $SFRs$, i.e., the star formation main sequence \citep[SFMS; e.g., ][]{Noeske2007, Elbaz2007, Speagle2014a}. Outliers above the SFMS are starburst galaxies \citep{Rodighiero2011}. 
We use these two correlations to test whether or not our subdivided LAEs are outliers in terms of dustiness and star-formation activity. Here, we include nebular emission in SED fitting unlike previous work for subdivided LAEs at $z\sim2$ \citep{Guaita2011} following our previous work for whole LAE sample \citep{Kusakabe2018a}.

\subsubsection{$IRX$}
The $IRX$ can be calculated from the UV attenuation $A_{1530}$ \citep[e.g.,][]{Meurer1999}. \citet{Buat2012} have found that high-$z$ galaxies ($z\simeq0.95-2.2$) follow the relation given in \citet{Overzier2011a}:  
\begin{equation}
\log_{10}{IRX} = \log_{10}(10^{0.4A_{1530}}-1 )\,-\,\log_{10}(0.595), 
\end{equation}
as shown in their figure 14 \footnote{This formula is derived with the total IR luminosity ($3$--$1000\mu{m}$, TIR) for local galaxies. According to the result in \citet{Buat2012}, we do not correct $IRXs$ to those with IR luminosity ($8$--$1000\mu{m}$) in the relation, unlike our previous work \citep{Kusakabe2018a}. }. We convert the $E(B-V)_\star$ of our subsamples into $IRX$ and compare them with two average relations at $z\sim2$ \citep[][]{Heinis2014, Bouwens2016}\footnote{\citet{Bouwens2016} have obtained a {\lq}consensus relation{\rq} from previous analyses for galaxies at $z \sim 2$--$3$ \citep{Reddy2010, Whitaker2014, Alvarez-Marquez2016}, which is consistent with their result using ALMA data. On the other hand, \citet{Heinis2014} derives a relation for UV-selected galaxies at $z \sim 1.5$ giving higher $IRXs$ than the {\lq}consensus relation{\rq} at low-stellar masses regime, however it is consistent wit a new result of star forming galaxies at $2<z<3$ with ALMA data \citet{McLure2018}.} as shown in figure \ref{fig:IRX}. At low-stellar masses with $M_\star \lesssim3$--$5\times10^{9}\, {\rm M_{\odot}}$, the average relation has not been defined well but it is probably located between the two.   

Figure \ref{fig:IRX} (a) shows the field-average values of our subsamples with the assumption of an SMC-like attenuation curve (red symbols), which are calculated from the results for the two fields shown in figure \ref{fig:IRX} (b) (orange and green symbols). The field-average results lie on an extrapolation of the relation for UV-selected galaxies at $z\sim1.5$ in \citet{Heinis2014}. Considering the relatively large uncertainties remaining in the two average relations, we conclude that our subdivided LAEs are not outliers but have normal dustinesses. This result is consistent with those obtained for all LAEs using Spitzer/MIPS $24\mu$m data by \citet{Kusakabe2015} and from SED fitting by \citet{Kusakabe2018a}. Note, however, that if we assume a Calzetti-like attenuation curve instead, our LAEs are expected to be dustier galaxies than ordinary galaxies at the same stellar masses as shown by pink symbols in figure \ref{fig:IRX} (c). In section \ref{subsubsec:scat}, we use the relation in \citet{Heinis2014} for the discussion of the origin of LAHs.

\subsubsection{Star formation mode}
At $z\sim2$, the SFMS has been determined well down to $M_\star\sim10^{10}\,{\rm M_\odot}$ \citep[e.g., ][]{Rodighiero2011,Whitaker2014,Tomczak2016,Shivaei2017}
since SFRs can be accurately measured from either rest-frame UV and FIR (or MIR) fluxes or H$\alpha$ and H$\beta$ emission-line fluxes.　Although these results are not consistent with each other as shown in figure \ref{fig:MS}, the true SFMS probably lies somewhere between the \citet{Tomczak2016} and \citet{Shivaei2017}'s results. Below $M_\star \sim10^{10}\,{\rm M_\odot}$, \citet{Santini2017} suggest that the SFMS continues down to $M_\star\sim10^{8}\,{\rm M_\odot}$ without changing its power-law slope. We compare the results for our LAEs with the extrapolated SFMS shown in \citet{Tomczak2016} and \citet{Shivaei2017} below.

Figure \ref{fig:MS} (a) shows the field-average values for the ten subsamples with an SMC-like attenuation curve (red symbols) while figure \ref{fig:MS} (b) the separate results for the two fields (orange and green symbols). All the field-average data points lie on the extrapolation of the SFMS in \citet{Tomczak2016}, being only slightly above the Shivaei et al. relation. This result is also consistent with those obtained for all LAEs by \citet{Kusakabe2015} and \citet{Kusakabe2018a}. We conclude that the majority of our subdivided LAEs are in a moderate star formation mode even after divided into two subsamples by various properties. 
In section \ref{subsubsec:scat}, we use the relation in \citet{Shivaei2017} for the discussion of the origin of LAHs.

We also compare our results to previous studies on individual LAEs and H$\alpha$ emitters (HAEs) at similar redshifts. For this comparison, we use the results based on a Calzetti attenuation curve (figure \ref{fig:MS} [c]) following these previous studies. We find in figure \ref{fig:MS} (d) that our ten subsamples (pink symbols) are distributed in the middle of individual LAEs with $M_\star$ and $SFR$ measurements \citep[$z\sim2$--$3$]{Hagen2016,Shimakawa2017a,Hashimoto2017a,Taniguchi2015}\footnote{ In \citet{Hagen2016} and \citet{Shimakawa2017a}, $M_\star$ are derived from SED fitting with the Calzetti curve and $SFR$ from the $IRX$--$\beta$ relation in \citet{Meurer1999}. On the other hand, \citet{Taniguchi2015} and \citet{Hashimoto2017a} derive both quantities from SED fitting with the Calzetti curve.}. 
In figure \ref{fig:MS} (e), our LAEs are found to be located at the lower-mass regime of NB-detected HAEs \citep{Tadaki2013, Matthee2016}. While the HAEs in \citet{Tadaki2013} (open cyan hexagons)
\footnote{They derive $M_\star$ from SED fitting with the Calzetti curve \citep[see][for more details]{Tadaki2017a}, while deriving $SFR$s from H$\alpha$ luminosities except for MIPS $24\mu$m detected objects whose SFRs are estimated from UV and MIPS photometry \citep[see also ][]{Tadaki2015}. 
Note that $SFR$s calculated from PACS data are not plotted here. }
lie on the SFMS, those in \citet{Matthee2016} (filled blue hexagons)
\footnote{When analyzing individual galaxies, they assume the Calzetti curve to derive $M_\star$ and assume $E(B-V)_\star=E(B-V)_{g}$ to correct H$\alpha$ luminosities (and hence $SFRs$) for dust extinction\citep[see SED fitting paper of HiZELS for more details,][]{Sobral2014}. However, when stacking, they use $A({\rm H}\alpha)=1$ mag to correct ${\rm H}\alpha$ luminosities for all subsamples.\label{ft:M2016}} are widely scattered along the horizontal direction around the SFMS because they are essentially H$\alpha$ luminosity selected. Some HAEs in \citet{Matthee2016} have similarly low stellar masses to our LAEs but with higher $SFRs$ due to this selection bias.

\section{Discussion}\label{sec:Discussions}
\begin{figure*}[ht] 
\begin{flushleft}
\includegraphics[width=0.9\linewidth]{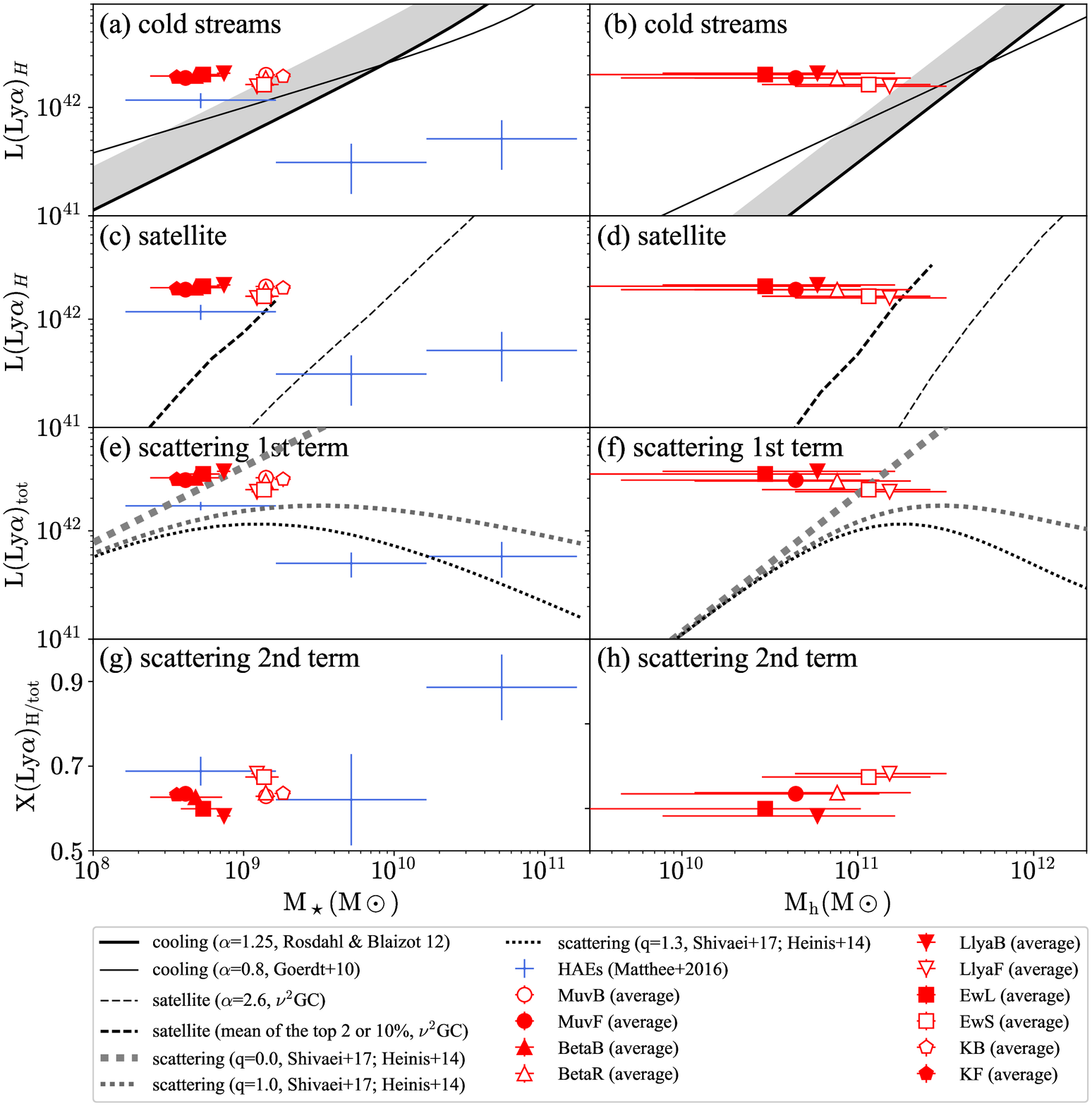}
\end{flushleft} 
  \caption{
  Test of the three LAH scenarios against the observed $L({\rm Ly\alpha})_{\rm H}$ and its mass dependence: 
(a) and (b) -- cold streams (cooling flows); (c) and (d) -- satellite star formation;  (e) and (h) -- resonant scattering. 
Thick and thin solid black lines in panels (a) and (b) show the  Ly$\alpha$ luminosity of cold streams by theoretical models with a  power law of $\alpha=1.25$ \citep{Rosdahl2012} and $\alpha=0.8$ \citep{Goerdt2010}, respectively. Gray shaded regions above the solid black lines roughly indicate the distribution of \citet{Rosdahl2012}'s simulated galaxies above the solid line, whose $L({\rm Ly\alpha})_{\rm H}$ reaches at most $\sim2.5$ times higher than the line. 
A thin dashed black line indicates the mean Ly$\alpha$ luminosity from the star formation in satellite galaxies of $\nu^2$GC \citep[][Ogura et al. in prep.]{Makiya2016, Shirakata2018}. A thick dashed black line indicates the mean of galaxies with the top $10$\% and $2$\% $L({\rm Ly\alpha})_{\rm H}$ at similar $M_\star$ and $M_{\rm h}$ in panels (c) and (d), respectively.  
Gray (thick), dark gray, and black (thin) dotted lines in panels (e) and (f) represent the Ly$\alpha$ luminosity escaping from the main body out to the CGM, with an absorption efficiency relative to UV continuum of $q=0.0, 1.0$ and $1.3$. We assume that all Ly$\alpha$ photons originate from star formation. 
Blue crosses indicate HAEs in \citet{Matthee2016}, whose Ly$\alpha$ luminosities are derived from $3''$ and $6''$-diameter aperture photometry (see footnotes \ref{ft:M2016} and \ref{ft:M2016ap}). 
Field average values of our ten subsamples are shown by symbols below: 
open (filled) circles for bright (faint) $M_{\rm UV}$, 
open (filled) triangles for red (blue) $\beta$, 
open (filled) inverted triangles for faint (bright) $L({\rm Ly\alpha})_{ps}$, 
open (filled) squares for small (large) $EW_{0,ps}({{\rm Ly}\alpha})$, 
and open (filled) pentagons for bright (faint) $m_{\rm K}$. 
Stellar parameters of our subsamples are derived with the assumption of an SMC-like attenuation curve. $M_{\rm h}$ are not calculated for the $K$-divided subsamples, 
and are not plotted for the bright $M_{\rm UV}$ and blue $\beta$ subsamples because of extremely large uncertainties.
All data are rescaled to a Salpeter IMF according to footnote \ref{ft:imf}. (Color online)
}
  \label{fig:Lhmodel}
\end{figure*}

\subsection{The origin of LAHs}\label{sec:d_lah}
As described in section \ref{sec:intro}, theoretical studies have suggested three physical origins of LAHs around high--$z$ star-forming galaxies: (a) cold streams (gravitational cooling), (b) star formation in satellite galaxies, and  (c) resonant scattering of Ly$\alpha$ photons in the CGM which have escaped from the central galaxy. 
In origins (a) and (b), the Ly$\alpha$ photons of LAHs are produced in situ, while in origin (c) they come from central galaxies. The difference between (a) and (b) is how to produce Ly$\alpha$ photons.　A flow chart and an illustration of these origins are shown in figure 6 in \citet{Mas-Ribas2017a} and figure 15 in \citet{Momose2016}, respectively. So far, observations have not yet identified the dominant origin(s) as explained below.

There are two observational studies on the origin of LAHs around star-forming galaxies. \citet{Leclercq2017} use $166$ LAEs at $z \sim 3$--$5$ detected with the MUSE, while \citet{Momose2016} are based on a stacking analysis of $\sim 3600$ $z\simeq2.2$ LAEs from a narrow-band survey, the same parent sample as we use in this study. \citet{Leclercq2017} have argued that a significant contribution from (b) star formation in satellite galaxies is somewhat unlikely since the UV component of MUSE-LAEs is compact and not spatially offset from the center of their LAH. However, they have not given a firm conclusion on the contributions from the remaining two origins. This is because while they have found a scaling relation of $L({\rm Ly\alpha})_{\rm H} \propto L_{\rm UV}^{0.45}$ which is not dissimilar to the scaling predicted from hydrodynamical simulations of cold streams by \citet{Rosdahl2012}, resonant scattering also prefers such a positive scaling relation if $f_{\rm esc}({\rm Ly\alpha})_{\rm tot}$ is constant. Moreover, they have also found that $\sim 80\%$ of their sample have a not-so-large total EW of ${\rm Ly\alpha}$ emission, $EW_{0,tot}({\rm Ly\alpha})\lesssim200$\AA, not exceeding the maximum dust-free $EW_{0}({\rm Ly\alpha})$ of population I\hspace{-.1em}I star formation, $\sim50$--$240$\AA,  with a solar metallicity and a Salpeter IMF \citep[e.g.,][]{Charlot1993,Malhotra2002}. If $EW_{0}({\rm Ly\alpha})$ is larger than $\sim 200$\AA, Ly$\alpha$ radiation from cold streams would be responsible for LAHs. 

\citet{Momose2016} have also found relatively low $EW_{0,tot}({\rm Ly}\alpha)$ and marginally ruled out the cold-stream origin based on a similar discussion to \citet{Leclercq2017}'s. In these two observational studies, $EW_{0,tot}({\rm Ly\alpha})$ are calculated by dividing the total Ly$\alpha$ luminosity by the UV luminosity of the {\it central} part. Therefore, the relatively low $EW_{0,tot}({\rm Ly\alpha})$ values do not necessarily mean that the net $EW_0$ of LAHs are also low; they would even be extremely high if LAHs do not have UV emission. Thus, the cold-stream scenario cannot be ruled out from the low $EW_{0,tot}({\rm Ly\alpha})$ values alone. The discussion using the $L({\rm Ly\alpha})_{\rm H}$--$L_{\rm UV}$ relation assumes $L_{\rm UV}\propto M_{\rm h}^{0.5}$ because the simulations have calculated $L({\rm Ly\alpha})_{\rm H}$ against $M_{\rm h}$. Since $L_{\rm UV}$ may not be a perfect tracer of $M_{\rm h}$, it is more desirable to use directly the $L({\rm Ly\alpha})_{\rm H}$-$M_{\rm h}$ relation, or the $L({\rm Ly\alpha})_{\rm H}$--$M_\star$ relation as a better substitute. In addition, comparing the normalization of the relation as well as its power-law slope can better constrain this scenario. With regard to (b) satellite star formation, independent observations are desirable to strengthen the conclusion by \citet{Leclercq2017} since \citet{Momose2016} have not been able to rule out this origin. Finally, if resonant scattering is the dominant origin, LAH luminosities have to be explained by the properties of the main body of galaxies such as $SFR$ and $E(B-V)$.

In section \ref{subsec:r_lah}, we find that the $L({\rm Ly\alpha})_{\rm H}$ and $L({\rm Ly\alpha})_{\rm tot}$ of our LAEs remain unchanged with increasing stellar mass. We also obtain a constant or increasing $X({\rm Ly\alpha})_{\rm H/tot}$ with $M_\star$ (see figure \ref{fig:Lhmodel}[g]). In the following subsections, we use these relations to discuss the three origins  with figure \ref{fig:Lhmodel}. We also use the results on HAEs obtained by \citet{Matthee2016}\footnote{They discuss the escape fraction using $L({\rm Ly\alpha})$ on $r=12$ kpc ($3''$ diameter) and $24$ kpc ($6''$) apertures. Although the average profile of their LAHs extends to $r=40$ kpc, we refer to $6''$ aperture luminosity as $L({\rm Ly\alpha})_{\rm tot}$ and to the difference in $3''$ and $6''$ aperture luminosities as $L({\rm Ly\alpha})_{\rm H}$. \label{ft:M2016ap}} to strengthen the discussion. We also briefly examine the fluorescence scenario in appendix \ref{sec:appendix_fluorescence}, following the very recent study on fluorescence emission for star-forming LAEs by \citet{Gallego2018}. 　

\subsubsection{(a) Cold streams}\label{coldstreams}

Theoretical studies and simulations suggest that high-$z$ ($z\gtrsim2$) galaxies obtain baryons through the accretion of relatively dense and cold ($\sim 10^4$ K) gas known as cold streams \citep[e.g.,][]{Fardal2001, Keres2005,Dekel2006}. The accreting gas releases the gravitational energy and emits Ly$\alpha$ photons, thus producing an extended Ly$\alpha$ halo without (extended) UV continuum emission \citep[e.g.,][]{Haiman2000,Furlanetto2005,Dijkstra2009,Lake2015}. 

The Ly$\alpha$ luminosity due to cold streams is suggested to increase with the $M_{\rm h}$ of host galaxies. A scaling of $L({\rm Ly\alpha})_{\rm H}\propto{M_{\rm h}}^{1.1}$-${M_{\rm h}}^{1.25}$ at $M_{\rm}=10^{10}$--$10^{13}\,M_\odot$ has been predicted by (zoom-in) cosmological hydrodynamical simulations in \citet{Faucher-Giguere2010} and \citet{Rosdahl2012}. \citet{Dijkstra2009} have obtained a similar correlation to \citet{Faucher-Giguere2010}'s from an analytic model which reproduces the Ly$\alpha$ luminosities, Ly$\alpha$ line widths, and number densities of observed LABs at $M_{\rm h}\gtrsim10^{11}M_\odot$. On the other hand, \citet{Goerdt2010} have derived a shallower power law slope $\sim0.8$ for LAB-hosting massive ($M_{\rm h} \sim 10^{12}$--$10^{13} M_\odot$) halos from high-resolution cosmological hydrodynamical adaptive mesh refinement simulations.

We examine if our subsamples are consistent with these theoretical predictions by comparing the power-law slope and amplitude of the $L({\rm Ly\alpha})_{\rm H}$-$M_{\rm h}$ relation. For a conservative discussion, we use \citet{Rosdahl2012}'s relation which gives the steepest slope and \citet{Goerdt2010}'s relation giving the shallowest slope as shown in figure \ref{fig:Lhmodel}(b) \footnote{We shift the relation shown in figure 8 in \citet{Rosdahl2012} at $z=3$ to $z=2$ by multiplying redshift-evolution term, $(1+z)^{1.3}$, given in figure 12 and equation 21 in \citet{Goerdt2010}. We also note that the relation at $z\sim3$ predicted in \citet{Faucher-Giguere2010} has a lower amplitude than that in \citet{Rosdahl2012} typically about a factor of two \citep[see appendix E in][for more details]{Rosdahl2012}.}:　
\begin{equation}
L({\rm Ly\alpha})_{\rm H}\sim 8\times10^{42} \left(\frac{ M_{\rm h}}{10^{12}M_\odot}\right)^{1.25}\left(\frac{1+z}{1+3}\right)^{1.3},
\label{eq:R12}
\end{equation}
\begin{equation}
L({\rm Ly\alpha})_{\rm H}=9.72\times10^{42}  \left(\frac{M_{\rm h}}{10^{12}M_\odot}\right)^{0.8} (1+z)^{1.3}.
\label{eq:G10}
\end{equation}
In figure \ref{fig:Lhmodel}(a), we convert $M_{\rm h}$ to $M_\star$ using the average relation between $M_\star$ and $M_{\rm h}$ at $z\sim2$ in \citet{Moster2013}\footnote{\citet{Kusakabe2018a} have found that our LAEs are on average slightly offset from the average relation to lower $M_{\rm h}$ values. Our discussion is unchanged if we instead use $M_{\rm h}$ reduced by this offset.}. The constant  $L({\rm Ly\alpha})_{\rm H}$ with $M_\star$ and $M_{\rm h}$ seen in the LAEs is inconsistent with the increasing $L({\rm Ly\alpha})_{\rm H}$ predicted by the theoretical models, although the uncertainties in our $M_{\rm h}$ estimates are large. The HAEs have also non-increasing $L({\rm Ly\alpha})_{\rm H}$ over two orders of magnitude in $M_\star$, highlighting the inconsistency found for the LAEs. As for amplitude, the LAEs shown by red filled (open) symbols have $\sim2$--$4$ ($\sim1$--$2$) times higher $L({\rm Ly\alpha})_{\rm H}$ than the two model predictions at the same $M_\star$ (figure \ref{fig:Lhmodel}[a]), and at least $\sim1$--$10$ ($\sim1$--$10$) times higher at the same $M_{\rm h}$ (figure \ref{fig:Lhmodel}[b]). Even when the individual distribution of \citet{Rosdahl2012}'s galaxies is considered, low-$M_\star$ LAEs (red filled symbols) have more than $10\,\sigma$ brighter $L({\rm Ly\alpha})_{\rm H}$ than the simulated galaxies with similar $M_\star$ (a gray shaded region). In other words, cold streams cannot produce as many Ly$\alpha$ photons in the CGM as observed. 

 Note that as mentioned in appendix \ref{sec: appendix_NBbias}, the $L({\rm Ly\alpha})_{\rm H}$ values of the faint $m_K$ and $M_{UV}$ subsamples are possibly overestimated since they miss small $EW({\rm Ly\alpha})$ (faint $L({\rm Ly\alpha})_{\rm C}$) sources due to the NB-selection bias. If we derive $L({\rm Ly\alpha})_{\rm H}$ conservatively from the $M_{UV}$--$L({\rm Ly\alpha})_{\rm H}$ relation for individual MUSE-LAEs without such a selection bias in \citet{Leclercq2017}, we obtain $\sim1.5$ times smaller $L({\rm Ly\alpha})_{\rm H}$, which results in a slightly positive correlation between $M_\star$ and $L({\rm Ly\alpha})_{\rm H}$. However, the power law index and the amplitude of the $M_\star$--$L({\rm Ly\alpha})_{\rm H}$ correlation of the $m_K$ subsamples is still shallower and higher than theoretical results at more than the $2\,\sigma$ and $10\,\sigma$ confidence levels, respectively (see more details in appendix \ref{sec:appendix_LAH}). Consequently, our study suggests that (a) cold streams are not the dominant origin of LAHs.

\subsubsection{(b) Satellite star formation}
Satellite galaxies emit Ly$\alpha$ photons through star formation. If satellite star formation significantly contributes to LAHs, they will involve an extended UV emission from the star formation \citep[e.g.,][]{Shimizu2011, Zheng2011, Lake2015, Mas-Ribas2017a}. Unfortunately, this emission is expected to be too diffuse to detect even by stacking of some $10^3$ objects as mentioned in \citet{Momose2016}.

The Ly$\alpha$ luminosity from satellite star formation can be interpreted as a function of the $M_{\rm h}$ and $M_\star$ of the central galaxy. In the local universe, the number of disk (i.e., star-forming) satellite galaxies is found to be described by a power law of the host halo mass of the central galaxy with a slope of $0.91\pm0.11$ for galaxies with $M_{\rm h}\sim10^{12}$--$10^{14}\,M_\odot$ \citep[see figure 14 and equation 6 in ][ see also figure 2 in \citet{Wang2014}]{Trentham2009}. At high redshifts, at least for massive central galaxies ($M_\star\sim10^{11}\,M_\odot$ at $z\sim1.4$), the radial number density profile of satellite galaxies is not significantly different from that at $z\sim0$ \citep[][]{Tal2013}. These local properties are reproduced by theoretical models \citep[e.g.,][]{Nickerson2013,Sales2014,Okamoto2010}. With an assumption that the total Ly$\alpha$ luminosity from satellite galaxies is proportional to the sum of their SFRs of satellite galaxies, $L({\rm Ly\alpha})_{\rm H}$ can be calculated from cosmological galaxy formation models. 

The ``New Numerical Galaxy Catalogue'' ($\nu^2$GC) is a cosmological galaxy formation model with semi-analytic approach \citep[][Ogura et al. in prep.]{Makiya2016, Shirakata2018}, which is based on a state-of-the-art $N$ body simulation performed by \citet{Ishiyama2015}. It can reproduce not only the present-day luminosity functions (LF) and HI mass function but also the evolution of the LFs and the cosmic star formation history \citep[][Ogura et al. in prep.]{Makiya2016, Shirakata2018}. We use model galaxies at $z\sim2.2$ in the $\nu^2$GC-S with a box size of $280\, h^{-1}$cMpc (LAE NB selection is not applied). The number of central galaxies is $\sim6\times10^{6}$. For each central galaxy, we calculate $L({\rm Ly\alpha})_{\rm H}$ by summing the SFRs of the satellites with an assumption of case B recombination. We find that the average $L({\rm Ly\alpha})_{\rm H}$ can be approximated as
\begin{equation}
L({\rm Ly\alpha})_{\rm H}\sim 8.3\times10^{42} \left(\frac{ M_{\rm h}}{10^{12}M_\odot}\right)^{2.58}
\label{eq:nu2GC-mh}
\end{equation}
at $M_{\rm h}\sim10^{10}$--$10^{12}\,M_\odot$ and 
\begin{equation}
L({\rm Ly\alpha})_{\rm H}\sim 1.9\times10^{42} \left(\frac{ M_\star}{10^{10}M_\odot}\right)^{1.36}
\label{eq:nu2GC-ms}
\end{equation}
at $M_{\star}\sim10^{8}$--$10^{10}\,M_\odot$ as shown with a thin black dashed line in figures \ref{fig:Lhmodel} (c) and (d). The power law of $M_{\rm h}$ for $L({\rm Ly\alpha})_{\rm H}$ is steeper than that for the observed number of disk satellite galaxies.

We focus on the amplitude and slope of the $L({\rm Ly\alpha})_{\rm H}$ -- mass relations. The LAEs shown by red symbols have more than $\sim1$ dex higher $L({\rm Ly\alpha})_{\rm H}$ than the mean of the model galaxies at the same $M_{\star}$ and $M_{\rm h}$. 
However, observations show that LAEs occupy only $\sim10$\% ($\sim2$\%) of all galaxies with the same $M_{\star}$ ($M_{\rm h}$) \citep{Kusakabe2018a}. For a conservative comparison, we limit the model galaxies to those with the top $10$\% ($2$\%) $L({\rm Ly\alpha})_{\rm H}$ at a fixed $M_{\star}$ ($M_{\rm h}$). We find that the mean $L({\rm Ly\alpha})_{\rm H}$ of these $L({\rm Ly\alpha})_{\rm H}$-bright model galaxies (thick dashed lines in figures \ref{fig:Lhmodel} [c] and [d]) 
is still about three times lower than the observed values. Moreover, the positive correlations of $L({\rm Ly\alpha})_{\rm H}$ with $M_\star$ and $M_{\rm h}$ seen for the model galaxies are incompatible with the constant $L({\rm Ly\alpha})_{\rm H}$ of our LAEs and with the decreasing $L({\rm Ly\alpha})_{\rm H}$ of the HAEs in \citet{Matthee2016}. These LAEs and HAEs span two orders of magnitude in $M_\star$. A non-increasing $L({\rm Ly\alpha})_{\rm H}$ over this wide mass range may be achieved if the Ly$\alpha$ photons from satellites of massive galaxies are heavily absorbed in the CGM, but the offset of $L({\rm Ly\alpha})_{\rm H}$ from our LAEs becomes larger. Such a heavy dust pollution in the CGM is probably unlikely.

 As described in the previous subsection, using \citet{Leclercq2017}'s $M_{\rm UV}$--$L({\rm Ly\alpha})_{\rm H}$ relation results in a slightly positive correlation. However, the power law index determined by the $m_K$ subsamples is still shallower than that of the model (see appendix \ref{sec:appendix_LAH} for detalis). In addition, it remains difficult for the model to explain the results of LAEs and HAEs in a unfied manner.  
From these results, we conclude that satellite star formation is unlikely to be the dominant origin.

\subsubsection{(c) Resonant scattering of Ly$\alpha$ photons in the CGM which are produced in central galaxies}\label{subsubsec:scat}
HI gas in the CGM can resonantly scatter Ly$\alpha$ photons which have escaped from the main body of the galaxy \citep[e.g., ][]{Laursen2007,Barnes2010, Zheng2011,Dijkstra2012,Verhamme2012}. However, there is no theoretical study that predicts $L({\rm Ly\alpha})_{\rm H}$ and its dependence on galaxy properties by solving the radiative transfer of Ly$\alpha$ photons in the CGM. In this subsection, we first describe the LAH luminosity of a galaxy assuming that all Ly$\alpha$ photons come from the main body. To do so, we introduce two parameters: the escape fraction out to the CGM and the scattering efficiency in the CGM. Then, we examine if resonant scattering can explain reasonably well the behavior of LAEs and HAEs shown in the previous section. Let $L({\rm Ly\alpha})_{int}$ be the total luminosity of Ly$\alpha$ photons produced in the main body. Some fraction of $L({\rm Ly\alpha})_{int}$ is absorbed by dust in the interstellar medium (ISM) and the rest escapes out into the CGM. With an assumption that dust absorption in the CGM is negligibly small, the escaping luminosity is equal to $L({\rm Ly\alpha})_{\rm tot}$ ($= L({\rm Ly\alpha})_{\rm C} + L({\rm Ly\alpha})_{\rm H}$), and the escape fraction into the CGM is calculated as $f_{\rm esc}({\rm Ly\alpha})_{\rm tot} = L({\rm Ly\alpha})_{\rm tot} / L({\rm Ly\alpha})_{int}$.
Then, a fraction, $X({\rm Ly\alpha})_{\rm H/tot}$, of the escaping  photons are scattered in the CGM, being extended as a LAH with $L({\rm Ly\alpha})_{\rm H}$.
Thus, $L({\rm Ly\alpha})_{\rm H}$ can be written as: 
\begin{eqnarray}
L({\rm Ly\alpha})_{\rm H}&=&L({\rm Ly\alpha})_{int}\,f_{\rm esc}({\rm Ly\alpha})_{\rm tot}\, X({\rm Ly\alpha})_{\rm H/tot}\\
&=&L({\rm Ly\alpha})_{\rm tot}\, X({\rm Ly\alpha})_{\rm H/tot}. 
\end{eqnarray}

In the following modeling, we assume that $L({\rm Ly\alpha})_{int}$ originates only from star formation, and express it as a function of $M_\star$ using the SFMS:
\begin{equation}
L({\rm Ly\alpha})_{int}  \,{\rm(erg\,s^{-1})}= SFR_{\rm MS} \,({\rm M_\odot\,yr^{-1}})/9.1 \times 10^{-43}.
\end{equation}
We then describe $f_{\rm esc}({\rm Ly\alpha})_{\rm tot}$ as a function of $M_\star$ using the $M_\star$--$IRX$ relation discussed in section \ref{subsec:r_sfirx}. The dust attenuation for 1216 \AA\ continuum, $A_{\rm 1216con}$, at a fixed $M_\star$ is calculated from $IRX (M_\star)$:    
\begin{equation}
A_{\rm 1216con}(M_\star)=2.5\log_{10}(0.595IRX(M_\star)+1.0)\, \left(\frac{\kappa_{1216}}{\kappa_{1500}}\right), 
\end{equation}
where $\kappa_{1216}$ and $\kappa_{1500}$ are the coefficients of the attenuation curve at $\lambda=1216$ \AA\ and $1500$ \AA, respectively. Introducing the relative efficiency of the attenuation of Ly$\alpha$ emission to the continuum at the same wavelength, $q=A_{\rm Ly\alpha}/A_{\rm 1216con}$ \citep[e.g.,][]{Finkelstein2008}, we can write $f_{\rm esc}({\rm Ly\alpha})_{\rm tot}$ as:
\begin{equation}
f_{\rm esc}({\rm Ly\alpha})_{\rm tot}=10^{-0.4\,q\,A_{\rm 1216con}(M_\star)},
\end{equation}
where $q=0$ and $q=1$ correspond to the case without attenuation of Ly$\alpha$ emission and with the same attenuation as that of continuum. 
We thus obtain: 
\begin{equation}
L({\rm Ly\alpha})_{\rm tot}(M_\star)=\left(\frac{SFR_{\rm MS} (M_\star)}{9.1\times10^{-43}}\right)10^{-0.4\,q\,A_{1216}(M_\star)}.
\end{equation}
We use \citet{Shivaei2017}'s SFMS and \citet{Heinis2014}'s $IRX$-$M_\star$ relation because our LAEs are on these relations (see section \ref{subsec:r_sfirx}). We also assume an SMC-like attenuation curve.

Shown in figure \ref{fig:Lhmodel}(e) are three calculations with $q=0.0,\, 1.0$, and $1.3$ (gray (thick), dark gray, and black (thin) dotted lines, respectively). The constant $L({\rm Ly\alpha})_{\rm tot}$ with increasing $M_\star$ seen in the LAEs is achieved if $q$ increases with $M_\star$. We note that all LAEs require $q<1$, with the less massive subsamples suggesting $q=0$, meaning that Ly$\alpha$ photons escape much more efficiently than UV photons. We do not compare the HAEs with these models since they do not follow well the SFMS and the $IRX$-$M_\star$ relation (see section \ref{subsec:r_sfirx}). As we show later, the HAEs can be explained by large $q$ values. Further discussion of $f_{\rm esc}({\rm Ly\alpha})_{\rm tot}$ and $q$ for our LAEs and the HAEs is given in section \ref{sec:d_fesc}. We also find that this result is unchanged even if we instead use a Calzetti attenuation curve, \citet{Tomczak2016}'s SFMS, and/or \citet{Bouwens2016}'s $IRX$-$M_\star$ relation.

The term $X({\rm Ly\alpha})_{\rm H/tot}$ can be interpreted as the efficiency of resonant scattering in the CGM. More massive galaxies may have a larger amount of HI gas in the halo and thus have a higher $X({\rm Ly\alpha})_{\rm H/tot}$ value. Figure \ref{fig:Lhmodel}(g) shows that this picture is consistent with our LAEs and \citet{Matthee2016}'s HAEs, because these two populations appear to follow a common, positive (although very shallow) correlation between $X({\rm Ly\alpha})_{\rm H/tot}$ and $M_\star$. This picture is also consistent with the $X({\rm Ly\alpha})_{\rm H/tot}$ -- $M_{\rm h}$ plot for our LAEs (figure \ref{fig:Lhmodel}[h]) within the large uncertainties in $M_{\rm h}$. In this case, the LAHs of our LAEs ($\lesssim40$ kpc in radius) are caused by HI gas roughly within the virial radius of hosting dark matter halos, $\sim20$--$50$ kpc, whose mass is estimated to be in the range $M_{\rm h}\sim10^{10}$--$10^{11}\,M_\odot$.
This relative extent of LAHs is close to those inferred for the LAHs of MUSE-LAEs by \citet{Leclercq2017}, typically $60$--$90\%$ of the virial radius, where they predict $M_{\rm h}$ from observed UV luminosities using the semi-analytic model of \citet{Garel2015}.　

Thus, in the resonant scattering scenario, the constant (or decreasing) $L({\rm Ly\alpha})_{\rm H}$ observed is achieved by a combination of increasing $L({\rm Ly\alpha})_{int}$, decreasing $f_{\rm esc}({\rm Ly\alpha})_{\rm tot}$, and (slightly) increasing $X({\rm Ly\alpha})_{\rm H/tot}$ with mass, and all three trends are explained reasonably well. Our study suggests that (c) resonant scattering is the dominant origin of the LAHs.

\subsubsection{Summary of the three comparisons}
It is found that resonant scattering most naturally explains the $L({\rm Ly\alpha})_{\rm H}$ and its dependence on galaxy properties seen in our LAEs and \citet{Matthee2016}'s HAEs. We, however, note that hydrodynamic cosmological simulations in \citet{Lake2015} show that scattered Ly$\alpha$ in the CGM can reach only out to $\sim15$ kpc, suggesting that cold streams or satellite star formation are also needed, although they slightly overestimate the observed radial Ly$\alpha$ profile at $15$ kpc (by a factor of 2). On the other hand, \citet{Xue2017} have found for LAEs at $z \sim 4$ that the radial profile of LAHs is very close to a predicted profile by \citet{Dijkstra2012} who have only considered resonant scattering. Theoretical models discussing the contribution of scattering to $f_{\rm esc}({\rm Ly\alpha})_{\rm tot}$ and $X({\rm Ly\alpha})_{\rm H/tot}$ as a function of $M_\star$ and $M_{\rm h}$ are needed for a more detailed comparison.  \citet{Mas-Ribas2017a} show that different origins give different spatial profiles of Ly$\alpha$, UV, and H$\alpha$ emission. According to the best-effort observations of Ly$\alpha$ and H$\alpha$ emission of LAEs in \citet{Sobral2017}, Ly$\alpha$ photons of LAEs at $z\sim2$ are found to escape over two times larger radii than H$\alpha$ photons, which implies (a) cold stream scenario or (c) resonant scattering scenario, although their results are based on
images with the PSF as large as $\sim2$ arcsecond (FWHM). Deep, spatially resolved observations of H$\alpha$ emission with James Webb Space Telescope (JWST) would provide us with important clues to the origin of LAHs.

\subsection{The origin high Ly$\alpha$ escape fractions}\label{sec:d_fesc}
\begin{figure*}[ht]
 \begin{flushleft}
      \includegraphics[width=0.9\linewidth]{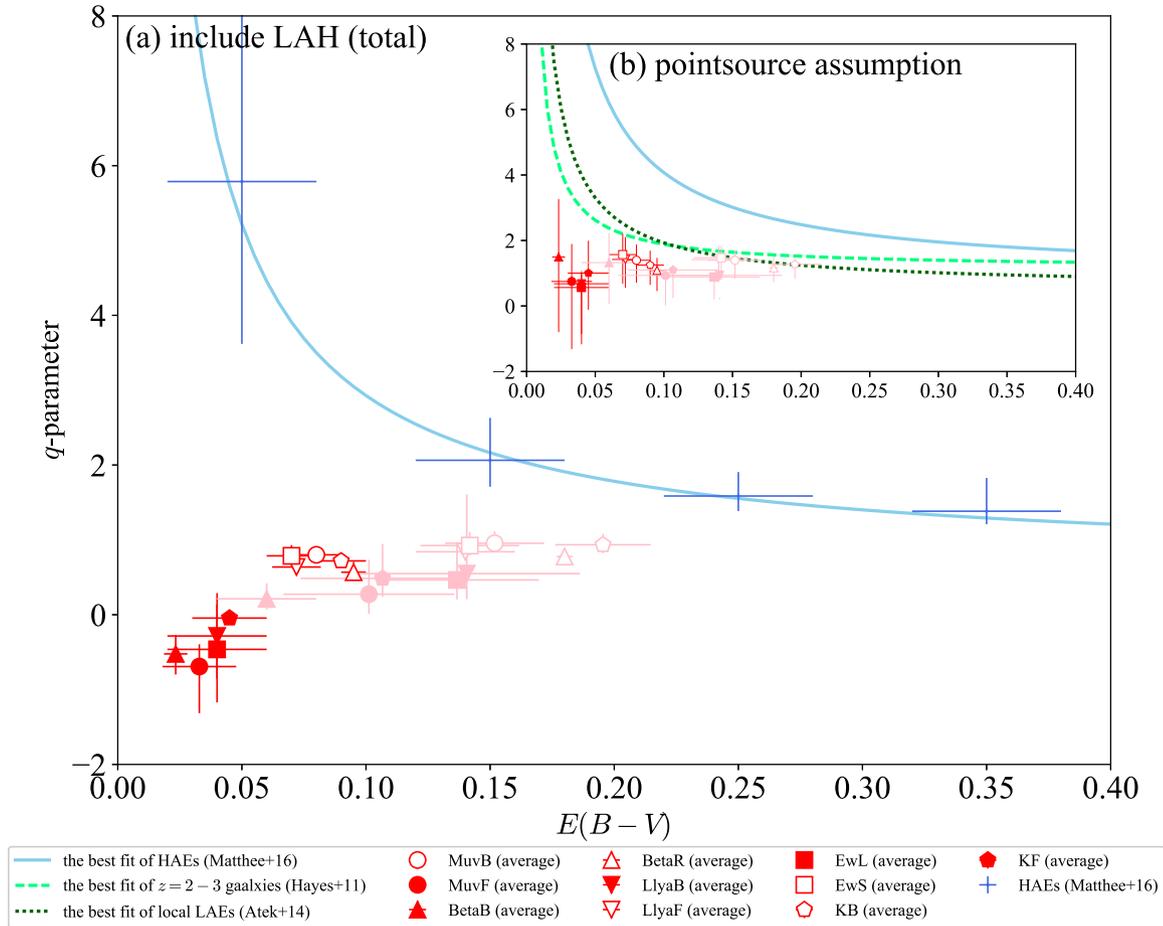}
\end{flushleft}      
  \caption{
  The $q$ parameter vs. $E(B-V)$. The LAH is included in  calculation of $q$ in panel (a), 
while not included in panel (b).
Blue crosses and a solid skyblue line show the values of $E(B-V)$-subdivided HAEs in \citet{Matthee2016} and the best fit two-parameter model to them as described in footnote \ref{ft:q_M16} (see also footnotes \ref{ft:M2016} and \ref{ft:M2016ap});  
$12$ kpc and $24$ kpc apertures are used in panels (a) and (b), respectively. 
A light green dashed line and a dark green dotted line represent the best fit relation for LAEs, HAEs, and UV-selected galaxies at $z\sim2$--$3$ in \citet{Hayes2011} and local LAEs in \citet{Atek2014a}, respectively. 
Field average values of our ten subsamples with an assumption of an SMC-like attenuation curve are shown by red symbols below: 
open (filled) circles for bright (faint) $M_{\rm UV}$, 
open (filled) triangles for red (blue) $\beta$, 
open (filled) inverted triangles for faint (bright) $L({\rm Ly\alpha})_{ps}$, 
open (filled) squares for small (large) $EW_{0,ps}({{\rm Ly}\alpha})$, 
and open (filled) pentagons for bright (faint) $m_{\rm K}$. 
Shown by pink symbols are the results with the Calzetti curve. $E(B-V)_\star=E(B-V)_{g}$ is assumed to derive $E(B-V)$. All data are rescaled to a Salpeter IMF according to footnote \ref{ft:imf}. (Color online)
}
  \label{fig:q}
\end{figure*}

By including $L({\rm Ly}\alpha)_{\rm H}$ in the total Ly$\alpha$ luminosity, we obtain very high $f_{\rm esc}({\rm Ly}\alpha)_{\rm tot}$ values for our LAEs as shown in section \ref{subsec:r_fesc}. These values are systematically higher than those obtained for LAEs in previous studies which have not considered $L({\rm Ly}\alpha)_{\rm H}$ \citep[e.g.,][]{Song2014, Hayes2011}.
They are also about one order of magnitude higher than those of HAEs with the same $M_\star$ and $E(B-V)$ (figure \ref{fig:fesc}), suggesting a large scatter in $f_{\rm esc}({\rm Ly}\alpha)_{\rm tot}$ among galaxies.

It is helpful to discuss $f_{\rm esc}({\rm Ly}\alpha)$ using $E(B-V)$, since additional mechanisms are needed to make $f_{\rm esc}({\rm Ly}\alpha)$ higher or lower than that expected from $E(B-V)$.  The attenuation of Ly$\alpha$ emission relative to that of continuum emission is evaluated by the $q$-parameter
\footnote{The $q$-parameter can be rewritten as:$q=\frac{-\log\left(f_{\rm esc}({\rm Ly}\alpha)\right)}{0.4E(B-V)\kappa_{1216}}$ $=\frac{\kappa}{k\kappa_{1216}} - \frac{\log{\rm C}}{0.4E(B-V)\kappa_{1216}}$, where $\kappa$ and $C$ are two parameters of a commonly used fitting formula of $f_{\rm esc}({\rm Ly}\alpha) = C 10^{-0.4E(B-V)\kappa}$ \citep[e.g.,][]{Hayes2011}. The two parameters are difficult to interpret physically, especially for a case with $C<1$.  \citet{Hayes2011} and \citet{Atek2014a} do not include $L({\rm Ly}\alpha)_{\rm H}$ to calculate the $f_{\rm esc}({\rm Ly}\alpha)$ and obtain $C=0.445$ with $\kappa=13.6$ and $C=0.22$ with $\kappa=6.67$, respectively. 
Although \citet{Matthee2016} include $L({\rm Ly}\alpha)_{\rm H}$ to calculate $f_{\rm esc}({\rm Ly}\alpha)_{\rm tot}$, their $C$ is less than $1$ ($C=0.08^{+0.02}_{-0.01}$ with $\kappa=7.64^{+1.38}_{-1.36})$, which is slightly larger than the value derived without $L({\rm Ly}\alpha)_{\rm H}$ ( $C=0.03^{+0.01}_{-0.01}$ with $\kappa=10.71^{+0.89}_{-1.01}$). Note that \citet{Atek2014a} uses Balmer decrements to estimate $E(B-V)_{gas}$, while other studies use SED fitting. \label{ft:q_M16}} \citep[e.g.,][]{Finkelstein2008,Finkelstein2009b}, 
as discussed in section \ref{subsubsec:scat}. Figure \ref{fig:q} shows $q$ as a function of $E(B-V)$ for our LAEs and \citet{Matthee2016}'s HAEs, which are divided into subsamples in accordance with $E(B-V)$. Regardless of the attenuation curve, the LAEs have small $q$ less than unity, which increases with $E(B-V)$. Remarkably, about a half of the subsamples, shown by red filled symbols, have $q<0$, meaning that the observed Ly$\alpha$ luminosity exceeds the one calculated from the SFR. On the other hand, the HAEs have larger $q$ ($>1$) decreasing with $E(B-V)$. The difference in $q$ between these two galaxy populations becomes larger at smaller $E(B-V)$. Note that if we calculate $q$ of our LAEs from $L({\rm Ly}\alpha)_{ps}$ instead of including $L({\rm Ly}\alpha)_{\rm H}$, we obtain higher values, $q\sim1$, 
being closer to the values found in previous studies \citep[e.g.,][]{Hayes2010,Nakajima2012}. 

Below, we discuss how LAEs can have low $q$ and hence high $f_{\rm esc}({\rm Ly}\alpha)_{\rm tot}$ than HAEs with the same $E(B-V)$, by grouping possible origins into three categories: 
(i) less efficient resonant scattering in a uniform ISM,
(ii) less efficient resonant scattering in a clumpy ISM,
and (iii) additional Ly$\alpha$ sources. We then discuss the difference in $q$ and $f_{\rm esc}({\rm Ly}\alpha)_{\rm tot}$ between the LAEs and HAEs. In this discussion, we assume that the contribution from cold streams and satellite galaxies to $L({\rm Ly\alpha})_{\rm H}$ is negligible.

\subsubsection{(i) Less efficient resonant scattering in a uniform ISM} \label{subsubsec:fesc_scat}

In a uniform ISM where dust and gas are well mixed, Ly$\alpha$ photons have a higher chance of dust absorption than continuum photons because of resonant scattering.  To reduce the efficiency of resonant scattering in a uniform ISM, one needs to reduce the column density of HI gas ($N_{\rm HI}$) or the scattering cross section ($\sigma_{\rm Ly\alpha}$) \citep[e.g.,][]{Duval2014, Garel2015}. 

First, it appears that LAEs indeed have lower $N_{\rm HI}$ than average galaxies with the same $M_\star$ (and hence the same $E(B-V)$ since average galaxies are expected to follow a common IRX-$M_\star$ relation). This is because \citet{Kusakabe2018a} suggest that LAEs at $z\sim 2$ have lower $M_{\rm h}$ than expected from the average $M_\star$-$M_{\rm h}$ relation. At a fixed $M_\star$, a lower $M_{\rm h}$ means a lower baryon mass and hence a lower gas mass, and it is reasonable to expect that galaxies with a lower gas mass have a lower $N_{\rm HI}$. The $N_{\rm HI}$ of LAEs is further reduced if they have a high ionizing parameter as suggested by e.g., \citet[][]{Nakajima2014}, \citet[][]{Song2014}, and \citet[][]{Nakajima2018} or have a relatively face-on inclination \citep[e.g.,][]{Verhamme2012, Yajima2012b, Behrens2014,Shibuya2014b, Kobayashi2016, Paulino-Afonso2018AA}.

The idea that LAEs have lower $N_{\rm HI}$ than average galaxies appears to be consistent with results based on observed Ly$\alpha$ profiles that LAEs have lower $N_{\rm HI}$ than LBGs \citep[e.g.,][]{Hashimoto2015, Verhamme2006}. This idea is also consistent with an anti-correlation between $M_{\rm HI}$ and $f_{\rm esc}({\rm Ly}\alpha)$ found for local galaxies, although their $f_{\rm esc}({\rm Ly}\alpha)$ values at a fixed $E(B-V)$ are lower than those of our LAEs \citep[Ly$\alpha$ Reference Sample][]{Hayes2013, Ostlin2014}.

The probability of the resonant scattering of Ly$\alpha$ photons is also reduced if the ISM is outflowing, 
because the gas sees redshifted Ly$\alpha$ photons \citep[e.g.,][]{Kunth1998, Verhamme2006}. This mechanism should work in LAEs because most LAEs have outflows \citep[e.g.,][]{Hashimoto2013, Shibuya2014a,Hashimoto2015, Guaita2017}. Outflowing gas is also needed to reproduce observed Ly$\alpha$ profiles characterized by a relatively broad, asymmetric shape with a redshifted peak. Note, however, that it is not clear whether LAEs have higher outflow velocities than average galaxies with the same $M_\star$ and $E(B-V)$.

To summarize, low HI column densities combined with some other mechanisms such as outflows appear to contribute to the high $f_{\rm esc}({\rm Ly}\alpha)_{\rm tot}$ seen in LAEs. However, none of these mechanisms can reduce $q$ below unity as long as a uniform ISM is assumed.

\subsubsection{(ii) Less efficient resonant scattering in a clumpy ISM} \label{subsubsec:fesc_geo}
Ly$\alpha$ photons are not attenuated by dust if dust is confined in HI clumps \citep[the clumpy ISMs;][]{Neufeld1991, Hansen2006} 
because Ly$\alpha$ photons are scattered on the surface of clumps before being absorbed by dust. \citet{Scarlata2009} find that the clumpy dust screen (ISMs) can reproduce observed line ratios of Ly$\alpha$ to H$\alpha$ (or $f_{\rm esc}({\rm Ly}\alpha)$), and H$\alpha$ to H$\beta$ (or$E(B-V)$) of local LAEs \citep[see also][]{Bridge2017}. It is, however, not clear what causes such a clumpy ISM geometry especially for LAEs. Indeed, \citet{Laursen2013} argue that any real ISM is unlikely to give $q<1$. \citet{Duval2014} also find that the clumpy ISM model \citep{Neufeld1991} can achieve $q<1$ only under unrealistic conditions: a large covering factor of clumps with high $E(B-V)$, a low HI content in interclump regions, and a uniform, constant, and slow outflow. 

\subsubsection{(iii) Additional Ly$\alpha$ sources }\label{subsubsec:fesc_add}
If galaxies have other Ly$\alpha$-photon sources in the main body besides star formation, the number of produced Ly$\alpha$ photons is larger than expected from the $SFR$, resulting in underestimation of $q$ and overestimation of $f_{\rm esc}({\rm Ly}\alpha)_{\rm tot}$. We discuss three candidate sources: AGNs, cold streams, and hard ionizing spectra.

First, the contribution of AGNs should be modest.
This is because we have removed all objects detected in either X-ray, UV, or radio regarding them as AGNs, and because the fraction of obscured AGNs (AGNs without detection in either X-ray, UV, or radio) in the remaining sample is estimated to be only $2\%$ \citep[see][]{Kusakabe2018a}. 

Second, \citet{Lake2015} have found from hydrodynamical simulations of galaxies with $M_{\rm h}=10^{11.5}\,M_\odot$ at $z\sim 3$ that the Ly$\alpha$ luminosity from cold streams in the central part of galaxies amounts to as high as $\sim45$\% of that from star formation. 
This result may apply to our LAEs to some degree.　

Third, if our LAEs have a hard ionizing spectrum (in other words, the production efficiency of ionizing photons compared to the UV luminosity, $\xi_{\rm ion}$, is large) as suggested in previous studies on higher-$z$ LAEs \citep[at $z\sim3$--$7$: e.g., ][]{Nakajima2016, Harikane2018, Nakajima2018} and brighter LAEs at $z\sim2.2$ \citep{Sobral2018arXiv}, the intrinsic number of ionizing photons is larger than that assumed in equation \ref{eq:sfr_lya}. A hard ionizing spectrum arises from a young age, a low metallicity, a stellar population with a contribution of massive binary systems, an increasing star formation history, and/or a top-heavy IMF. If our LAEs have $\sim0.2$ dex harder $\xi_{\rm ion}$ than the assumed fiducial value ($\log_{10}(\xi_{\rm ion}/{\rm Hz\, erg^{-1}})\sim25.11$), they have $f_{\rm esc}({\rm Ly}\alpha)_{\rm tot}$ lower than unity even in the case of an SMC-like curve. A much harder $\xi_{\rm ion}$ by $\sim0.4$--$1$ dex would even help to explain the difference in $f_{\rm esc}({\rm Ly}\alpha)_{\rm tot}$ between LAEs and HAEs seen in figure \ref{fig:fesc} (right) in the case of the Calzetti curve. To infer $\xi_{\rm ion}$ for our sample, we adopt an empirical relation presented by \citet{Sobral2018arXiv} in their figure 2\footnote{Their $f_{\rm esc}({\rm Ly}\alpha)$ is derived from H$\alpha$ luminosity with dust attenuation correction, $0.9$ mag \citep[see also][]{Sobral2017}, and Ly$\alpha$ flux measured as a point source with a $3''$-diameter aperture. }. This relation implies a higher $f_{\rm esc}({\rm Ly}\alpha)$ and a harder $\xi_{\rm ion}$ for LAEs with a larger $EW_{0,ps}({{\rm Ly}\alpha})$. Using this relation, we indeed
obtain a harder $\xi_{\rm ion}$ of $\log_{10}(\xi_{\rm ion}/{\rm Hz\, erg^{-1}})\sim25.3$ for our large-EW LAE subsample whose typical $EW_{0,ps}({{\rm Ly}\alpha})$ is $\sim70$\AA. This value is also comparable to those found for $z\sim3$ LAEs in \citet{Nakajima2018}. In this case, their total Ly$\alpha$ escape fraction, $f_{\rm esc}({\rm Ly}\alpha)_{\rm tot}$, would become smaller than unity ($\sim0.6$--$0.8$) based on equation \ref{eq:m16}, suggesting that an additional Ly$\alpha$ source is not necessarily needed. However, the same relation gives a modest $\xi_{\rm ion}$ of $\log_{10}(\xi_{\rm ion}/{\rm Hz\, erg^{-1}})\sim25.1$ for the small-EW LAE subsample ($EW_{0,ps}({{\rm Ly}\alpha})\sim25$\AA\,), resulting in $f_{\rm esc}({\rm Ly}\alpha)_{\rm tot}\sim0.1$--$0.3$ which remains significantly higher than those of HAEs with the same  $M_\star$/$SFR$/$E(B-V)$. These calculations imply that it remains uncertain whether or not LAEs , especially those with a small $EW_{0}({{\rm Ly}\alpha})$, typically have a hard ionizing spectrum. They also imply that another mechanism is possibly needed (in addition to hard ionizing spectra) to fully explain the large $f_{\rm esc}({\rm Ly}\alpha)_{\rm tot}$ including the systematic difference from HAEs.

In any case, the very low $q$ values ($\lesssim0$) seen in about half of our LAEs (red filled objects) indicate a non-neglible contribution from additional Ly$\alpha$ sources. \citet{Song2014} have also found several bright LAEs with $q<0$ as shown in their figure 14, where $q$ would decrease more if they include $L(\rm Ly\alpha)_{\rm H}$ in the calculation of $f_{\rm esc}(\rm Ly\alpha)_{\rm tot}$. 

\subsubsection{Summary of the mechanisms affecting the $q$-parameter}
The origin of very high $f_{\rm esc}({\rm Ly}\alpha)_{\rm tot}$ and very low $q$ found for LAEs is a long-standing problem. This study makes this problem more serious by including $L({\rm Ly\alpha})_{\rm H}$ in the calculation of these parameters. Remarkably, all of our subsamples have $q<1$ and a half of them reach $q<0$.

Low $N_{\rm HI}$ and small $\sigma_{\rm Ly\alpha}$ should help to increase $f_{\rm esc}({\rm Ly}\alpha)_{\rm tot}$ and reduce $q$ to some degree. However, additional mechanisms are needed to reduce $q$ less than unity, as highlighted by the very low $q$ values, with some being negative, found for our LAE subsamples. Cold streams in the main body of LAEs and hard ionizing spectra are candidate mechanisms while a clumpy ISM may be unlikely.
The $q$ value of galaxies is probably determined by the balance between the efficiency of resonant scattering and additional Ly$\alpha$-photon sources. Spectroscopic observations of LAEs' H$\alpha$ luminosities would provide more accurate measurements of $f_{\rm esc}({\rm Ly}\alpha)_{\rm tot}$ ($q$-parameters). They will also enable us to evaluate the spectral hardness from the UV to H$\alpha$ luminosity ratio and to constrain the contribution of cold streams from the Ly$\alpha$ to H$\alpha$ luminosity ratio.

Our LAEs have much lower $q$ values than the HAEs in the lowest-$E(B-V)$ bin, which indicates that not all galaxies with small $E(B-V)$ (or equivalently, small $M_\star$) can be LAEs. A possible reason for this large difference is that our LAEs have lower $M_{\rm h}$ and hence lower $M_{\rm HI}$. \citet{Matthee2016}'s HAEs in the lowest-$M_\star$ bin ($M_\star\sim3\times10^{9}$--$8\times10^{9}\,M_\odot$) used for clustering analysis by \citet[][]{Cochrane2018} reside in massive dark matter halos of $M_{\rm h}\sim7^{+9}_{-4}\times10^{12}\,M_\odot$ \citep{Cochrane2018}, which is one dex larger than the average $M_{\rm h}$ of our LAEs
\footnote{
We calculate this $M_{\rm h}$ value from the correlation length given in \citet{Cochrane2018} in the same manner as for our LAEs. Their $r_0$ and $M_{\rm h}$ are higher than those averaged over all the LAEs ($r_0=2.30^{+ 0.36}_{- 0.41}\,h^{-1}\,Mpc$, i.e., $M_{\rm h}=$$3.2^{+ 4.7}_{- 2.5}\times10^{10}\,M_\odot$), although their median $M_\star$ ($\sim6\times10^{9}\,M_\odot$) is slightly higher than our average value ($\sim1\times10^{9}\,M_\odot$). }.
It would imply that the lowest-$E(B-V)$ HAEs in \citet{Matthee2016} have higher $M_{\rm h}$ than our LAEs, since the lowest-$E(B-V)$ HAEs should largely overlap with the lowest-$M_\star$ HAEs. Furthermore, a large fraction of low-$M_\star$ ($M_\star\lesssim10^{9}\,M_\star$) HAEs are expected to be starburst galaxies as shown in figure \ref{fig:MS}, implying a large amount of gas (including HI) to fuel the starburst. 
However, the higher $M_{\rm h}$ also imply that they have brighter $L({\rm Ly\alpha})$ from  cold streams (in the main body). 
If the higher $M_{\rm HI}$ can reduce the $L({\rm Ly\alpha})$ produced from both star formation and cold streams sufficiently, the higher $q$ values of the HAEs can be reproduced.

\section{Conclusions}\label{sec:conclusion}
We have studied the dependence of LAH luminosity on stellar properties and dark matter halo mass using $\sim900$ star forming LAEs at $z\sim2$ to identify the dominant origin of LAHs. To do so, we have divided the whole sample into ten subsamples in accordance with five physical quantities ($m_{\rm K}$, $M_{\rm UV}$, $\beta$, $L({\rm Ly\alpha})$ and $EW_0({\rm Ly}\alpha)$), some of which are expected to correlate with $M_\star$ and 
$M_{\rm h}$. We have estimated for each subsample the LAH luminosity from  a stacked observational relation obtained by \citet{Momose2016}. We have used the obtained dependence of LAH luminosity to test three candidate origins: cold streams, satellite star formation, and resonant scattering. We have also derived 
total Ly$\alpha$ escape fractions and $q$ values by including the halo component, and discussed how LAEs can have high escape fractions. Our main results are as follows. 

\begin{enumerate}
\item We compare \citet{Momose2016}'s observational $L({\rm Ly}\alpha)_{\rm C}$--$L({\rm Ly}\alpha)_{\rm H}$ relation obtained from stacking analysis of essentially the same sample as ours, with the distribution of individual LAEs by VLT/MUSE in \citet{Leclercq2017}. We find that their observational relation agrees well with the average trend of individual LAEs as shown in figure \ref{fig:lah}, ensuring the use of the relation for our analysis.

\item Our LAEs are found to lie on an extrapolation of the $M_\star$--$IRX$ relation at $z\sim1.5$ in \citet{Heinis2014} and that of the SFMS at $z\sim2$ in \citet{Shivaei2017} if an SMC-like attenuation curve is assumed (shown in figures \ref{fig:IRX} and \ref{fig:MS}). These results are used in the discussion of the origin of LAHs. 

\item The ten subdivided LAE samples are found to have similar $L({\rm Ly}\alpha)_{\rm H}\sim2\times10^{42}\,{\rm erg\,s^{-1}}$ and $L({\rm Ly}\alpha)_{\rm tot}\sim2\times10^{42}$--$4\times10^{42}\,{\rm erg\,s^{-1}}$ (shown in figure \ref{fig:lah}). 
Their $L({\rm Ly}\alpha)_{\rm H}$ and $L({\rm Ly}\alpha)_{\rm tot}$ remain almost unchanged or even decrease when $M_\star$ increases by factor $2$--$5$. They are also nearly independent of $SFR$, $E(B-V)_\star$, and $M_{\rm h}$, although the uncertainties in $M_{\rm h}$ are large. The HAEs in \citet{Matthee2016} also have non-increasing $L({\rm Ly}\alpha)_{\rm H}$ and $L({\rm Ly}\alpha)_{\rm tot}$. 
These results are inconsistent with the cold stream scenario and the satellite star formation scenario both of which predict a nearly linear scaling of $L({\rm Ly}\alpha)_{\rm H}$ with mass (figure \ref{fig:Lhmodel}). Specifically, the power law slope of the $M_\star$--$L({\rm Ly}\alpha)_{\rm H}$ relation for the $K$-divided subsamples, the most stellar-mass sensitive subsamples, is shallower than predictions with more than the $2\,\sigma$ confidence level.
The former scenario also fails to reproduce the bright $L({\rm Ly}\alpha)_{\rm H}$ of low-mass subsamples  at, e.g., a more than the $10\sigma$ level for the faint $m_K$ subsample. The most likely is the resonant scattering scenario because it can naturally explain these results. 
\item 
The $f_{\rm esc}({\rm Ly}\alpha)_{\rm tot}$ of all ten subsamples is higher than $\sim30$\%, with some exceeding $100$\%, with very low $q$ values of $-1 \lesssim q \lesssim 1$. Using the Calzetti curve instead of an SMC-like curve makes $f_{\rm esc}({\rm Ly}\alpha)_{\rm tot}$ lower but still in a range of $10$--$100$\% with $q\sim0$--$1$. 
The $f_{\rm esc}({\rm Ly}\alpha)_{\rm tot}$ of the LAEs anti-correlates with $M_\star$, $SFR$, and $E(B-V)$ regardless of the assumed attenuation curve (figure \ref{fig:fesc}).
Their $f_{\rm esc}({\rm Ly}\alpha)_{\rm tot}$ and $q$ are higher and lower, respectively, than those of HAEs with similar $M_\star$ and $E(B-V)$. 
The very low $q$ values of the LAEs suggest the existence of an additional Ly$\alpha$ source in the main body; Ly$\alpha$ emission from cold streams and hard ionizing spectra are possible candidates.
The difference in $q$ between the LAEs and HAEs is possibly caused by a different balance between resonant scattering and additional Ly$\alpha$-photon source(s). 
\end{enumerate}

In the near future, we will obtain much better $M_{\rm h}$ estimates for $\sim9000$ LAEs with new $NB387$ data from $\simeq 25$ deg$^2$ taken with Hyper Suprime-Cam \citep[SILVERRUSH; ][]{Ouchi2018,Shibuya2018a} as part of a large imaging survey program \citep{Aihara2018a}. It will enable us to compare observed relations of $L({\rm Ly}\alpha)_{\rm tot}$ with theoretical predictions more directly.

\section*{Acknowledgments}
We thank the anonymous referee for constructive comments and suggestions. We are grateful to Kazuyuki Ogura and Masahiro Nagashima for kindly providing $\nu^2$GC data and giving helpful comments. We are grateful to Yoshiaki Ono for giving insightful comments and suggestions on SED fitting. We would like to express our gratitude to Jorryt Matthee and Ken-ichi Tadaki for kindly providing their data plotted in figures \ref{fig:fesc}, and \ref{fig:MS}--\ref{fig:q} and figure \ref{fig:MS}, respectively. We thank David Sobral for giving insightful comments and suggestions. We acknowledge Ryosuke Goto, Akira Konno, Ryota Kawamata, Taku Okamura, Kazushi Irikura, Ryota Kakuma, and Makoto Ando for constructive discussions at meetings. This work is based on observations taken by the Subaru Telescope which is operated by the National Astronomical Observatory of Japan. The authors wish to recognize and acknowledge the very significant cultural role and reverence that the summit of Maunakea has always had within the indigenous Hawaiian community. Based on data products produced by TERAPIX and the Cambridge Astronomy Survey Unit on behalf of the UltraVISTA consortium. This research made use of IRAF, which is distributed by NOAO, which is operated by AURA under a cooperative agreement with the National Science Foundation and of Python packages for Astronomy: Astropy \citep[][]{TheAstropyCollaboration2013}, Colossus, CosmoloPy and PyRAF, which is produced by the Space Telescope Science Institute, which is operated by AURA for NASA. H.K acknowledges support from the JSPS through the JSPS Research Fellowship for Young Scientists. This work is supported in part by KAKENHI (16K05286) Grant-in-Aid for Scientific Research (C) through the JSPS.


\begin{appendix}
\section{NB selection bias}\label{sec: appendix_NBbias}

In this Appendix, we first describe the NB-selection bias of our LAE sample, and then discuss the effect of this bias on the obtained $M_\star$--$L({\rm Ly}\alpha)_{\rm H}$ relations. As shown in figures \ref{fig:param} (g) and (h), our sample misses UV-faint LAEs ($M_{UV}\gtrsim-19$ mag) with faint $L({\rm Ly}\alpha)_{ps}$ and small $EW_{0,\,ps}({\rm Ly}\alpha)$. This selection bias has the following effects on subsample properties. 

\begin{description}
 \item[$M_{UV}$ and $m_K$ subsamples]
 The UV-faint ($M_{UV}>-19.2$ mag) subsample is biased toward  brighter $L({\rm Ly}\alpha)_{ps}$ and larger $EW_{0,\,ps}({\rm Ly}\alpha)$. The $K$-faint subsample ($m_K>25.0$ mag) is probably biased similarly. Although the $L({\rm Ly}\alpha)_{\rm H}$ of these subsamples is probably overestimated, we find in appendix \ref{sec:appendix_LAH} that it does not change our results. This selection bias probably does not change $M_\star$ values since $m_K$ and $M_{UV}$ are a good tracer of $M_\star$. The bright $m_K$ and $M_{UV}$ subsamples are almost free from this bias. 

 \item[$\beta$ subsamples]
 Galaxies with fainter UV luminosities generally have smaller $\beta$ \citep[e.g.,][]{Alavi2014}. Although our $\beta$ subsamples are probably biased to some degree, it is difficult to evaluate the effects on $M_\star$ and $L({\rm Ly}\alpha)_{\rm H}$ estimates quantitatively. However, the effects should be smaller than those on the UV and $K$ subsamples, since the $M_{UV}$--$\beta$ correlation has a large scatter (see figure \ref{fig:param}[f]).

 \item[$L({\rm Ly}\alpha$)$_{ps}$ and $EW_{0, ps}$(Ly$\alpha$) subsamples]
 The faint $L({\rm Ly}\alpha)$ and small $EW({\rm Ly}\alpha)$ subsamples are biased toward bright UV magnitudes. Although their $L({\rm Ly}\alpha)_{\rm H}$ values are probably not affected by the selection bias, their $M_\star$ values are expected to be overestimated to some degree. The bright $L({\rm Ly}\alpha)$ and large $EW({\rm Ly}\alpha)$ subsamples are not biased. If the $M_\star$ of the faint $L({\rm Ly}\alpha)$ and small $EW({\rm Ly}\alpha)$ subsamples decreases, the power-law slope of the $M_\star$--$L({\rm Ly}\alpha)_{\rm H}$ relation becomes shallower, enlarging the descrepancy from the models of cold streams and satellite star formation.

 \end{description}

In appendix \ref{sec:appendix_LAH}, we use the MUSE sample to evaluate the robustness of $L({\rm Ly}\alpha)_{\rm H}$ estimates for our faint $m_K$ and $M_{UV}$ subsamples. The MUSE sample is complementary to our sample, because it is essentially UV-limited but contains much fewer objects than ours.

\section{Issues on ZP offsets of broadband images}\label{sec:appendix_zp}
In section \ref{subsec:ulya}, we find that the distribution of $M_{\rm UV}$, $\beta$, $L({\rm Ly\alpha})_{ps}$, and $EW_{0, ps}({\rm Ly}\alpha)$ is different between the SXDS and COSMOS fields. LAEs in the SXDS field tend to have bluer $\beta$, fainter $L({\rm Ly\alpha})_{ps}$ and weaker $EW_{0, ps}({\rm Ly}\alpha)$ compared with those in the COSMOS fields. This is possibly because of systematic offsets of the ZP magnitudes of the optical broadband images adopted in the original papers \citep{Capak2007,Taniguchi2007,Furusawa2008}. According to \citet{Yagi2013} and \citet{Skelton2014}, the amount of ZP offsets of the $B$, $V$, and $R$ images in the SXDS field is expected to be $\sim0.0$--$0.2$ mag. \citet{Capak2007}, \citet{Ilbert2009}, and \citet{Skelton2014} suggest that the amount of ZP offsets of the $B$, $V$, and $R$ images in the COSMOS field is $\sim0.0$--$0.2$ mag. If the relative ZP offset of $B$ to $V$ images is $0.04$ mag \citep[as suggested for SXDS-North subfield in][]{Yagi2013}, the corresponding shift of $\beta$ is $0.18$. If the case with a larger relative offset of $0.148$ mag \citep[as suggested for COSMOS field in][]{Ilbert2009}, the corresponding shift of $\beta$ is as large as $0.66$. 

More seriously, papers in a given field often claim opposite shift directions, and we can not give a firm conclusion. For instance, in SXDS field, the ZP corrections based on \citet{Yagi2013} make $\beta$ redder, while those based on \citet{Skelton2014} make $\beta$ bluer. In this paper, we use the original ZPs following \citet{Kusakabe2018a} and include ZP uncertainties in the flux-density errors in the calculations given in sections \ref{subsec:ulya} and \ref{sec:sed}. In the following paragraphs, we roughly estimate possible offsets of $\beta$, $L({\rm Ly\alpha})_{ps}$, and $EW_{0, ps}({\rm Ly}\alpha)$ due to such ZP offsets.

The shifts of $\beta$ have a larger effect on $L({\rm Ly\alpha})_{ps}$ and $EW_{0, ps}({\rm Ly}\alpha)$ for smaller-$EW_{0, ps}({\rm Ly}\alpha)$ objects, since the contribution of the continuum emission around rest-frame $1216$\AA\ in an $NB387$ image is larger than that for larger-$EW_{0, ps}({\rm Ly}\alpha)$ objects as shown in equations \ref{eq:uvlya1} and \ref{eq:uvlya2}. To evaluate this effect quantitatively, we consider a simple case that the ZPs of the broadband images in one of the two fields are shifted resulting in a difference of $\sim0.4$ in $\beta$, which is found in figure \ref{fig:param}. 

For instance, a large-$EW_{0, ps}({\rm Ly}\alpha)$ object with $M_{\rm UV}=-19$ mag, $\beta=-2.0$, and $EW_{0, ps}({\rm  Ly}\alpha)=100$ \AA\ has $NB387=24.38$ mag. If the $\beta$ is overestimated to be $-1.6$ due to ZP shifts of broad-bands images (with a correct $M_{\rm UV}$ estimation at rest-frame $1600$\AA\ of $-19$ mag), the corresponding flux density at rest-frame $1216$\AA\ is underestimated. It results in relatively small shifts of $L({\rm Ly\alpha})_{ps}$ and $EW_{0, ps}({\rm Ly}\alpha)$, $\sim3$ \% and $\sim15$ \% overestimation, respectively, for the same $NB387$ magnitude ($24.38$ mag). On the other hand, the same shift of $\beta$ causes larger overestimation of $L({\rm Ly\alpha})_{ps}$ and $EW_{0, ps}({\rm Ly}\alpha)$, $\sim14$ \% and $\sim28$ \%, respectively, for a smaller-$EW_{0, ps}({\rm Ly}\alpha)$ object with $M_{\rm UV}=-19$ mag, $\beta=-2.0$, and $EW_{0, ps}({\rm  Ly}\alpha)=20$ \AA\ (and with a fixed $NB387$, $25.44$ mag). In an extreme case of a very-small-$EW_{0, ps}({\rm Ly}\alpha)$ object with $EW_{0, ps}({\rm Ly}\alpha)=10$\AA\footnote{Our LAEs are selected from narrow band selection criteria with $20$--$30$ \AA\ cut for galaxy SEDs with $30$ Myr without dust attenuation
(i.e. a fixed $\beta$) as shown in figure \ref{fig:param} in \citet{Konno2016}. The limiting EW depends on $\beta$ as is common in narrow band selections of LAEs. Therefore, EWs of some of
our LAEs are derived to be smaller than $20$--$30$ \AA.}, a large overestimation of $\beta=-1.34$ (0.66 offset) can overestimate $L({\rm Ly\alpha})_{ps}$ and $EW_{0, ps}({\rm Ly}\alpha)$ by $\sim46$ \% and $\sim75$\%, respectively. This $L({\rm Ly\alpha})_{ps}$ offset reaches $\sim0.2$ dex, which corresponds to the difference in the minimum $L({\rm Ly\alpha})_{ps}$ in figure \ref{fig:param}.

As shown in figure \ref{fig:param}, since the $EW_{0, ps}({\rm Ly}\alpha)$ of a stacked LAE for each subsample is larger than $20$ \AA, such ZP offsets do not change our results, though $L({\rm Ly\alpha})_{ps}$ and $EW_{0, ps}({\rm Ly}\alpha)$ of some of our individual LAEs (before stacking) with a small $EW_{0, ps}({\rm Ly}\alpha)$ are possibly under/overestimated.

\section{Robustness of $L({\rm Ly}\alpha)_H$ estimates}\label{sec:appendix_LAH}

\begin{figure}[ht]
\begin{flushleft}
      \includegraphics[width=0.9\linewidth]{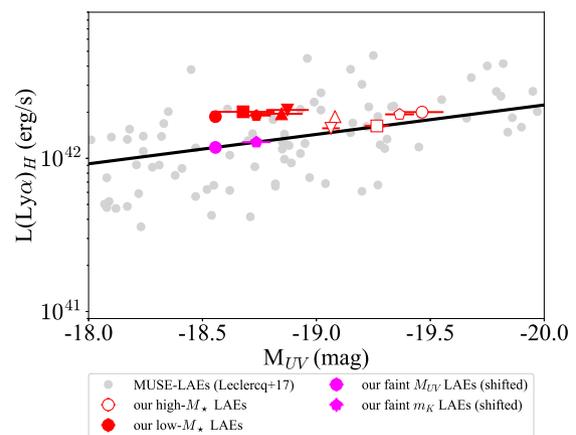}
\end{flushleft}      
  \caption{  
   $L({\rm Ly\alpha})_{\rm H}$ as a function of $M_{\rm UV}$. 
   Grey points represent MUSE-LAEs at $z\sim3-6$ and a black solid line the best fit of a linear function to them \citep{Leclercq2017}. The field average values of our ten subsamples using the stacked relation (equation \ref{eq:m16}) are shown by red symbols below: 
open (filled) circles for bright (faint) $M_{\rm UV}$, 
open (filled) triangles for red (blue) $\beta$, 
open (filled) inverted triangles for faint (bright) $L({\rm Ly\alpha})_{ps}$, 
open (filled) squares for small (large) $EW_{0,ps}({{\rm Ly}\alpha})$, 
and open (filled) pentagons for bright (faint) $m_{\rm K}$.  Results using \citet{Leclercq2017}'s relation for two subsamples are shown by filled magenta symbols: a circle for the faint $M_{\rm UV}$ subsample and a pentagon for the faint $m_K$ subsample. (Color online)
}
  \label{fig:muv_lh}
\end{figure}

\begin{figure}[ht] 
\begin{flushleft}
\begin{flushleft}
      \includegraphics[width=0.95\linewidth]{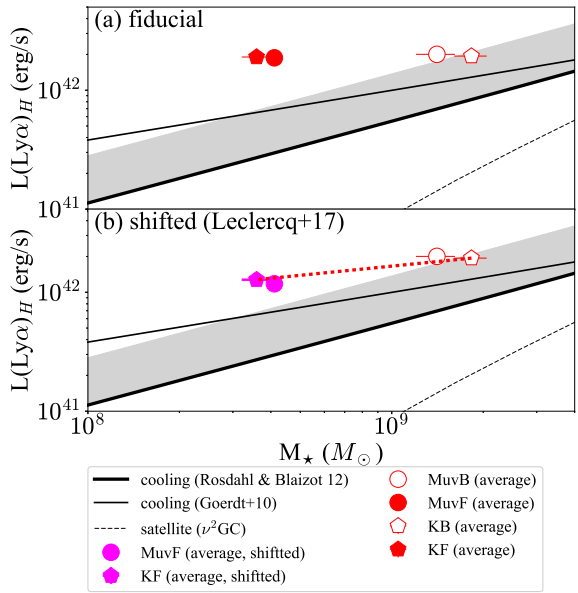}
\end{flushleft}      
\end{flushleft}      
  \caption{ 
  $L({\rm Ly\alpha})_{\rm H}$ vs. $M_\star$: (a) the fiducial results and (b) NB-selection bias corrected results using the $M_{UV}$--$L({\rm Ly\alpha})_{\rm H}$ relation in \citet{Leclercq2017}. The field average values of our $M_{UV}$ and $m_K$ subsamples using the stacked relation (equation \ref{eq:m16}) are shown by red symbols below: 
open (filled) circles for bright (faint) $M_{\rm UV}$, and open (filled) pentagons for bright (faint) $m_{\rm K}$.  Results using \citet{Leclercq2017}'s relation for two subsamples are shown by filled magenta symbols: a circle for the faint $M_{\rm UV}$ subsample and a pentagon for the faint $m_K$ subsample. 
 Thick and thin solid black lines show the Ly$\alpha$ luminosities from cold streams (cooling flows) by theoretical models in \citet{Rosdahl2012} and \citet{Goerdt2010}, respectively, which are converted from original $M_\star$--$L({\rm Ly\alpha})_{\rm H}$ relations using the $M_\star$--$M_{\rm h}$ relation in \citet{Moster2013}. Gray shaded regions above the solid black lines roughly indicate the distribution of \citet{Rosdahl2012}'s simulated galaxies above the solid line, whose $L({\rm Ly\alpha})_{\rm H}$ reaches at most $\sim2.5$ times higher than the line. A black dashed line shows the Ly$\alpha$ luminosities from satellite star formation calculated by a theoretical model \citep[$\nu^2$GC][Ogura et al. in prep.]{Makiya2016, Shirakata2018}. A dotted red line in panel (b) shows the slope determined by the $m_K$ subsamples.  (Color online)
}
  \label{fig:NBbias}
\end{figure}

We first examine the robustness of the stacked relation (equation \ref{eq:m16}) in \citet{Momose2016}. We then evaluate the effects of the NB-selection bias on the $m_K$ and $M_{UV}$ subsamples. 

 To test the robustness of $L({\rm Ly\alpha})_{\rm H}$ values derived from equation \ref{eq:m16}, we calculate $L({\rm Ly\alpha})_{\rm H}$ from $EW_{0,ps}(L_{\rm Ly\alpha})$ using another stacked relation presented in \citet{Momose2016}, an anti-correlation between ${\rm X}(L_{\rm Ly\alpha})_{\rm tot/C}$ and $EW_{0,ps}(L_{\rm Ly\alpha})$. We find that using this relation gives nearly the same $L({\rm Ly\alpha})_{\rm H}$ values as those derived from equation \ref{eq:m16}, with differences being at most $0.09$ dex. 

To evaluate the effects of the NB-selection bias on $L({\rm Ly\alpha})_{\rm H}$ for the faint $m_K$ and $M_{UV}$ subsamples, we re-estimate $L({\rm Ly\alpha})_{\rm H}$ with a complementary result of the MUSE-LAEs in \citet{Leclercq2017} which is essentially free from this kind of bias: the relation between $M_{UV}$ and $L({\rm Ly\alpha})_{\rm H}$. They have found $L({\rm Ly\alpha})_{\rm H}$ anti-correlates with $L_{\rm UV}$ (see their figure 16). As shown in figure \ref{fig:muv_lh}, our high-$M_\star$ LAEs (red filled objects), which are not affected by the NB-bias, are consistent with the best-fit relation of MUSE-LAEs (black solid line), while the faint $m_K$ and $M_{UV}$ subsamples are found to lie alightly above the relation. As a result, the power-law slopes of the $m_K$ and $M_{UV}$ subsamples become positive as shown in figure \ref{fig:NBbias}. However, they are still shallow. For example, the $m_K$-divided subsamples give a power-law index of $0.26\pm0.05$, which is more than $2\,\sigma$ shallower than those of the cold streams models in \citet{Goerdt2010} and \citet{Rosdahl2012}, $\sim0.38$ and $\sim0.75$, respectively. This slope is also more than $20\,\sigma$ shallower than that of the satellite star formation model, $\sim1.36$. Moreover, the $L({\rm Ly\alpha})_{\rm H}$ values of the faint $m_K$ and $M_{UV}$ subsamples also remain higher than predicted from the cold streams models at a $>10\,\sigma$ level. We conclude that the conclusions obtained in section 6 are robust.

\section{Estimated Ly$\alpha$ luminosities}\label{sec:appendix_Llya}
\begin{table}
\tbl{Ly$\alpha$ luminosities for the subsamples.
}{
\begin{tabular}{lccc}
\hline
subsamples & $L({\rm Ly}\alpha)_{\rm C}$ &  $L({\rm Ly}\alpha)_{\rm H}$ & $L({\rm Ly}\alpha)_{\rm tot}$\\
 & $10^{41}\,L_\odot$ & $10^{41}\,L_\odot$ & $10^{41}\,L_\odot$ \\
& (1) & (2) & (3)  \\\hline 
 \multicolumn{4}{c}{SXDS}\\
 bright UV& $7.7^{+2.3}_{-1.5}$ & $15.9^{+2.0}_{-1.7}$ & $23.6^{+4.3}_{-3.2}$  \\  
faint UV & $9.4^{+1.8}_{-0.8}$ & $17.5^{+1.4}_{-0.7}$ & $26.9^{+3.2}_{-1.5}$  \\  
blue $\beta$ & $9.1^{+2.1}_{-1.2}$ & $17.3^{+1.6}_{-1.1}$ & $26.4^{+3.7}_{-2.4}$  \\  
red $\beta$ & $8.5^{+1.8}_{-0.9}$ & $16.8^{+1.5}_{-0.9}$ & $25.3^{+3.4}_{-1.8}$  \\  
bright Ly$\alpha$ & $13.8^{+2.4}_{-1.1}$ & $20.2^{+0.8}_{-0.5}$ & $34.0^{+3.3}_{-1.5}$  \\  
faint Ly$\alpha$ & $6.2^{+1.9}_{-1.0}$ & $14.3^{+2.1}_{-1.4}$ & $20.5^{+4.1}_{-2.4}$  \\  
large EW & $12.5^{+2.1}_{-0.8}$ & $19.6^{+0.9}_{-0.4}$ & $32.1^{+3.1}_{-1.2}$  \\  
small EW & $6.6^{+2.1}_{-1.3}$ & $14.7^{+2.2}_{-1.7}$ & $21.3^{+4.2}_{-3.0}$  \\  
bright $K$ & $7.9^{+2.3}_{-1.2}$ & $16.1^{+2.0}_{-1.3}$ & $24.0^{+4.4}_{-2.5}$  \\  
faint $K$& $9.1^{+1.9}_{-0.9}$ & $17.3^{+1.5}_{-0.8}$ & $26.3^{+3.4}_{-1.7}$  \\ \hline 
 \multicolumn{4}{c}{COSMOS}\\
bright UV & $14.7^{+3.0}_{-1.4}$ & $20.6^{+0.8}_{-0.6}$ & $35.3^{+3.8}_{-2.0}$  \\  
faint UV & $11.9^{+2.0}_{-0.6}$ & $19.2^{+1.0}_{-0.4}$ & $31.1^{+2.9}_{-1.0}$  \\  
blue $\beta$ & $13.5^{+2.4}_{-1.0}$ & $20.1^{+0.9}_{-0.5}$ & $33.5^{+3.2}_{-1.5}$  \\  
red $\beta$ & $12.4^{+2.3}_{-0.9}$ & $19.5^{+1.0}_{-0.5}$ & $31.9^{+3.3}_{-1.5}$  \\  
bright Ly$\alpha$ & $15.7^{+2.5}_{-0.9}$ & $20.9^{+0.5}_{-0.3}$ & $36.6^{+3.1}_{-1.2}$  \\  
faint Ly$\alpha$ & $8.1^{+1.8}_{-0.8}$ & $16.4^{+1.6}_{-0.8}$ & $24.5^{+3.4}_{-1.7}$  \\  
large EW & $14.3^{+2.4}_{-0.7}$ & $20.4^{+0.7}_{-0.3}$ & $34.7^{+3.1}_{-0.9}$  \\  
small EW & $8.9^{+2.2}_{-1.2}$ & $17.1^{+1.7}_{-1.1}$ & $26.0^{+3.9}_{-2.3}$  \\  
bright $K$ & $13.4^{+2.7}_{-1.1}$ & $20.0^{+1.0}_{-0.6}$ & $33.4^{+3.6}_{-1.7}$  \\  
faint $K$& $12.6^{+2.1}_{-0.8}$ & $19.6^{+0.9}_{-0.4}$ & $32.2^{+3.0}_{-1.2}$  \\  \hline
\end{tabular}
}
\label{tbl:Lya}
\tabnote{Note. (1) Ly$\alpha$ luminosity at the central part derived by multiplying $L({\rm Ly}\alpha)_{ps}$ by $0.77$; (2) Ly$\alpha$ luminosity of the LAH derived from equation \ref{eq:m16}; (3) total Ly$\alpha$ luminosity derived from equation \ref{eq:m16}.}
\end{table}

In table \ref{tbl:Lya}, we show the three kinds of Ly$\alpha$ luminosities for individual subsamples. Note that the typical $1\sigma$ uncertainties in the individual data points in \citet{Momose2016}'s $L({\rm Ly\alpha})_{\rm C}$ - $L({\rm Ly\alpha})_{\rm H}$ relation are propagated to uncertainties in $L({\rm Ly\alpha})_{\rm H}$ and $L({\rm Ly\alpha})_{\rm tot}$ of $\sim22\%$ and $\sim16\%$, respectively (see figure \ref{fig:Lh})

\begin{table*}
\tbl{ Best-fit parameters of SED fitting for the two fields.}{
\scalebox{1.0}[0.75]{ 
\begin{tabular}{lccccc}
\hline
subsample &$M_{\star}$ & $E(B-V)_{\star}$ & Age & $SFR$ & $\chi^2_r$\\ 
 & ($10^8 M_{\odot}$) & (mag) & (Myr) & ($M_{\odot}$yr$^{-1}$) & \\
 & (1) & (2) & (3) & (4) & (5) \\\hline
 \multicolumn{6}{c}{SXDS field/ SMC-like attenuation curve } \\
bright UV & $12.5^{+4.5}_{-2.1}$& $0.07^{+0.01}_{-0.02}$ & $ 255^{+ 198}_{-  74}$& $ 5.9^{+ 0.9}_{- 1.3}$ & $0.538$\\
faint UV & $4.1^{+1.4}_{-1.5}$& $0.02^{+0.02}_{-0.01}$ & $ 321^{+ 188}_{- 178}$& $ 1.5^{+ 0.5}_{- 0.2}$ & $0.139$\\
blue $\beta$ & $7.1^{+2.4}_{-1.8}$& $0.02^{+0.01}_{-0.01}$ & $ 404^{+ 236}_{- 149}$& $ 2.1^{+ 0.4}_{- 0.3}$ & $0.588$\\
red $\beta$ & $14.9^{+3.2}_{-3.8}$& $0.10^{+0.02}_{-0.01}$ & $ 286^{+ 118}_{- 125}$& $ 6.2^{+ 1.8}_{- 0.8}$ & $2.282$\\ 
bright Ly$\alpha$ & $6.9^{+1.6}_{-2.3}$& $0.02^{+0.02}_{-0.01}$ & $ 453^{+ 187}_{- 226}$& $ 1.9^{+ 0.5}_{- 0.2}$ & $0.289$\\ 
faint Ly$\alpha$ & $11.5^{+4.3}_{-2.0}$& $0.06^{+0.01}_{-0.02}$ & $ 360^{+ 280}_{- 105}$& $ 3.9^{+ 0.6}_{- 0.8}$ & $1.461$\\ 
large EW & $4.5^{+1.6}_{-1.5}$& $0.02^{+0.02}_{-0.01}$ & $ 360^{+ 211}_{- 180}$& $ 1.5^{+ 0.5}_{- 0.2}$ & $0.255$\\ 
small EW & $11.7^{+4.3}_{-2.0}$& $0.06^{+0.01}_{-0.02}$ & $ 321^{+ 250}_{-  94}$& $ 4.4^{+ 0.7}_{- 0.9}$ & $0.775$\\
bright $K$ & $21.5^{+5.5}_{-5.3}$& $0.08^{+0.01}_{-0.02}$ & $ 453^{+ 265}_{- 167}$& $ 5.8^{+ 1.0}_{- 1.1}$ & $0.680$\\ 
faint $K$ &  $3.8^{+1.3}_{-1.4}$& $0.03^{+0.02}_{-0.02}$ & $ 203^{+ 158}_{- 112}$& $ 2.2^{+ 0.8}_{- 0.5}$ & $0.692$\\ \hline
 \multicolumn{6}{c}{SXDS field/ the Calzetti attenuation curve } \\
bright UV & $12.0^{+3.0}_{-3.8}$& $0.13^{+0.03}_{-0.03}$ & $ 143^{+ 112}_{-  79}$& $ 9.7^{+ 4.5}_{- 2.7}$ & $0.902$ \\ 
faint UV & $3.1^{+2.3}_{-1.8}$& $0.06^{+0.05}_{-0.05}$ & $ 161^{+ 348}_{- 128}$& $ 2.3^{+ 2.3}_{- 1.0}$ &  $0.114$ \\ 
blue $\beta$ & $6.7^{+2.7}_{-2.4}$& $0.04^{+0.05}_{-0.03}$ & $ 321^{+ 320}_{- 207}$& $ 2.5^{+ 1.8}_{- 0.7}$ &  $0.581$ \\ 
red $\beta$ & $16.0^{+3.7}_{-4.0}$& $0.18^{+0.02}_{-0.02}$ & $ 161^{+  94}_{-  70}$& $11.6^{+ 3.5}_{- 2.4}$ & $2.978$ \\ 
bright Ly$\alpha$ & $5.2^{+3.3}_{-3.3}$& $0.07^{+0.06}_{-0.06}$ & $ 203^{+ 438}_{- 174}$& $ 3.0^{+ 4.0}_{- 1.4}$ & $0.268$ \\ 
faint Ly$\alpha$ &  $10.9^{+3.2}_{-2.7}$& $0.12^{+0.02}_{-0.04}$ & $ 203^{+ 202}_{-  89}$& $ 6.4^{+ 1.9}_{- 2.1}$ & $1.550$ \\ 
large EW &$2.8^{+3.3}_{-1.9}$& $0.09^{+0.04}_{-0.08}$ & $ 102^{+ 469}_{-  85}$& $ 3.1^{+ 2.8}_{- 1.8}$ & $0.212$ \\ 
small EW & $11.1^{+2.9}_{-2.9}$& $0.12^{+0.03}_{-0.03}$ & $ 181^{+ 141}_{-  90}$& $ 7.2^{+ 3.0}_{- 2.0}$ & $1.016$ \\ 
bright $K$ & $16.1^{+5.8}_{-4.2}$& $0.17^{+0.03}_{-0.04}$ & $ 143^{+ 143}_{-  72}$& $13.0^{+ 5.5}_{- 4.4}$ & $1.012$ \\ 
faint $K$ &  $3.6^{+1.3}_{-1.5}$& $0.06^{+0.03}_{-0.03}$ & $ 143^{+ 143}_{-  91}$& $ 2.9^{+ 1.5}_{- 0.8}$ & $0.673$ \\ \hline\hline
 \multicolumn{6}{c}{COSMOS field/ SMC-like attenuation curve } \\
bright UV & $16.8^{+5.9}_{-2.9}$& $0.09^{+0.01}_{-0.02}$ & $ 227^{+ 177}_{-  66}$& $ 8.8^{+ 1.3}_{- 1.9}$ & $0.377$ \\ 
faint UV & $4.2^{+2.7}_{-1.8}$& $0.05^{+0.02}_{-0.02}$ & $ 227^{+ 282}_{- 137}$& $ 2.2^{+ 0.9}_{- 0.5}$ &   $0.244$ \\ 
blue $\beta$ & $2.3^{+2.5}_{-1.8}$& $0.03^{+0.02}_{-0.02}$ & $ 114^{+ 246}_{- 105}$& $ 2.3^{+ 2.7}_{- 0.7}$ & $0.458$ \\ 
red $\beta$ &$13.1^{+4.8}_{-2.3}$& $0.09^{+0.01}_{-0.02}$ & $ 286^{+ 223}_{-  84}$& $ 5.5^{+ 0.8}_{- 1.2}$ &  $0.560$ \\ 
bright Ly$\alpha$ & $8.5^{+2.8}_{-2.7}$& $0.06^{+0.02}_{-0.01}$ & $ 286^{+ 167}_{- 143}$& $ 3.6^{+ 1.1}_{- 0.5}$ & $0.257$ \\  
faint Ly$\alpha$ & $13.5^{+4.6}_{-3.3}$& $0.08^{+0.01}_{-0.01}$ & $ 360^{+ 211}_{- 133}$& $ 4.6^{+ 0.8}_{- 0.6}$ & $2.238$ \\ 
large EW & $8.1^{+2.8}_{-2.6}$& $0.06^{+0.02}_{-0.01}$ & $ 321^{+ 188}_{- 160}$& $ 3.1^{+ 1.0}_{- 0.4}$ & $0.311$ \\ 
small EW & $19.5^{+7.3}_{-3.5}$& $0.08^{+0.01}_{-0.02}$ & $ 404^{+ 315}_{- 118}$& $ 5.9^{+ 0.8}_{- 1.2}$ & $3.052$ \\ 
bright $K$ & $16.7^{+3.5}_{-4.2}$& $0.10^{+0.02}_{-0.01}$ & $ 227^{+  94}_{-  99}$& $ 8.7^{+ 2.5}_{- 1.1}$ & $0.208$\\ 
faint $K$ & $2.9^{+2.2}_{-2.4}$& $0.06^{+0.02}_{-0.02}$ & $ 114^{+ 172}_{- 107}$& $ 3.0^{+ 4.2}_{- 0.8}$ & $0.278$ \\ \hline
 \multicolumn{6}{c}{COSMOS field/ the Calzetti attenuation curve } \\
bright UV & $15.5^{+5.5}_{-5.6}$& $0.17^{+0.02}_{-0.03}$ & $ 102^{+ 101}_{-  57}$& $17.4^{+ 6.7}_{- 5.1}$ & $1.185$ \\ 
faint UV & $2.6^{+2.5}_{-1.1}$& $0.13^{+0.02}_{-0.05}$ & $  57^{+ 170}_{-  32}$& $ 5.1^{+ 2.1}_{- 2.4}$ &   $0.213$ \\ 
blue $\beta$ & $1.7^{+2.6}_{-1.0}$& $0.08^{+0.02}_{-0.06}$ & $  47^{+ 239}_{-  35}$& $ 3.9^{+ 1.9}_{- 2.1}$ &  $0.413$ \\ 
red $\beta$ & $10.9^{+5.1}_{-3.4}$& $0.18^{+0.02}_{-0.03}$ & $ 102^{+ 126}_{-  52}$& $12.2^{+ 4.2}_{- 3.8}$ & $1.305$ \\  
bright Ly$\alpha$ & $3.8^{+3.0}_{-1.0}$& $0.17^{+0.01}_{-0.04}$ & $  35^{+  79}_{-  12}$& $11.8^{+ 2.4}_{- 5.0}$ & $0.377$ \\ 
faint Ly$\alpha$ & $13.4^{+3.5}_{-3.5}$& $0.16^{+0.03}_{-0.03}$ & $ 181^{+ 141}_{-  90}$& $ 8.7^{+ 3.7}_{- 2.4}$ &  $2.609$ \\ 
large EW & $4.4^{+3.4}_{-1.4}$& $0.16^{+0.02}_{-0.04}$ & $  57^{+ 123}_{-  28}$& $ 8.6^{+ 3.1}_{- 3.5}$ & $0.368$ \\  
small EW & $19.3^{+5.1}_{-4.8}$& $0.16^{+0.02}_{-0.03}$ & $ 203^{+ 158}_{-  89}$& $11.2^{+ 3.3}_{- 3.0}$ & $3.267$ \\ 
bright $K$ & $9.7^{+4.2}_{-1.4}$& $0.21^{+0.02}_{-0.02}$ & $  40^{+  41}_{-  11}$& $26.3^{+ 6.0}_{- 7.0}$ & $1.057$ \\ 
faint $K$ &  $1.7^{+1.2}_{-0.7}$& $0.13^{+0.02}_{-0.01}$ & $  26^{+  38}_{-  14}$& $ 6.7^{+ 1.9}_{- 1.7}$ &  $0.279$ \\  \hline
\end{tabular}
}
}\label{tbl:sed_field}
\tabnote{Note. (1) Stellar mass; (2) color excess; (3) age; (4) $SFR$; and (5) reduced $\chi^2$ value.
Metallicity is fixed to $0.2{\rm Z}_\odot$, redshift to 2.18, and $f_{\rm esc}^{\rm ion}$ to 0.2. 
} 
\end{table*}

\begin{figure*}[ht] 
      \includegraphics[width=1.0\linewidth]{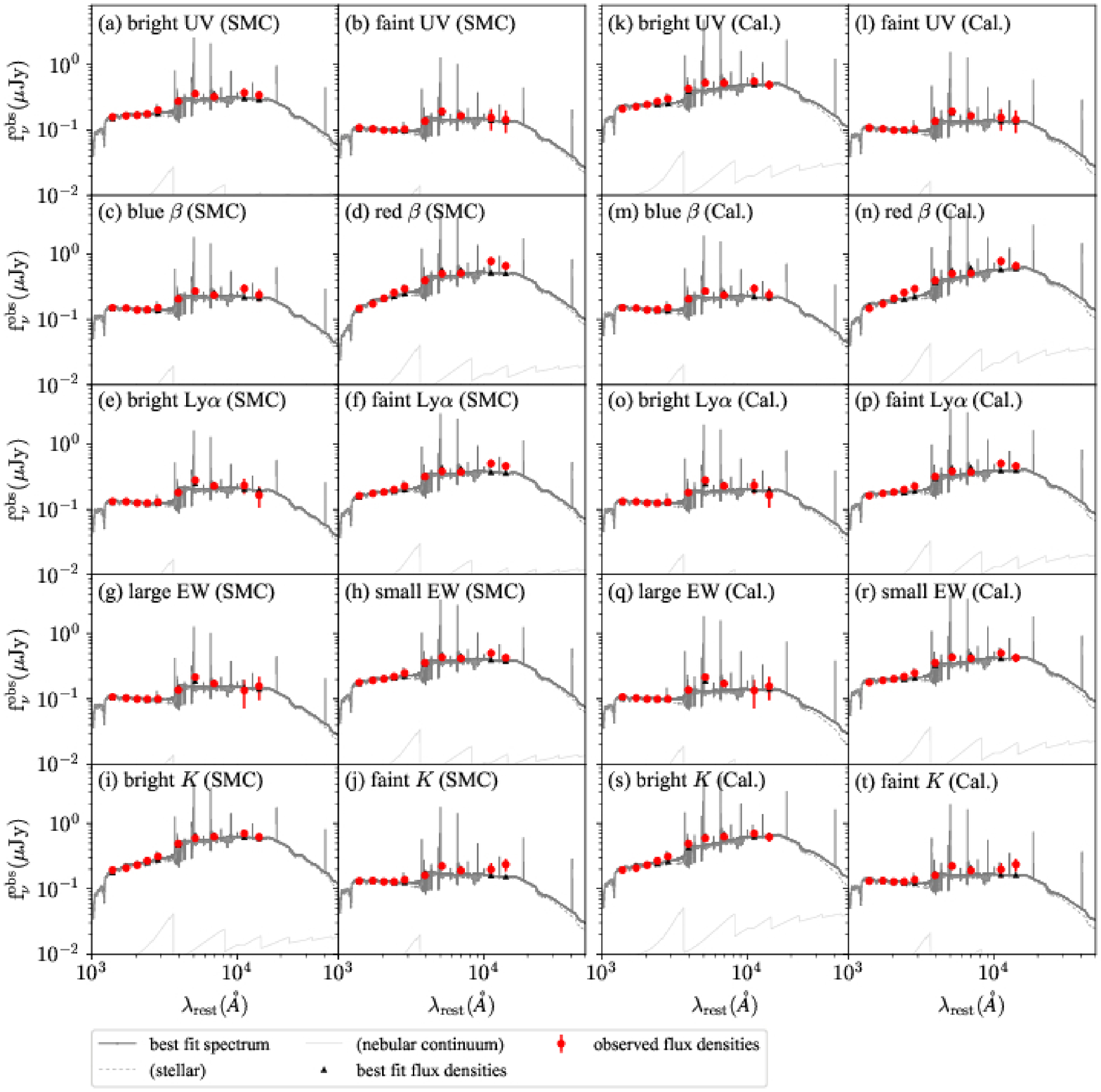}  
  \caption{  
 Best-fit SEDs for the ten subsamples of the SXDS field. Panels (a)--(j) show results with an assumption of SMC-like attenuation curve: (a) bright UV, (b) faint UV, (c) blue $\beta$, (d) red $\beta$, (e) bright Ly$\alpha$, (f) faint Ly$\alpha$, (g) large $EW_{0}({{\rm Ly}\alpha})$, (h) small $EW_{0}({{\rm Ly}\alpha})$, (i) bright $K$, and (j) faint $K$, while panels (k)--(t) show those with an assumption of Calzetti curve. For each panel, a gray solid line, a light gray solid line and a light gray dotted line show the best-fit model spectrum, its nebular continuum component and its stellar continuum component, respectively. Red filled circles and black filled triangles represent the observed flux densities and the flux densities calculated from the best-fit spectrum, respectively. (Color online)
}
  \label{fig:sed_sxds}
\end{figure*}

\begin{figure*}[ht] 
      \includegraphics[width=1.0\linewidth]{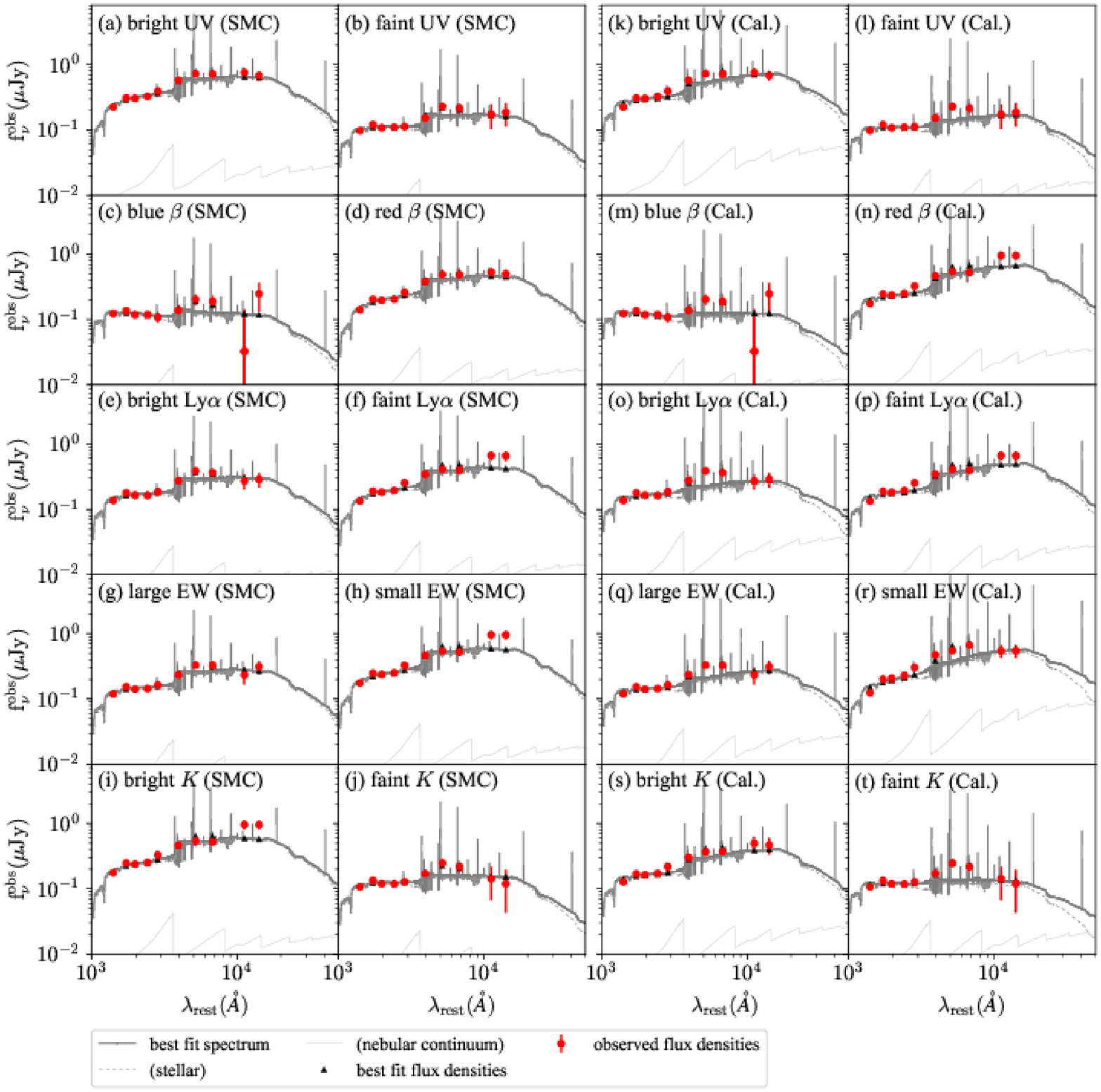}  
  \caption{  Best-fit SEDs for the ten subsamples of the COSMOS field. Panels (a)--(j) show results with an assumption of SMC-like attenuation curve: (a) bright UV, (b) faint UV, (c)  blue $\beta$, (d) red $\beta$, (e) bright Ly$\alpha$, (f) faint Ly$\alpha$, (g) bright $K$, (g) large $EW_{0}({{\rm Ly}\alpha})$, (h) small $EW_{0}({{\rm Ly}\alpha})$, (i) bright $K$, and (j) faint $K$, while panels (k)--(t) show those with an assumption of Calzetti curve. For each panel, a gray solid line, light gray solid line and a light gray dotted line show the best-fit model spectrum, its nebular continuum component and its stellar continuum component, respectively. Red filled circles and black filled triangles represent the observed flux densities and the flux densities calculated from the best-fit spectrum, respectively. (Color online) }
  \label{fig:sed_cosmos}
\end{figure*}

\begin{figure*}[ht]
\begin{flushleft}
      \includegraphics[width=0.9\linewidth]{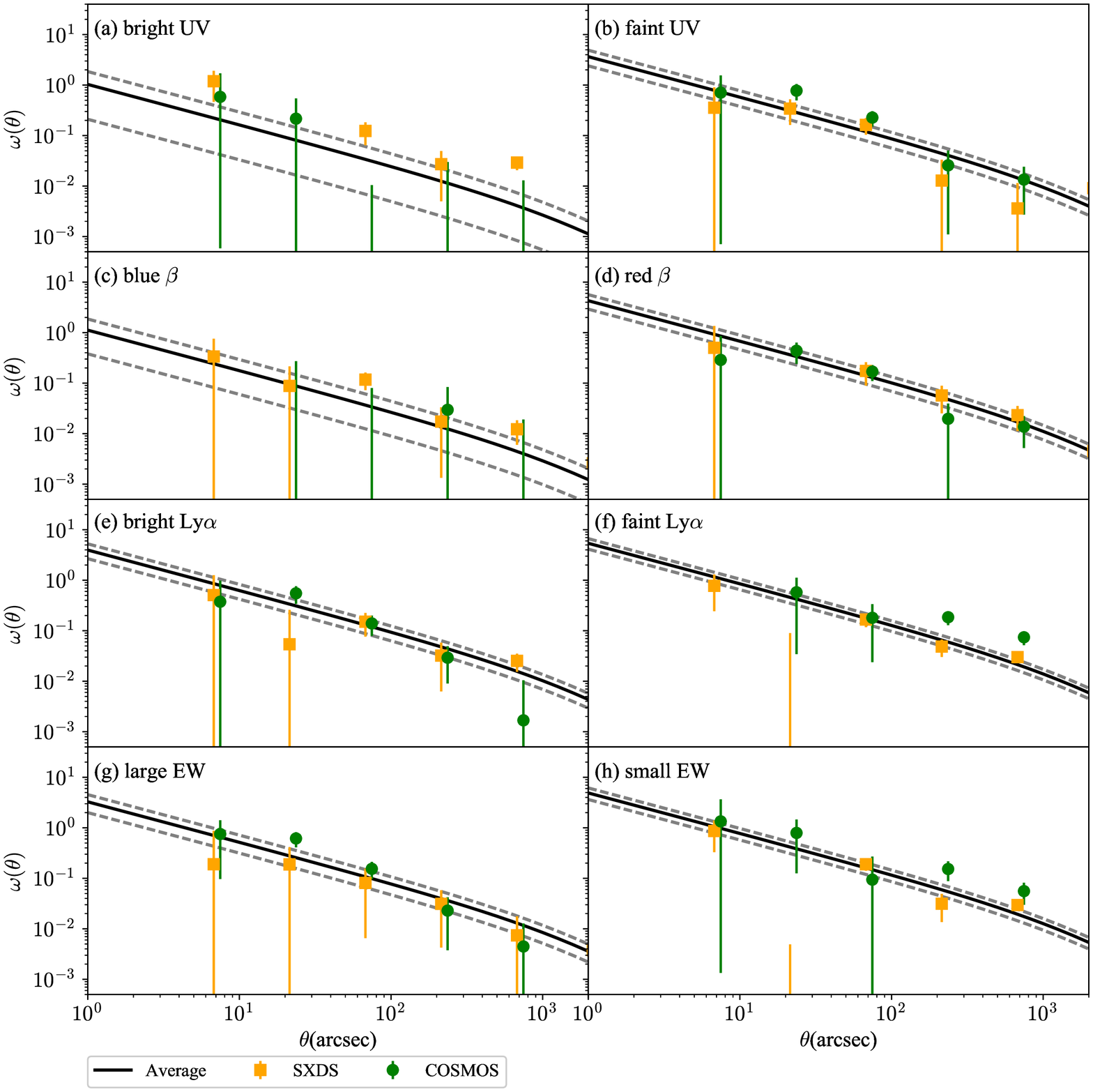}
\end{flushleft}      
  \caption{  ACF measurements for the eight subsamples: (a) bright UV, (b) faint UV, (c)  blue $\beta$, (d) red $\beta$, (e) bright Ly$\alpha$, (f) faint Ly$\alpha$, (g) large $EW_{0}({{\rm Ly}\alpha})$, and (h) small $EW_{0}({{\rm Ly}\alpha})$,. For each panel,  orange filled squares and green filled circles represent measurements in the SXDS and COSMOS fields.  A black solid line and gray dotted line indicate the field-average best-fit ACFs with fixed $\beta =0.8$, whose fitting range is $40$--$1000''$. we slightly shift all data points along the abscissa by a value depending on the field for presentation purpose. (Color online)
}

  \label{fig:ACF}
\end{figure*}

\section{Results of the SED fitting}\label{sec:appendix_sed}
Figures \ref{fig:sed_sxds} and \ref{fig:sed_cosmos} show the best-fit SEDs for the ten subsamples for each field,  and table \ref{tbl:sed_average} the best-fit stellar parameters.

\section{The best fit ACFs}\label{sec:appendix_acf}
We show the best-fit ACFs for the two fields and their field-average values in figure \ref{fig:ACF}. We do not perform clustering analysis for the $K$-divided subsamples as described in section \ref{subsec:criteria}. We do not plot the $M_{\rm h}$ of the UV bright and blue $\beta$ subsamples (figures \ref{fig:Lh} and \ref{fig:Lhmodel}), since they are not constrained well. This is partly because at small $r_0$ values like those of these two subsamples, $M_{\rm h}$ depends very sensitively on $r_0$ according to the bias model \citep[see appendix B in][]{Khostovan2018}. The differences in the ACF measurement between the two fields have been discussed in \citet{Kusakabe2018a}.

\section{The fluorescence scenario}\label{sec:appendix_fluorescence} 
Some of the ionizing photons produced in central galaxies are converted to fluorescence Ly$\alpha$ emission due to recombination of neutral hydrogen gas in the CGM. We do not include the fluorescence scenario in the discussion of LAHs (section \ref{sec:d_lah}), since this scenario has been favored for QSOs' LAHs \citep[e.g.,][]{Hennawi2013, Cantalupo2014}. Recently, however, the fluorescence emission of star forming LAEs has been discussed with MUSE data \citep{Gallego2018, Leclercq2017}. 
In this appendix, we briefly examine this scenario on the basis of
the minimum amount of ionizing photons, $N({\rm ion})$ (photon s$^{-1}$), and  hense $\xi_{\rm ion}$, required to maintain the LAH of LAEs with fluorescence while reproducing the nebular emission of the main bodies. 

Our LAEs in a sub region in the SXDS field have been observed with a narrow band targeting H$\alpha$ emission \citep{Nakajima2012}. The stacked H$\alpha$ luminosity as a point source is $L({\rm H}\alpha)_{ps, tot}\sim 8.4\times10^{41}\,L_\odot$ with dust attenuation correction \citep[$E(B-V)\sim0.1$, ][]{Kusakabe2015}. The number of ionizing photons that are produced by star formation and then converted to nebular emission in the LAEs is $N({\rm ion})_{H\alpha, corr}=L({\rm H}\alpha)_{ps, tot}/1.36\times10^{-12}\sim6.1\times10^{53}$ photon s$^{-1}$ following \citet{Kennicutt1998}. 

The LAH luminosity is calculated as $L({\rm Ly}\alpha)_{\rm H}\sim 2.0\times10^{42}\,L_\odot$ from the stacked $L({\rm Ly}\alpha)_{ps}\sim 1.8\times10^{42}\,L_\odot$. It is converted to the number of Ly$\alpha$ photons, $N({\rm Ly\alpha})_{\rm LAH}=L({\rm Ly}\alpha)_{\rm H}/h\nu_{\rm Ly\alpha}\sim1.3\times10^{53}$ photon s$^{-1}$ with the Planck constant ($h$) and the frequency of Ly$\alpha$ ($\nu_{\rm Ly\alpha}$). The fraction of recombinations which results in Ly$\alpha$ photons in the optically thick case (case B) is $\eta_{\rm thick}=0.66$, which is larger than the fraction for the optically thin limit, $\eta_{\rm thin}=0.42$ \citep{Osterbrock2006, Hennawi2013}. The minimum number of ionizing photons that escape from the ISM to the CGM required to maintain the observed LAHs is $N({\rm ion})_{\rm LAH}=N({\rm Ly\alpha})_{\rm LAH}/\eta_{\rm thick}\sim1.8\times10^{53}$ photon s$^{-1}$.  It is notable that the LAH luminosity (surface brightness, more accurately) is independent of the luminosity of ionizing radiation in the highly-ionized, optical thin regime \citep{Hennawi2013} which requires larger number of $N({\rm ion})_{\rm LAH}$ than that in optically thick case at a fixed hydrogen gas distribution. 
Ionizing radiation is attenuated by dust in the ISM before escaping out to the CGM. The dust-attenuation corrected $N({\rm ion})_{\rm LAH}$, $N({\rm ion})_{\rm LAH, corr}$, is estimated roughly to be $1.1\times10^{54}$ photon s$^{-1}$ with an underestimated correction with $\kappa\sim20$ from an SMC-like attenuation curve at $\sim1000$\AA\, ($>912$ \AA).  This part gives the largest uncertainty in the whole calculation. 

The minimum value of the intrinsic $N({\rm ion})$ produced in the galaxy is $N({\rm ion})_{H\alpha, corr}+N({\rm ion})_{\rm LAH, corr}\sim1.8\times10^{54}$ photon s$^{-1}$. Here, we do not consider ionizing photons escaping out to the IGM, although LAEs (at $z\sim3$) are found to have high escape fractions of $\sim10$--$30$\% \citep[e.g.,][]{Nestor2013,Fletcher2018arXiv}. 

 Our LAEs are estimated to have the total $SFR=5.7\,M_\odot {\rm yr}^{-1}$ on average from SED fitting \citep{Kusakabe2015}. With the fiducial $\xi_{ion}$ value of $\sim1.3\times10^{25}$ Hz erg$^{-1}$ \citep{Kennicutt1998, Sobral2018arXiv}, this SFR is converted into $N({\rm ion})\sim5.3\times10^{53}$ photon s$^{-1}$, which is three times lower than the minimum required value obtained above.  In order for $N({\rm ion})$ to reach $\sim1.8\times10^{54}$ photon s$^{-1}$, $\xi_{ion}$ must be as high as $\sim4\times10^{25}$ Hz erg$^{-1}$. 
 The minimum value of the required $\xi_{ion}$ is higher than the estimated $\xi_{ion}$ ($\sim2\times10^{25}$ Hz erg$^{-1}$ ) for the large-EW LAEs ($EW_{0,ps}({{\rm Ly}\alpha})\sim70$\AA\,) from \citet{Sobral2018arXiv}'s relation (see section \ref{subsubsec:fesc_add}). Note that the required $\xi_{ion}$ is consistent with a high $\xi_{ion}$ estimated for LAEs at $z\sim3$ in \citet{Nakajima2018arXiv}. We can also estimate the minimum value of the escape fraction of ionizing photons from the ISM as $\sim10$\% from $N({\rm ion})_{\rm LAH}$/($N({\rm ion})_{\rm LAH, corr}$+$N({\rm ion})_{H\alpha, corr}$). This is larger than $2\%$ for LAEs at $z\sim3.5$ in \citet{Gallego2018}. 

The fluorescence scenario requires a high $\xi_{ion}$ and a high escape fraction of ionizing photons for the LAEs in subregion of SXDS field with H$\alpha$ observation even without including ionizing photons escaping out to the IGM. However, we can not exclude the fluorescence scenario completely because of the lack of direct observations of $\xi_{ion}$ and a high escape fraction of ionizing photons. Further observational and theoretical studies are required to examine the fluorescence scenario for star forming galaxies. 

\end{appendix}
\bibliographystyle{apj}

\begin{thebibliography}{}
\expandafter\ifx\csname natexlab\endcsname\relax\def\natexlab#1{#1}\fi

\bibitem[{Aihara {et~al.}(2018)Aihara, Arimoto, Armstrong, Arnouts, Bahcall,
  Bickerton, Bosch, Bundy, Capak, Chan, Chiba, Coupon, Egami, Enoki, Finet,
  Fujimori, Fujimoto, Furusawa, Furusawa, Goto, Goulding, Greco, Greene, Gunn,
  Hamana, Harikane, Hashimoto, Hattori, Hayashi, Hayashi, He{\l}miniak,
  Higuchi, Hikage, Ho, Hsieh, Huang, Huang, Ikeda, Imanishi, Inoue, Iwasawa,
  Iwata, Jaelani, Jian, Kamata, Karoji, Kashikawa, Katayama, Kawanomoto, Kayo,
  Koda, Koike, Kojima, Komiyama, Konno, Koshida, Koyama, Kusakabe, Leauthaud,
  Lee, Lin, Lin, Lupton, Mandelbaum, Matsuoka, Medezinski, Mineo, Miyama,
  Miyatake, Miyazaki, Momose, More, More, Moritani, Moriya, Morokuma, Mukae,
  Murata, Murayama, Nagao, Nakata, Niida, Niikura, Nishizawa, Obuchi, Oguri,
  Oishi, Okabe, Okamoto, Okura, Ono, Onodera, Onoue, Osato, Ouchi, Price, Pyo,
  Sako, Sawicki, Shibuya, Shimasaku, Shimono, Shirasaki, Silverman, Simet,
  Speagle, Spergel, Strauss, Sugahara, Sugiyama, Suto, Suyu, Suzuki, Tait,
  Takada, Takata, Tamura, Tanaka, Tanaka, Tanaka, Tanaka, Terai, Terashima,
  Toba, Tominaga, Toshikawa, Turner, Uchida, Uchiyama, Umetsu, Uraguchi, Urata,
  Usuda, Utsumi, Wang, Wang, Wong, Yabe, Yamada, Yamanoi, Yasuda, Yeh,
  Yonehara, \& Yuma}]{Aihara2018a}
Aihara, H., Arimoto, N., Armstrong, R., {et~al.} 2018, Publications of the
  Astronomical Society of Japan, 70, 1

\bibitem[{Alavi {et~al.}(2014)Alavi, Siana, Richard, Stark, Scarlata, Teplitz,
  Freeman, Dominguez, Rafelski, Robertson, \& Kewley}]{Alavi2014}
Alavi, A., Siana, B., Richard, J., {et~al.} 2014, The Astrophysical Journal,
  780, 143

\bibitem[{{\'{A}}lvarez-M{\'{a}}rquez
  {et~al.}(2016){\'{A}}lvarez-M{\'{a}}rquez, Burgarella, Heinis, Buat, Lo~Faro,
  B{\'{e}}thermin, L{\'{o}}pez-Fort{\'{i}}n, Cooray, Farrah, Hurley, Ibar,
  Ilbert, Koekemoer, Lemaux, P{\'{e}}rez-Fournon, Rodighiero, Salvato, Scott,
  Taniguchi, Vieira, \& Wang}]{Alvarez-Marquez2016}
{\'{A}}lvarez-M{\'{a}}rquez, J., Burgarella, D., Heinis, S., {et~al.} 2016,
  Astronomy {\&} Astrophysics, 587, A122

\bibitem[{Ando {et~al.}(2006)Ando, Ohta, Iwata, Akiyama, Aoki, \&
  Tamura}]{Ando2006}
Ando, M., Ohta, K., Iwata, I., {et~al.} 2006, The Astrophysical Journal, 645,
  L9

\bibitem[{Ando {et~al.}(2007)Ando, Ohta, Iwata, Akiyama, Aoki, \&
  Tamura}]{Ando2007}
---. 2007, Publications of the Astronomical Society of Japan, 59, 717

\bibitem[{Atek {et~al.}(2008)Atek, Kunth, Hayes, {\"{O}}stlin, \&
  Mas-Hesse}]{Atek2008}
Atek, H., Kunth, D., Hayes, M., {\"{O}}stlin, G., \& Mas-Hesse, J.~M. 2008,
  Astronomy and Astrophysics, 488, 491

\bibitem[{Atek {et~al.}(2014)Atek, Kunth, Schaerer, Miguel Mas-Hesse, Hayes,
  {\"{O}}stlin, \& Kneib}]{Atek2014a}
Atek, H., Kunth, D., Schaerer, D., {et~al.} 2014, Astronomy {\&} Astrophysics,
  561, A89

\bibitem[{Barnes \& Haehnelt(2010)}]{Barnes2010}
Barnes, L.~A., \& Haehnelt, M.~G. 2010, Monthly Notices of the Royal
  Astronomical Society, 403, 870

\bibitem[{Behrens \& Braun(2014)}]{Behrens2014}
Behrens, C., \& Braun, H. 2014, Astronomy and Astrophysics, 572, A74

\bibitem[{Blanc {et~al.}(2011)Blanc, Adams, Gebhardt, Hill, Drory, Hao, Bender,
  Ciardullo, Finkelstein, Fry, Gawiser, Gronwall, Hopp, Jeong, Kelzenberg,
  Komatsu, MacQueen, Murphy, Roth, Schneider, \& Tufts}]{Blanc2011}
Blanc, G.~a., Adams, J.~J., Gebhardt, K., {et~al.} 2011, The Astrophysical
  Journal, 736, 31

\bibitem[{Bouwens {et~al.}(2016)Bouwens, Aravena, Decarli, Walter, da~Cunha,
  Labb{\'{e}}, E.~Bauer, Bertoldi, Carilli, Chapman, Daddi, Hodge, J.~Ivison,
  Karim, Le~Fevre, Magnelli, Ota, Riechers, R.~Smail, van~der Werf, Weiss, Cox,
  Elbaz, Gonzalez-Lopez, Infante, Oesch, Wagg, \& Wilkins}]{Bouwens2016}
Bouwens, R.~J., Aravena, M., Decarli, R., {et~al.} 2016, The Astrophysical
  Journal, 833, 72

\bibitem[{Bridge {et~al.}(2017)Bridge, Hayes, Melinder, {\"{O}}stlin, Gronwall,
  Ciardullo, Atek, Cannon, Gronke, Guaita, Hagen, Herenz, Kunth, Laursen,
  Mas-Hesse, \& Pardy}]{Bridge2017}
Bridge, J.~S., Hayes, M., Melinder, J., {et~al.} 2017, The Astrophysical
  Journal, 852, 9

\bibitem[{Brocklehurst(1971)}]{Brocklehurst1971}
Brocklehurst, M. 1971, MNRAS, 153, 471

\bibitem[{Bruzual \& Charlot(2003)}]{Bruzual2003}
Bruzual, G., \& Charlot, S. 2003, Monthly Notices of the Royal Astronomical
  Society, 344, 1000

\bibitem[{Buat {et~al.}(2012)Buat, Noll, Burgarella, Giovannoli, Charmandaris,
  Pannella, Hwang, Elbaz, Dickinson, Magdis, Reddy, \& Murphy}]{Buat2012}
Buat, V., Noll, S., Burgarella, D., {et~al.} 2012, Astronomy {\&} Astrophysics,
  545, A141

\bibitem[{Calzetti {et~al.}(2000)Calzetti, Armus, Bohkin, Kinnery, Koornneef,
  \& Storchi-bergmann}]{Calzetti2000}
Calzetti, D., Armus, L., Bohkin, R., {et~al.} 2000, ApJ, 533, 682

\bibitem[{Cantalupo {et~al.}(2014)Cantalupo, Arrigoni-battaia, Prochaska,
  Hennawi, \& Madau}]{Cantalupo2014}
Cantalupo, S., Arrigoni-battaia, F., Prochaska, J.~X., Hennawi, J.~F., \&
  Madau, P. 2014, Nature, 506, 63

\bibitem[{Cantalupo {et~al.}(2005)Cantalupo, Porciani, Lilly, \&
  Miniati}]{Cantalupo2005}
Cantalupo, S., Porciani, C., Lilly, S.~J., \& Miniati, F. 2005, The
  Astrophysical Journal, 628, 61

\bibitem[{Capak {et~al.}(2004)Capak, Cowie, Hu, Barger, Dickinson, Fernandez,
  Giavalisco, Komiyama, Kretchmer, McNally, Miyazaki, Okamura, \&
  Stern}]{Capak2004}
Capak, P., Cowie, L.~L., Hu, E.~M., {et~al.} 2004, The Astronomical Journal,
  127, 180

\bibitem[{Capak {et~al.}(2007)Capak, Aussel, Ajiki, McCracken, Mobasher,
  Scoville, Shopbell, Taniguchi, Thompson, Tribiano, Sasaki, Blain, Brusa,
  Carilli, Comastri, Carollo, Cassata, Colbert, Ellis, Elvis, Giavalisco,
  Green, Guzzo, Hasinger, Ilbert, Impey, Jahnke, Kartaltepe, Kneib, Koda,
  Koekemoer, Komiyama, Leauthaud, Le~Fevre, Lilly, Liu, Massey, Miyazaki,
  Murayama, Nagao, Peacock, Pickles, Porciani, Renzini, Rhodes, Rich, Salvato,
  Sanders, Scarlata, Schiminovich, Schinnerer, Scodeggio, Sheth, Shioya, Tasca,
  Taylor, Yan, \& Zamorani}]{Capak2007}
Capak, P., Aussel, H., Ajiki, M., {et~al.} 2007, The Astrophysical Journal
  Supplement Series, 172, 99

\bibitem[{Chabrier \& Chabrier(2003)}]{Chabrier2003}
Chabrier, G., \& Chabrier, G. 2003, PASP, 115, 763

\bibitem[{Charlot \& Fall(1993)}]{Charlot1993}
Charlot, S., \& Fall, S.~M. 1993, Astrophysical Journal v.415, 415, 580

\bibitem[{Cochrane {et~al.}(2018)Cochrane, Best, Sobral, Smail, Geach, Stott,
  \& Wake}]{Cochrane2018}
Cochrane, R.~K., Best, P.~N., Sobral, D., {et~al.} 2018, Monthly Notices of the
  Royal Astronomical Society, 475, 3730

\bibitem[{Daddi {et~al.}(2004)Daddi, Cimatti, Renzini, Fontana, Mignoli,
  Pozzetti, Tozzi, \& Zamorani}]{Daddi2004}
Daddi, E., Cimatti, a., Renzini, a., {et~al.} 2004, The Astrophysical Journal,
  617, 746

\bibitem[{Dekel \& Birnboim(2006)}]{Dekel2006}
Dekel, A., \& Birnboim, Y. 2006, Monthly Notices of the Royal Astronomical
  Society, 368, 2

\bibitem[{Diemer \& Kravtsov(2015)}]{Diemer2015}
Diemer, B., \& Kravtsov, A.~V. 2015, The Astrophysical Journal, 799, 108

\bibitem[{Dijkstra \& Kramer(2012)}]{Dijkstra2012}
Dijkstra, M., \& Kramer, R. 2012, Monthly Notices of the Royal Astronomical
  Society, 424, 1672

\bibitem[{Dijkstra \& Loeb(2009)}]{Dijkstra2009}
Dijkstra, M., \& Loeb, A. 2009, Monthly Notices of the Royal Astronomical
  Society, 400, 1109

\bibitem[{Duval {et~al.}(2014)Duval, Schaerer, {\"{O}}stlin, \&
  Laursen}]{Duval2014}
Duval, F., Schaerer, D., {\"{O}}stlin, G., \& Laursen, P. 2014, Astronomy {\&}
  Astrophysics, 562, A52

\bibitem[{Eisenstein \& Hu(1998)}]{Eisenstein1998}
Eisenstein, D.~J., \& Hu, W. 1998, The Astrophysical Journal, 496, 605

\bibitem[{Eisenstein \& Hu(1999)}]{Eisenstein1999}
---. 1999, the Astrophysical Journal, 511, 5

\bibitem[{Elbaz {et~al.}(2007)Elbaz, Daddi, Borgne, Dickinson, Alexander,
  Chary, \& Starck}]{Elbaz2007}
Elbaz, D., Daddi, E., Borgne, D.~L., {et~al.} 2007, A{\&}A, 468, 33

\bibitem[{Erb {et~al.}(2006)Erb, Steidel, Shapley, Pettini, Reddy, \&
  Adelberger}]{Erb2006b}
Erb, D.~K., Steidel, C.~C., Shapley, A.~E., {et~al.} 2006, The Astrophysical
  Journal, 647, 128

\bibitem[{Fardal {et~al.}(2001)Fardal, Katz, Gardner, Hernquist, Weinberg, \&
  Dav{\'{e}}}]{Fardal2001}
Fardal, M.~A., Katz, N., Gardner, J.~P., {et~al.} 2001, Astrophysical Journal,
  562, 605

\bibitem[{Faucher-Gigu{\`{e}}re {et~al.}(2010)Faucher-Gigu{\`{e}}re,
  Kere{\v{s}}, Dijkstra, Hernquist, \& Zaldarriaga}]{Faucher-Giguere2010}
Faucher-Gigu{\`{e}}re, C.~A., Kere{\v{s}}, D., Dijkstra, M., Hernquist, L., \&
  Zaldarriaga, M. 2010, Astrophysical Journal, 725, 633

\bibitem[{Feldmeier {et~al.}(2013)Feldmeier, Hagen, Ciardullo, Gronwall,
  Gawiser, Guaita, Hagen, Bond, Acquaviva, Blanc, Orsi, \&
  Kurczynski}]{Feldmeier2013}
Feldmeier, J.~J., Hagen, A., Ciardullo, R., {et~al.} 2013, Astrophysical
  Journal, 776, 75

\bibitem[{Finkelstein {et~al.}(2009)Finkelstein, Rhoads, Malhotra, \&
  Grogin}]{Finkelstein2009b}
Finkelstein, S.~L., Rhoads, J.~E., Malhotra, S., \& Grogin, N. 2009, The
  Astrophysical Journal, 691, 465

\bibitem[{Finkelstein {et~al.}(2008)Finkelstein, Rhoads, Malhotra, Grogin, \&
  Wang}]{Finkelstein2008}
Finkelstein, S.~L., Rhoads, J.~E., Malhotra, S., Grogin, N., \& Wang, J. 2008,
  ApJ, 678, 655

\bibitem[{Fletcher {et~al.}(2018)Fletcher, Robertson, Nakajima, Ellis, Stark,
  \& Inoue}]{Fletcher2018arXiv}
Fletcher, T.~J., Robertson, B.~E., Nakajima, K., {et~al.} 2018, arXiv:,
  1806.01741, doi:10.13005/bbra/1095

\bibitem[{Fudamoto {et~al.}(2017)Fudamoto, Oesch, Schinnerer, Groves, Karim,
  Magnelli, Sargent, Cassata, Lang, Liu, F{\`{e}}vre, Smol{\v{c}}i{\'{c}}, \&
  Tasca}]{Fudamoto2017}
Fudamoto, Y., Oesch, P.~A., Schinnerer, E., {et~al.} 2017, MNRAS, 472, 483

\bibitem[{Furlanetto {et~al.}(2005)Furlanetto, Schaye, Springel, \&
  Hernquist}]{Furlanetto2005}
Furlanetto, S.~R., Schaye, J., Springel, V., \& Hernquist, L. 2005, The
  Astrophysical Journal, 622, 7

\bibitem[{Furusawa {et~al.}(2008)Furusawa, Kosugi, Akiyama, Takata, Sekiguchi,
  Tanaka, Iwata, Kajisawa, Yasuda, Doi, Ouchi, Simpson, Shimasaku, Yamada,
  Furusawa, Morokuma, Ishida, Aoki, Fuse, Imanishi, Iye, Karoji, Kobayashi,
  Kodama, Komiyama, Maeda, Miyazaki, Mizumoto, Nakata, \&
  Noumaru}]{Furusawa2008}
Furusawa, H., Kosugi, G., Akiyama, M., {et~al.} 2008, ApJs, 176, 1

\bibitem[{Gallego {et~al.}(2018)Gallego, Cantalupo, Lilly, Marino, Pezzulli,
  Schaye, Wisotzki, Bacon, Inami, Akhlaghi, Tacchella, Richard, Bouche,
  Steinmetz, \& Carollo}]{Gallego2018}
Gallego, S.~G., Cantalupo, S., Lilly, S., {et~al.} 2018, Monthly Notices of the
  Royal Astronomical Society, 475, 3854

\bibitem[{Garel {et~al.}(2015)Garel, Blaizot, Guiderdoni, Michel-Dansac, Hayes,
  \& Verhamme}]{Garel2015}
Garel, T., Blaizot, J., Guiderdoni, B., {et~al.} 2015, Monthly Notices of the
  Royal Astronomical Society, 450, 1279

\bibitem[{Giacconi {et~al.}(2001)Giacconi, Rosati, \& Tozzi}]{Giacconi2001}
Giacconi, R., Rosati, P., \& Tozzi, P. 2001, The Astrophysical Journal, 551,
  624

\bibitem[{Goerdt {et~al.}(2010)Goerdt, Dekel, Sternberg, Ceverino, Teyssier, \&
  Primack}]{Goerdt2010}
Goerdt, T., Dekel, A., Sternberg, A., {et~al.} 2010, Monthly Notices of the
  Royal Astronomical Society, 407, 613

\bibitem[{Gordon {et~al.}(2003)Gordon, Clayton, Misselt, Landolt, \&
  Wolff}]{Gordon2003}
Gordon, K.~D., Clayton, G.~C., Misselt, K.~A., Landolt, A.~U., \& Wolff, M.~J.
  2003, ApJ, 594, 279

\bibitem[{Goto {et~al.}(2009)Goto, Utsumi, Furusawa, Miyazaki, \&
  Komiyama}]{Goto2009}
Goto, T., Utsumi, Y., Furusawa, H., Miyazaki, S., \& Komiyama, Y. 2009, Monthly
  Notices of the Royal Astronomical Society, 400, 843

\bibitem[{Guaita {et~al.}(2010)Guaita, Gawiser, Padilla, Francke, Bond,
  Gronwall, Ciardullo, Feldmeier, Sinawa, Blanc, \& Virani}]{Guaita2010}
Guaita, L., Gawiser, E., Padilla, N., {et~al.} 2010, The Astrophysical Journal,
  714, 255

\bibitem[{Guaita {et~al.}(2011)Guaita, Acquaviva, Padilla, Gawiser, Bond,
  Ciardullo, Treister, Kurczynski, Gronwall, Lira, \& Schawinski}]{Guaita2011}
Guaita, L., Acquaviva, V., Padilla, N., {et~al.} 2011, The Astrophysical
  Journal, 733, 114

\bibitem[{Guaita {et~al.}(2017)Guaita, Talia, Pentericci, Verhamme, Cassata,
  Lemaux, Orlitova, Ribeiro, Schaerer, Zamorani, Garilli, Le~Brun,
  Le~F{\`{e}}vre, Maccagni, Tasca, Thomas, Vanzella, Zucca, Amorin, Bardelli,
  Castellano, Grazian, Hathi, Koekemoer, \& Marchi}]{Guaita2017}
Guaita, L., Talia, M., Pentericci, L., {et~al.} 2017, Astronomy {\&}
  Astrophysics, 606, A19

\bibitem[{Hagen {et~al.}(2016)Hagen, Zeimann, Behrens, Ciardullo, Gebhardt,
  Gronwall, Bridge, Fox, Schneider, Trump, Blanc, Chiang, Chonis, Finkelstein,
  Hill, Jogee, \& Gawiser}]{Hagen2016}
Hagen, A., Zeimann, G.~R., Behrens, C., {et~al.} 2016, The Astrophysical
  Journal, 817, 79

\bibitem[{Haiman {et~al.}(2000)Haiman, Spaans, \& Quataert}]{Haiman2000}
Haiman, Z., Spaans, M., \& Quataert, E. 2000, Astrophysical Journal, 537, L5

\bibitem[{Hansen \& Peng~Oh(2006)}]{Hansen2006}
Hansen, M., \& Peng~Oh, S. 2006, Monthly Notices of the Royal Astronomical
  Society, 367, 979

\bibitem[{Harikane {et~al.}(2018)Harikane, Ouchi, Shibuya, Kojima, Zhang, Itoh,
  Ono, Higuchi, Inoue, Chevallard, Capak, Nagao, Onodera, Faisst, Martin,
  Rauch, Bruzual, Charlot, Davidzon, Fujimoto, Hilmi, Ilbert, Lee, Matsuoka,
  Silverman, \& Toft}]{Harikane2018}
Harikane, Y., Ouchi, M., Shibuya, T., {et~al.} 2018, The Astrophysical Journal,
  859, 84

\bibitem[{Hashimoto {et~al.}(2013)Hashimoto, Ouchi, Shimasaku, Ono, Nakajima,
  Rauch, \& Okamura}]{Hashimoto2013}
Hashimoto, T., Ouchi, M., Shimasaku, K., {et~al.} 2013, The Astrophysical
  Journal, 765, 70

\bibitem[{Hashimoto {et~al.}(2015)Hashimoto, Verhamme, Ouchi, Shimasaku,
  Schaerer, Nakajima, Shibuya, Rauch, Ono, \& Goto}]{Hashimoto2015}
Hashimoto, T., Verhamme, A., Ouchi, M., {et~al.} 2015, The Astrophysical
  Journal, 812, 157

\bibitem[{Hashimoto {et~al.}(2017)Hashimoto, Ouchi, Shimasaku, Schaerer,
  Nakajima, Shibuya, Ono, Rauch, \& Goto}]{Hashimoto2017a}
Hashimoto, T., Ouchi, M., Shimasaku, K., {et~al.} 2017, MNRAS, 465, 1543

\bibitem[{Hayashino {et~al.}(2004)Hayashino, Matsuda, Tamura, Yamauchi, Yamada,
  Ajiki, Fujita, Murayama, Nagao, Ohta, Okamura, Ouchi, Shimasaku, Shioya, \&
  Taniguchi}]{Hayashino2004}
Hayashino, T., Matsuda, Y., Tamura, H., {et~al.} 2004, The Astrophysical
  Journal Letters, 128, 2073

\bibitem[{Hayes {et~al.}(2005)Hayes, {\"{O}}stlin, Mas-Hesse, Kunth, Leitherer,
  \& Petrosian}]{Hayes2005}
Hayes, M., {\"{O}}stlin, G., Mas-Hesse, J., {et~al.} 2005, Astronomy and
  Astrophysics, 438, 71

\bibitem[{Hayes {et~al.}(2011)Hayes, Schaerer, {\"{O}}stlin, Mas-Hesse, Atek,
  \& Kunth}]{Hayes2011}
Hayes, M., Schaerer, D., {\"{O}}stlin, G., {et~al.} 2011, The Astrophysical
  Journal, 730, 8

\bibitem[{Hayes {et~al.}(2010)Hayes, {\"{O}}stlin, Schaerer, Mas-Hesse,
  Leitherer, Atek, Kunth, Verhamme, de~Barros, \& Melinder}]{Hayes2010}
Hayes, M., {\"{O}}stlin, G., Schaerer, D., {et~al.} 2010, Nature, 464, 562

\bibitem[{Hayes {et~al.}(2013)Hayes, {\"{O}}stlin, Schaerer, Verhamme,
  Mas-Hesse, Adamo, Atek, Cannon, Duval, Guaita, Herenz, Kunth, Laursen,
  Melinder, Orlitov{\'{a}}, Ot{\'{i}}-Floranes, \& Sandberg}]{Hayes2013}
---. 2013, The Astrophysical Journal, 765, L27

\bibitem[{Hayes {et~al.}(2014)Hayes, {\"{O}}stlin, Duval, Sandberg, Guaita,
  Melinder, Adamo, Schaerer, Verhamme, Orlitov{\'{a}}, Mas-Hesse, Cannon, Atek,
  Kunth, Laursen, Ot{\'{i}}-Floranes, Pardy, Rivera-Thorsen, \&
  Herenz}]{Hayes2014}
Hayes, M., {\"{O}}stlin, G., Duval, F., {et~al.} 2014, ApJ, 782, 6

\bibitem[{Heinis {et~al.}(2014)Heinis, Buat, B{\'{e}}thermin, Bock, Burgarella,
  Conley, Cooray, Farrah, Ilbert, Magdis, Marsden, Oliver, Rigopoulou, Roehlly,
  Schulz, Symeonidis, Viero, Xu, \& Zemcov}]{Heinis2014}
Heinis, S., Buat, V., B{\'{e}}thermin, M., {et~al.} 2014, Monthly Notices of
  the Royal Astronomical Society, 437, 1268

\bibitem[{Hennawi \& Prochaska(2013)}]{Hennawi2013}
Hennawi, J.~F., \& Prochaska, J.~X. 2013, The Astrophysical Journal, 766, 58

\bibitem[{Ilbert {et~al.}(2009)Ilbert, Capak, Salvato, Aussel, McCracken,
  Sanders, Scoville, Kartaltepe, Arnouts, Floc'h, Mobasher, Taniguchi,
  Lamareille, Leauthaud, Sasaki, Thompson, Zamojski, Zamorani, Bardelli,
  Bolzonella, Bongiorno, Brusa, Caputi, Carollo, Contini, Cook, Coppa,
  Cucciati, de~la Torre, de~Ravel, Franzetti, Garilli, Hasinger, Iovino,
  Kampczyk, Kneib, Knobel, Kovac, Le~Borgne, Le~Brun, F{\`{e}}vre, Lilly,
  Looper, Maier, Mainieri, Mellier, Mignoli, Murayama, Pell{\`{o}}, Peng,
  P{\'{e}}rez-Montero, Renzini, Ricciardelli, Schiminovich, Scodeggio, Shioya,
  Silverman, Surace, Tanaka, Tasca, Tresse, Vergani, \& Zucca}]{Ilbert2009}
Ilbert, O., Capak, P., Salvato, M., {et~al.} 2009, The Astrophysical Journal,
  690, 1236

\bibitem[{Ishiyama {et~al.}(2015)Ishiyama, Enoki, Kobayashi, Makiya, Nagashima,
  \& Oogi}]{Ishiyama2015}
Ishiyama, T., Enoki, M., Kobayashi, M. A.~R., {et~al.} 2015, Publications of
  the Astronomical Society of Japan, 67, 61

\bibitem[{Keel {et~al.}(1999)Keel, Cohen, Windhorst, \& Waddington}]{Keel1999}
Keel, W.~C., Cohen, S.~H., Windhorst, R.~A., \& Waddington, I. 1999, The
  Astrophysical Journal Letters, 118, 2547

\bibitem[{Kennicutt(1998)}]{Kennicutt1998}
Kennicutt, R.~C. 1998, {STAR FORMATION IN GALAXIES ALONG THE HUBBLE SEQUENCE}

\bibitem[{Kere{\v{s}} {et~al.}(2005)Kere{\v{s}}, Katz, Weinberg, \&
  Dav{\'{e}}}]{Keres2005}
Kere{\v{s}}, D., Katz, N., Weinberg, D.~H., \& Dav{\'{e}}, R. 2005, Monthly
  Notices of the Royal Astronomical Society, 363, 2

\bibitem[{Khostovan {et~al.}(2018)Khostovan, Sobral, Mobasher, Best, Smail,
  Matthee, Darvish, Nayyeri, Hemmati, \& Stott}]{Khostovan2018}
Khostovan, A.~A., Sobral, D., Mobasher, B., {et~al.} 2018, Monthly Notices of
  the Royal Astronomical Society, 478, 2999

\bibitem[{Kobayashi {et~al.}(2016)Kobayashi, Murata, Koekemoer, Murayama,
  Taniguchi, Kajisawa, Shioya, Scoville, Nagao, \& Capak}]{Kobayashi2016}
Kobayashi, M. A.~R., Murata, K.~L., Koekemoer, A.~M., {et~al.} 2016, The
  Astrophysical Journal, 819, 0

\bibitem[{Kojima {et~al.}(2017)Kojima, Ouchi, Nakajima, Shibuya, Harikane, \&
  Ono}]{Kojima2017}
Kojima, T., Ouchi, M., Nakajima, K., {et~al.} 2017, PASJ, 69, 44

\bibitem[{Kollmeier {et~al.}(2010)Kollmeier, Zheng, Dav{\'{e}}, Gould, Katz,
  Miralda-Escud{\'{e}}, \& Weinberg}]{Kollmeier2010}
Kollmeier, J.~A., Zheng, Z., Dav{\'{e}}, R., {et~al.} 2010, Astrophysical
  Journal, 708, 1048

\bibitem[{Konno {et~al.}(2016)Konno, Ouchi, Nakajima, Duval, Kusakabe, Ono, \&
  Shimasaku}]{Konno2016}
Konno, A., Ouchi, M., Nakajima, K., {et~al.} 2016, The Astrophysical Journal,
  823, 20

\bibitem[{Koprowski {et~al.}(2018)Koprowski, Coppin, Geach, McLure, Almaini,
  Blain, Bremer, Bourne, Chapman, Conselice, Dunlop, Farrah, Hartley, Karim,
  Knudsen, Michalowski, Scott, Simpson, Smith, \& van~der Werf}]{Koprowski2018}
Koprowski, M.~P., Coppin, K.~E., Geach, J.~E., {et~al.} 2018, Monthly Notices
  of the Royal Astronomical Society, 479, 4355

\bibitem[{Kornei {et~al.}(2010)Kornei, Shapley, Erb, Steidel, Reddy, Pettini,
  \& Bogosavljevi{\'{c}}}]{Kornei2010}
Kornei, K.~a., Shapley, A.~E., Erb, D.~K., {et~al.} 2010, The Astrophysical
  Journal, 711, 693

\bibitem[{Kroupa(2001)}]{Kroupa2001}
Kroupa, P. 2001, Monthly Notices of the Royal Astronomical Society, 322, 231

\bibitem[{Kunth {et~al.}(2003)Kunth, Leitherer, Mas-Hesse, {\"{O}}stlin, \&
  Petrosian}]{Kunth2003}
Kunth, D., Leitherer, C., Mas-Hesse, J.~M., {\"{O}}stlin, G., \& Petrosian, A.
  2003, The Astrophysical Journal, 597, 263

\bibitem[{Kunth {et~al.}(1998)Kunth, Mas-Hesse, Terlevich, Terlevich, Lequeux,
  \& Fall}]{Kunth1998}
Kunth, D., Mas-Hesse, J.~M., Terlevich, E., {et~al.} 1998, Astronomy and
  Astrophysics, 334, 11

\bibitem[{Kusakabe {et~al.}(2015)Kusakabe, Shimasaku, Nakajima, \&
  Ouchi}]{Kusakabe2015}
Kusakabe, H., Shimasaku, K., Nakajima, K., \& Ouchi, M. 2015, The Astrophysical
  Journal, 800, L29

\bibitem[{Kusakabe {et~al.}(2018)Kusakabe, Shimasaku, Ouchi, Nakajima, Goto,
  Hashimoto, Konno, Harikane, Silverman, \& Capak}]{Kusakabe2018a}
Kusakabe, H., Shimasaku, K., Ouchi, M., {et~al.} 2018, Publications of the
  Astronomical Society of Japan, 70, 1

\bibitem[{Lai {et~al.}(2008)Lai, Huang, Fazio, Gawiser, Ciardullo, Damen,
  Franx, Gronwall, Labbe, Magdis, \& Dokkum}]{Lai2008}
Lai, K., Huang, J.-s., Fazio, G., {et~al.} 2008, ApJ, 674, 70

\bibitem[{Laigle {et~al.}(2016)Laigle, McCracken, Ilbert, Hsieh, Davidzon,
  Capak, Hasinger, Silverman, Pichon, Coupon, Aussel, Borgne, Caputi, Cassata,
  Chang, Civano, Dunlop, Fynbo, Kartaltepe, Koekemoer, Fevre, Floc'h,
  Leauthaud, Lilly, Lin, Marchesi, Milvang-Jensen, Salvato, Sanders, Scoville,
  Smolcic, Stockmann, Taniguchi, Tasca, Toft, Vaccari, \& Zabl}]{Laigle2016}
Laigle, C., McCracken, H.~J., Ilbert, O., {et~al.} 2016, The Astrophysical
  Journal Supplement Series, 224, 1

\bibitem[{Lake {et~al.}(2015)Lake, Zheng, Cen, Sadoun, Momose, \&
  Ouchi}]{Lake2015}
Lake, E., Zheng, Z., Cen, R., {et~al.} 2015, The Astrophysical Journal, 806, 46

\bibitem[{Landy \& Szalay(1993)}]{Landy1993}
Landy, S.~D., \& Szalay, A.~S. 1993, The Astrophysical Journal, 412, 64

\bibitem[{Laursen {et~al.}(2013)Laursen, Duval, \& {\"{O}}stlin}]{Laursen2013}
Laursen, P., Duval, F., \& {\"{O}}stlin, G. 2013, Astrophysical Journal, 766,
  124

\bibitem[{Laursen \& Sommer-Larsen(2007)}]{Laursen2007}
Laursen, P., \& Sommer-Larsen, J. 2007, The Astrophysical Journal, 657, L69

\bibitem[{Lawrence {et~al.}(2007)Lawrence, Warren, Almaini, Edge, Hambly,
  Jameson, Lucas, Casali, Adamson, Dye, Emerson, Foucaud, Hewett, Hirst,
  Hodgkin, Irwin, Lodieu, McMahon, Simpson, Smail, Mortlock, \&
  Folger}]{Lawrence2007}
Lawrence, A., Warren, S.~J., Almaini, O., {et~al.} 2007, Monthly Notices of the
  Royal Astronomical Society, 379, 1599

\bibitem[{Leclercq {et~al.}(2017)Leclercq, Bacon, Wisotzki, Mitchell, Garel,
  Verhamme, Blaizot, Hashimoto, Herenz, Conseil, Cantalupo, Inami, Contini,
  Richard, Maseda, Schaye, Marino, Akhlaghi, Brinchmann, \&
  Carollo}]{Leclercq2017}
Leclercq, F., Bacon, R., Wisotzki, L., {et~al.} 2017, Astronomy {\&}
  Astrophysics, 608, A8

\bibitem[{Madau(1995)}]{Madau1995}
Madau, P. 1995, The Astrophysical Journal, 441, 18

\bibitem[{Makiya {et~al.}(2016)Makiya, Enoki, Ishiyama, Kobayashi, Nagashima,
  Okamoto, Okoshi, Oogi, \& Shirakata}]{Makiya2016}
Makiya, R., Enoki, M., Ishiyama, T., {et~al.} 2016, Publications of the
  Astronomical Society of Japan, 68, 25

\bibitem[{Malhotra \& Rhoads(2002)}]{Malhotra2002}
Malhotra, S., \& Rhoads, J.~E. 2002, ApJ, 565, L71

\bibitem[{Malkan {et~al.}(2017)Malkan, Cohen, Maruyama, Kashikawa, Ly,
  Ishikawa, Shimasaku, Hayashi, \& Motohara}]{Malkan2017}
Malkan, M.~A., Cohen, D.~P., Maruyama, M., {et~al.} 2017, The Astrophysical
  Journal, 850, 5

\bibitem[{Mas-Ribas \& Dijkstra(2016)}]{Mas-Ribas2016}
Mas-Ribas, L., \& Dijkstra, M. 2016, The Astrophysical Journal, 822, 84

\bibitem[{Mas-Ribas {et~al.}(2017)Mas-Ribas, Dijkstra, Hennawi, Trenti, Momose,
  \& Ouchi}]{Mas-Ribas2017a}
Mas-Ribas, L., Dijkstra, M., Hennawi, J.~F., {et~al.} 2017, The Astrophysical
  Journal, 841, 19

\bibitem[{Matsuda {et~al.}(2011)Matsuda, Yamada, Hayashino, Yamauchi, Nakamura,
  Morimoto, Ouchi, Ono, Kousai, Nakamura, Horie, Fujii, Umemura, \&
  Mori}]{Matsuda2011}
Matsuda, Y., Yamada, T., Hayashino, T., {et~al.} 2011, Monthly Notices of the
  Royal Astronomical Society: Letters, 410, 13

\bibitem[{Matsuda {et~al.}(2012)Matsuda, Yamada, Hayashino, Yamauchi, Nakamura,
  Morimoto, Ouchi, Ono, Umemura, \& Mori}]{Matsuda2012}
---. 2012, Monthly Notices of the Royal Astronomical Society, 425, 878

\bibitem[{Matthee {et~al.}(2016)Matthee, Sobral, Oteo, Best, Smail,
  R¨ottgering, \& Paulino-Afonso}]{Matthee2016}
Matthee, J., Sobral, D., Oteo, I., {et~al.} 2016, Monthly Notices of the Royal
  Astronomical Society, 458, 449

\bibitem[{McCracken {et~al.}(2012)McCracken, Milvang-Jensen, Dunlop, Franx,
  Fynbo, Le~F{\`{e}}vre, Holt, Caputi, Goranova, Buitrago, Emerson, Freudling,
  Hudelot, L{\'{o}}pez-Sanjuan, Magnard, Mellier, M{\o}ller, Nilsson,
  Sutherland, Tasca, \& Zabl}]{McCracken2012}
McCracken, H.~J., Milvang-Jensen, B., Dunlop, J., {et~al.} 2012, Astronomy {\&}
  Astrophysics, 544, A156

\bibitem[{McLure {et~al.}(2018)McLure, Dunlop, Cullen, Bourne, Best, Khochfar,
  Bowler, Biggs, Geach, Scott, Michalowski, Rujopakarn, van Kampen,
  Kirkpatrick, \& Pope}]{McLure2018}
McLure, R.~J., Dunlop, J.~S., Cullen, F., {et~al.} 2018, Monthly Notices of the
  Royal Astronomical Society, 476, 3991

\bibitem[{Mehta {et~al.}(2018)Mehta, Scarlata, Capak, Davidzon, Faisst, Hsieh,
  Ilbert, Jarvis, Laigle, Phillips, Silverman, Strauss, Tanaka, Bowler, Coupon,
  Foucaud, Hemmati, Masters, McCracken, Mobasher, Ouchi, Shibuya, \&
  Wang}]{Mehta2018}
Mehta, V., Scarlata, C., Capak, P., {et~al.} 2018, The Astrophysical Journal
  Supplement Series, 235, 36

\bibitem[{Meurer {et~al.}(1999)Meurer, Heckman, \& Calzetti}]{Meurer1999}
Meurer, G.~R., Heckman, T., \& Calzetti, D. 1999, ApJ, 521, 64

\bibitem[{Momose {et~al.}(2018)Momose, Goto, Utsumi, Hashimoto, Chiang, Kim,
  Kashikawa, \& Shimasaku}]{Momose2018arXiv}
Momose, R., Goto, T., Utsumi, Y., {et~al.} 2018, arXiv:, 1809.10916

\bibitem[{Momose {et~al.}(2014)Momose, Ouchi, Nakajima, Ono, Shibuya,
  Shimasaku, Yuma, Mori, \& Umemura}]{Momose2014}
Momose, R., Ouchi, M., Nakajima, K., {et~al.} 2014, MNRAS442, 442, 110

\bibitem[{Momose {et~al.}(2016)Momose, Ouchi, Nakajima, Ono, Shibuya,
  Shimasaku, Yuma, Mori, \& Umemura}]{Momose2016}
---. 2016, Monthly Notices of the Royal Astronomical Society, 457, 2318

\bibitem[{Mori \& Umemura(2006)}]{Mori2006}
Mori, M., \& Umemura, M. 2006, Nature, 440, 644

\bibitem[{Moster {et~al.}(2013)Moster, Naab, \& White}]{Moster2013}
Moster, B.~P., Naab, T., \& White, S. D.~M. 2013, Monthly Notices of the Royal
  Astronomical Society, 428, 3121

\bibitem[{Nakajima {et~al.}(2016)Nakajima, Ellis, Iwata, Inoue, Kusakabe,
  Ouchi, \& Robertson}]{Nakajima2016}
Nakajima, K., Ellis, R.~S., Iwata, I., {et~al.} 2016, The Astrophysical
  Journal, 831, L9

\bibitem[{Nakajima {et~al.}(2018{\natexlab{a}})Nakajima, Fletcher, Ellis,
  Robertson, \& Iwata}]{Nakajima2018arXiv}
Nakajima, K., Fletcher, T., Ellis, R.~S., Robertson, B.~E., \& Iwata, I.
  2018{\natexlab{a}}, ArXiv:1801.03085, 13, 1

\bibitem[{Nakajima \& Ouchi(2014)}]{Nakajima2014}
Nakajima, K., \& Ouchi, M. 2014, Monthly Notices of the Royal Astronomical
  Society, 442, 900

\bibitem[{Nakajima {et~al.}(2013)Nakajima, Ouchi, Shimasaku, Hashimoto, Ono, \&
  Lee}]{Nakajima2013}
Nakajima, K., Ouchi, M., Shimasaku, K., {et~al.} 2013, The Astrophysical
  Journal, 769, 3

\bibitem[{Nakajima {et~al.}(2012)Nakajima, Ouchi, Shimasaku, Ono, Lee, Foucaud,
  Ly, Dale, Salim, Finn, Almaini, \& Okamura}]{Nakajima2012}
---. 2012, The Astrophysical Journal, 745, 12

\bibitem[{Nakajima {et~al.}(2018{\natexlab{b}})Nakajima, Schaerer,
  Le~F{\`{e}}vre, Amor{\'{i}}n, Talia, Lemaux, Tasca, Vanzella, Zamorani,
  Bardelli, Grazian, Guaita, Hathi, Pentericci, \& Zucca}]{Nakajima2018}
Nakajima, K., Schaerer, D., Le~F{\`{e}}vre, O., {et~al.} 2018{\natexlab{b}},
  Astronomy {\&} Astrophysics, 612, A94

\bibitem[{Nestor {et~al.}(2013)Nestor, Shapley, Kornei, Steidel, \&
  Siana}]{Nestor2013}
Nestor, D.~B., Shapley, A.~E., Kornei, K.~a., Steidel, C.~C., \& Siana, B.
  2013, The Astrophysical Journal, 765, 47

\bibitem[{Neufeld(1991)}]{Neufeld1991}
Neufeld, d.~A. 1991, ApJ, 370, 85

\bibitem[{Nickerson {et~al.}(2013)Nickerson, Stinson, Couchman, Bailin, \&
  Wadsley}]{Nickerson2013}
Nickerson, S., Stinson, G., Couchman, H. M.~P., Bailin, J., \& Wadsley, J.
  2013, Monthly Notices of the Royal Astronomical Society, 429, 452

\bibitem[{Noeske {et~al.}(2007)Noeske, Kassin, Weiner, Faber, Koo, Lotz,
  Harker, Bundy, Metevier, Phillips, Cooper, Croton, Konidaris, \&
  Willmer}]{Noeske2007}
Noeske, K.~G., Kassin, S.~a., Weiner, B.~J., {et~al.} 2007, ApJ, 10, 35

\bibitem[{Okamoto {et~al.}(2010)Okamoto, Frenk, Jenkins, \&
  Theuns}]{Okamoto2010}
Okamoto, T., Frenk, C.~S., Jenkins, A., \& Theuns, T. 2010, Monthly Notices of
  the Royal Astronomical Society, 406, 208

\bibitem[{Oke \& Gunn(1983)}]{Oke1983}
Oke, J.~B., \& Gunn, J.~E. 1983, The Astrophysical Journal, 266, 713

\bibitem[{Ono {et~al.}(2010)Ono, Shimasaku, Dunlop, Farrah, McLure, \&
  Okamura}]{Ono2010b}
Ono, Y., Shimasaku, K., Dunlop, J., {et~al.} 2010, The Astrophysical Journal,
  724, 1524

\bibitem[{Osterbrock \& Ferland(2006)}]{Osterbrock2006}
Osterbrock, D.~E., \& Ferland, G.~J. 2006, {Astrophysics of gaseous nebulae and
  active galactic nuclei}

\bibitem[{{\"{O}}stlin {et~al.}(2009){\"{O}}stlin, Hayes, Kunth, Mas-Hesse,
  Leitherer, Petrosian, \& Atek}]{Ostlin2009}
{\"{O}}stlin, G., Hayes, M., Kunth, D., {et~al.} 2009, The Astronomical
  Journal, 138, 923

\bibitem[{{\"{O}}stlin {et~al.}(2014){\"{O}}stlin, Hayes, Duval, Sandberg,
  Rivera-Thorsen, Marquart, Orlitov{\'{a}}, Adamo, Melinder, Guaita, Atek,
  Cannon, Gruyters, Herenz, Kunth, Laursen, Mas-Hesse, Micheva,
  Ot{\'{i}}-Floranes, Pardy, Roth, Schaerer, \& Verhamme}]{Ostlin2014}
{\"{O}}stlin, G., Hayes, M., Duval, F., {et~al.} 2014, The Astrophysical
  Journal, 797, 11

\bibitem[{Oteo {et~al.}(2015)Oteo, Sobral, Ivison, Smail, Best, Cepa, \&
  P??rez-Garc??a}]{Oteo2015}
Oteo, I., Sobral, D., Ivison, R.~J., {et~al.} 2015, Monthly Notices of the
  Royal Astronomical Society, 452, 2018

\bibitem[{Ouchi {et~al.}(2003)Ouchi, Shimasaku, Furusawa, Miyazaki, Doi,
  Hamabe, Hayashino, Kimura, Kodaira, Komiyama, Matsuda, Miyazaki, Nakata,
  Okamura, Sekiguchi, Shioya, Tamura, Taniguchi, Yagi, \& Yasuda}]{Ouchi2003}
Ouchi, M., Shimasaku, K., Furusawa, H., {et~al.} 2003, The Astrophysical
  Journal, 582, 60

\bibitem[{Ouchi {et~al.}(2018)Ouchi, Harikane, Shibuya, Shimasaku, Taniguchi,
  Konno, Kobayashi, Kajisawa, Nagao, Ono, Inoue, Umemura, Mori, Hasegawa,
  Higuchi, Komiyama, Matsuda, Nakajima, Saito, \& Wang}]{Ouchi2018}
Ouchi, M., Harikane, Y., Shibuya, T., {et~al.} 2018, Publications of the
  Astronomical Society of Japan, 70, 1

\bibitem[{Overzier {et~al.}(2011)Overzier, Heckman, Wang, Armus, Buat, Howell,
  Meurer, Seibert, Siana, Basu-Zych, Charlot, Gon{\c{c}}alves, Martin, Neill,
  Rich, Salim, \& Schiminovich}]{Overzier2011a}
Overzier, R.~a., Heckman, T.~M., Wang, J., {et~al.} 2011, The Astrophysical
  Journal, 726, L7

\bibitem[{Paulino-Afonso {et~al.}(2018)Paulino-Afonso, Sobral, Ribeiro,
  Matthee, Santos, Calhau, Forshaw, Johnson, Merrick, P{\'{e}}rez, \&
  Sheldon}]{Paulino-Afonso2018AA}
Paulino-Afonso, A., Sobral, D., Ribeiro, B., {et~al.} 2018, Monthly Notices of
  the Royal Astronomical Society, doi:10.1093/mnras/sty281

\bibitem[{Pirzkal {et~al.}(2007)Pirzkal, Malhotra, Rhoads, \& Xu}]{Pirzkal2007}
Pirzkal, N., Malhotra, S., Rhoads, J.~E., \& Xu, C. 2007, The Astrophysical
  Journal, 667, 49

\bibitem[{{Planck Collaboration}(2016)}]{Planckcollaboration2016}
{Planck Collaboration}. 2016, A{\&}A, 594, 13

\bibitem[{Reddy {et~al.}(2010)Reddy, Erb, Pettini, Steidel, \&
  Shapley}]{Reddy2010}
Reddy, N.~a., Erb, D.~K., Pettini, M., Steidel, C.~C., \& Shapley, A.~E. 2010,
  The Astrophysical Journal, 712, 1070

\bibitem[{Reddy {et~al.}(2018)Reddy, Oesch, Bouwens, Montes, Illingworth,
  Steidel, van Dokkum, Atek, Carollo, Cibinel, Holden, Labb{\'{e}}, Magee,
  Morselli, Nelson, \& Wilkins}]{Reddy2018}
Reddy, N.~A., Oesch, P.~A., Bouwens, R.~J., {et~al.} 2018, The Astrophysical
  Journal, 853, 56

\bibitem[{Rodighiero {et~al.}(2011)Rodighiero, Daddi, Baronchelli, Cimatti,
  Renzini, Aussel, Popesso, Lutz, Andreani, Berta, Cava, Elbaz, Feltre,
  Fontana, F{\"{o}}rster~Schreiber, Franceschini, Genzel, Grazian, Gruppioni,
  Ilbert, Le~Floch, Magdis, Magliocchetti, Magnelli, Maiolino, McCracken,
  Nordon, Poglitsch, Santini, Pozzi, Riguccini, Tacconi, Wuyts, \&
  Zamorani}]{Rodighiero2011}
Rodighiero, G., Daddi, E., Baronchelli, I., {et~al.} 2011, The Astrophysical
  Journal, 739, L40

\bibitem[{Rosdahl \& Blaizot(2012)}]{Rosdahl2012}
Rosdahl, J., \& Blaizot, J. 2012, Monthly Notices of the Royal Astronomical
  Society, 423, 344

\bibitem[{Sales {et~al.}(2014)Sales, Vogelsberger, Genel, Torrey, Nelson,
  Rodriguez-Gomez, Wang, Pillepich, Sijacki, Springel, \&
  Hernquist}]{Sales2014}
Sales, L.~V., Vogelsberger, M., Genel, S., {et~al.} 2014, Monthly Notices of
  the Royal Astronomical Society: Letters, 447, L6

\bibitem[{Salpeter(1955)}]{Salpeter1955}
Salpeter, E.~E. 1955, Astrophysical Journal, 121, 161

\bibitem[{Santini {et~al.}(2017)Santini, Fontana, Castellano, Criscienzo,
  Merlin, Amorin, Cullen, Daddi, Dickinson, Dunlop, Grazian, Lamastra, McLure,
  Micha{\l}owski, Pentericci, \& Shu}]{Santini2017}
Santini, P., Fontana, A., Castellano, M., {et~al.} 2017, The Astrophysical
  Journal, 847, 76

\bibitem[{Scarlata {et~al.}(2009)Scarlata, Colbert, Teplitz, Panagia, Hayes,
  Siana, Rau, Francis, Caon, Pizzella, \& Bridge}]{Scarlata2009}
Scarlata, C., Colbert, J., Teplitz, H.~I., {et~al.} 2009, The Astrophysical
  Journal, 704, L98

\bibitem[{Schlegel {et~al.}(1998)Schlegel, Finkbeiner, \& Davis}]{Schlegel1998}
Schlegel, D. J. D.~J., Finkbeiner, D. P. D.~P., \& Davis, M. 1998, The
  Astrophysical Journal, 500, 525

\bibitem[{Scoville {et~al.}(2007)Scoville, Abraham, Aussel, Barnes, Benson,
  Blain, Calzetti, Comastri, Capak, Carilli, Carlstrom, Carollo, Colbert, \&
  Daddi}]{Scoville2007}
Scoville, N., Abraham, R.~G., Aussel, H., {et~al.} 2007, ApJs, 172, 38

\bibitem[{Shibuya {et~al.}(2014{\natexlab{a}})Shibuya, Ouchi, Nakajima, Yuma,
  Hashimoto, Shimasaku, Mori, \& Umemura}]{Shibuya2014b}
Shibuya, T., Ouchi, M., Nakajima, K., {et~al.} 2014{\natexlab{a}}, The
  Astrophysical Journal, 785, 64

\bibitem[{Shibuya {et~al.}(2014{\natexlab{b}})Shibuya, Ouchi, Nakajima,
  Hashimoto, Ono, Rauch, Gauthier, Shimasaku, Goto, Mori, \&
  Umemura.}]{Shibuya2014a}
---. 2014{\natexlab{b}}, The Astrophysical Journal, 788, 74

\bibitem[{Shibuya {et~al.}(2018)Shibuya, Ouchi, Konno, Higuchi, Harikane, Ono,
  Shimasaku, Taniguchi, Kobayashi, Kajisawa, Nagao, Furusawa, Goto, Kashikawa,
  Komiyama, Kusakabe, Lee, Momose, Nakajima, Tanaka, Wang, \&
  Yuma}]{Shibuya2018a}
Shibuya, T., Ouchi, M., Konno, A., {et~al.} 2018, Publications of the
  Astronomical Society of Japan, 70, 1

\bibitem[{Shimakawa {et~al.}(2017)Shimakawa, Kodama, Shibuya, Kashikawa,
  Tanaka, Matsuda, Tadaki, Koyama, Hayashi, Suzuki, \&
  Yamamoto}]{Shimakawa2017a}
Shimakawa, R., Kodama, T., Shibuya, T., {et~al.} 2017, MNRAS, 468, 1123

\bibitem[{Shimizu {et~al.}(2011)Shimizu, Yoshida, \& Okamoto}]{Shimizu2011}
Shimizu, I., Yoshida, N., \& Okamoto, T. 2011, Monthly Notices of the Royal
  Astronomical Society, 418, 2273

\bibitem[{Shirakata {et~al.}(2018)Shirakata, Okamoto, Kawaguchi, Nagashima,
  Ishiyama, Makiya, Kobayashi, Enoki, Oogi, \& Okoshi}]{Shirakata2018}
Shirakata, H., Okamoto, T., Kawaguchi, T., {et~al.} 2018, Monthly Notices of
  the Royal Astronomical Society, doi:10.1093/mnras/sty2958

\bibitem[{Shivaei {et~al.}(2017)Shivaei, Reddy, Shapley, Siana, Kriek,
  Mobasher, Coil, Freeman, Sanders, Price, Azadi, \& Zick}]{Shivaei2017}
Shivaei, I., Reddy, N., Shapley, A., {et~al.} 2017, The Astrophysical Journal,
  837, 157

\bibitem[{Simon(2007)}]{Simon2007}
Simon, P. 2007, Astronomy {\&} Astrophysics, 473, 711

\bibitem[{Skelton {et~al.}(2014)Skelton, Whitaker, Momcheva, Brammer, van
  Dokkum, Labb{\'{e}}, Franx, van~der Wel, Bezanson, Da~Cunha, Fumagalli,
  F{\"{o}}rster~Schreiber, Kriek, Leja, Lundgren, Magee, Marchesini, Maseda,
  Nelson, Oesch, Pacifici, Patel, Price, Rix, Tal, Wake, \&
  Wuyts}]{Skelton2014}
Skelton, R.~E., Whitaker, K.~E., Momcheva, I.~G., {et~al.} 2014, The
  Astrophysical Journal Supplement Series, 214, 24

\bibitem[{Sobral {et~al.}(2014)Sobral, Best, Smail, Mobasher, Stott, \&
  Nisbet}]{Sobral2014}
Sobral, D., Best, P.~N., Smail, I., {et~al.} 2014, Monthly Notices of the Royal
  Astronomical Society, 437, 3516

\bibitem[{Sobral \& Matthee(2018)}]{Sobral2018arXiv}
Sobral, D., \& Matthee, J. 2018, ArXiv:, 1803.08923

\bibitem[{Sobral {et~al.}(2017)Sobral, Matthee, Best, Stroe, R{\"{o}}ttgering,
  Oteo, Smail, Morabito, \& Paulino-Afonso}]{Sobral2017}
Sobral, D., Matthee, J., Best, P., {et~al.} 2017, Monthly Notices of the Royal
  Astronomical Society, 466, 1242

\bibitem[{Song {et~al.}(2014)Song, Finkelstein, Gebhardt, Hill, Drory, Ashby,
  Blanc, Bridge, Chonis, Ciardullo, Fabricius, Fazio, Gawiser, Gronwall, Hagen,
  Huang, Jogee, Livermore, Salmon, Schneider, Willner, \& Zeimann}]{Song2014}
Song, M., Finkelstein, S.~L., Gebhardt, K., {et~al.} 2014, The Astrophysical
  Journal, 791, 3

\bibitem[{Speagle {et~al.}(2014)Speagle, Steinhardt, Capak, \&
  Silverman}]{Speagle2014a}
Speagle, J.~S., Steinhardt, C.~L., Capak, P.~L., \& Silverman, J.~D. 2014, The
  Astrophysical Journal Supplement Series, 214, 15

\bibitem[{Steidel {et~al.}(2011)Steidel, Bogosavljevi{\'{c}}, Shapley,
  Kollmeier, Reddy, Erb, \& Pettini}]{Steidel2011}
Steidel, C.~C., Bogosavljevi{\'{c}}, M., Shapley, A.~E., {et~al.} 2011, The
  Astrophysical Journal, 736, 160

\bibitem[{Tadaki {et~al.}(2013)Tadaki, Kodama, Tanaka, Hayashi, Koyama, \&
  Shimakawa}]{Tadaki2013}
Tadaki, K.-i., Kodama, T., Tanaka, I., {et~al.} 2013, The Astrophysical
  Journal, 778, 114

\bibitem[{Tadaki {et~al.}(2015)Tadaki, Kohno, Kodama, Ikarashi, Aretxaga,
  Berta, Caputi, Dunlop, Hatsukade, Hayashi, Hughes, Ivison, Izumi, Koyama,
  Lutz, Makiya, Matsuda, Nakanishi, Rujopakarn, Tamura, Umehata, Wang, Wilson,
  Wuyts, Yamaguchi, \& Yun}]{Tadaki2015}
Tadaki, K.~I., Kohno, K., Kodama, T., {et~al.} 2015, Astrophysical Journal
  Letters, 811, L3

\bibitem[{Tadaki {et~al.}(2017)Tadaki, Genzel, Kodama, Wuyts, Wisnioski,
  Schreiber, Burkert, Lang, Tacconi, Lutz, Belli, Davies, Hatsukade, Hayashi,
  Herrera-Camus, Ikarashi, Inoue, Kohno, Koyama, Mendel, Nakanishi, Shimakawa,
  Suzuki, Tamura, Tanaka, {\"{U}}bler, \& Wilman}]{Tadaki2017a}
Tadaki, K.-i., Genzel, R., Kodama, T., {et~al.} 2017, The Astrophysical
  Journal, 834, 135

\bibitem[{Tal {et~al.}(2013)Tal, van Dokkum, Franx, Leja, Wake, \&
  Whitaker}]{Tal2013}
Tal, T., van Dokkum, P.~G., Franx, M., {et~al.} 2013, The Astrophysical
  Journal, 769, 31

\bibitem[{Taniguchi \& Shioya(2000)}]{Taniguchi2000}
Taniguchi, Y., \& Shioya, Y. 2000, Astrophysical Journal Letters v.489, 532,
  L13

\bibitem[{Taniguchi {et~al.}(2007)Taniguchi, Scoville, Murayama, Sanders,
  Mobasher, Aussel, Capak, Ajiki, Miyazaki, Komiyama, Shioya, Nagao, Sasaki,
  Koda, Carilli, Giavalisco, Guzzo, Hasinger, Impey, LeFevre, Lilly, Renzini,
  Rich, Schinnerer, Shopbell, Kaifu, Karoji, Arimoto, Okamura, \&
  Ohta}]{Taniguchi2007}
Taniguchi, Y., Scoville, N., Murayama, T., {et~al.} 2007, The Astrophysical
  Journal Supplement Series, 172, 9

\bibitem[{Taniguchi {et~al.}(2015)Taniguchi, Kajisawa, Kobayashi, Nagao,
  Shioya, Scoville, Sanders, Capak, Koekemoer, Toft, McCracken, Le~F{\`{e}}vre,
  Tasca, Sheth, Renzini, Lilly, Carollo, Kova{\v{c}}, Ilbert, Schinnerer, Fu,
  Tresse, Griffiths, \& Civano}]{Taniguchi2015}
Taniguchi, Y., Kajisawa, M., Kobayashi, M. A.~R., {et~al.} 2015, The
  Astrophysical Journal, 809, L7

\bibitem[{{The Astropy Collaboration} {et~al.}(2013){The Astropy
  Collaboration}, Robitaille, Tollerud, Greenfield, Droettboom, Bray, Aldcroft,
  Davis, Ginsburg, Price-Whelan, Kerzendorf, Conley, Crighton, Barbary, Muna,
  Ferguson, Grollier, Parikh, Nair, G{\"{u}}nther, Deil, Woillez, Conseil,
  Kramer, Turner, Singer, Fox, Weaver, Zabalza, Edwards, Azalee~Bostroem,
  Burke, Casey, Crawford, Dencheva, Ely, Jenness, Labrie, Lim, Pierfederici,
  Pontzen, Ptak, Refsdal, Servillat, \&
  Streicher}]{TheAstropyCollaboration2013}
{The Astropy Collaboration}, Robitaille, T.~P., Tollerud, E.~J., {et~al.} 2013,
  Astronomy {\&} Astrophysics, 558, A33

\bibitem[{Tinker {et~al.}(2010)Tinker, Robertson, Kravtsov, Klypin, Warren,
  Yepes, \& Gottl{\"{o}}ber}]{Tinker2010}
Tinker, J.~L., Robertson, B.~E., Kravtsov, A.~V., {et~al.} 2010, The
  Astrophysical Journal, 724, 878

\bibitem[{Tomczak {et~al.}(2016)Tomczak, Quadri, Tran, Labbe, Straatman,
  Papovich, Glazebrook, Allen, Brammer, Cowley, Dickinson, Elbaz, Inami,
  Kacprzak, Morrison, Nanayakkara, Persson, Rees, Salmon, Schreiber, Spitler,
  \& Whitaker}]{Tomczak2016}
Tomczak, A.~R., Quadri, R.~F., Tran, K.-V.~H., {et~al.} 2016, The Astrophysical
  Journal, 817, 118

\bibitem[{Totsuji \& Kihara(1969)}]{Totsuji1969}
Totsuji, H., \& Kihara, T. 1969, Publications of the Astronomical Society of
  Japan, 21, 221

\bibitem[{Trentham \& Tully(2009)}]{Trentham2009}
Trentham, N., \& Tully, R.~B. 2009, Monthly Notices of the Royal Astronomical
  Society, 398, 722

\bibitem[{Vargas {et~al.}(2014)Vargas, Bish, Acquaviva, Gawiser, Finkelstein,
  Ciardullo, Ashby, Feldmeier, Ferguson, Gronwall, Guaita, Hagen, Koekemoer,
  Kurczynski, Newman, \& Padilla}]{Vargas2014}
Vargas, C.~J., Bish, H., Acquaviva, V., {et~al.} 2014, The Astrophysical
  Journal, 783, 26

\bibitem[{Verhamme {et~al.}(2012)Verhamme, Dubois, Blaizot, Garel, Bacon,
  Devriendt, Guiderdoni, \& Slyz}]{Verhamme2012}
Verhamme, A., Dubois, Y., Blaizot, J., {et~al.} 2012, A{\&}A, 546, 111

\bibitem[{Verhamme {et~al.}(2006)Verhamme, Schaerer, \& Maselli}]{Verhamme2006}
Verhamme, A., Schaerer, D., \& Maselli, A. 2006, Astronomy and Astrophysics,
  460, 397

\bibitem[{Wang {et~al.}(2014)Wang, Sales, Henriques, \& White}]{Wang2014}
Wang, W., Sales, L.~V., Henriques, B. M.~B., \& White, S. D.~M. 2014, Monthly
  Notices of the Royal Astronomical Society, 442, 1363

\bibitem[{Whitaker {et~al.}(2014)Whitaker, Franx, Leja, van Dokkum, Henry,
  Skelton, Fumagalli, Momcheva, Brammer, Labbe, Nelson, \&
  Rigby}]{Whitaker2014}
Whitaker, K.~E., Franx, M., Leja, J., {et~al.} 2014, ApJ, 795, 104

\bibitem[{Wisotzki {et~al.}(2016)Wisotzki, Bacon, Blaizot, Brinchmann, Herenz,
  Schaye, Bouch{\'{e}}, Cantalupo, Contini, Carollo, Caruana, Courbot,
  Emsellem, Kamann, Kerutt, Leclercq, Lilly, Patr{\'{i}}cio, Sandin, Steinmetz,
  Straka, Urrutia, Verhamme, Weilbacher, \& Wendt}]{Wisotzki2016}
Wisotzki, L., Bacon, R., Blaizot, J., {et~al.} 2016, A{\&}A, 587, 98

\bibitem[{Wisotzki {et~al.}(2018)Wisotzki, Bacon, Brinchmann, Cantalupo,
  Richter, Schaye, Schmidt, Urrutia, Weilbacher, Akhlaghi, Bouch{\'{e}},
  Contini, Guiderdoni, Herenz, Inami, Kerutt, Leclercq, Marino, Maseda,
  Monreal-Ibero, Nanayakkara, Richard, Saust, Steinmetz, \&
  Wendt}]{Wisotzki2018}
Wisotzki, L., Bacon, R., Brinchmann, J., {et~al.} 2018, Nature, 562, 229

\bibitem[{Xue {et~al.}(2017)Xue, Lee, Dey, Reddy, Hong, Prescott, Inami,
  Jannuzi, \& Gonzalez}]{Xue2017}
Xue, R., Lee, K.-S., Dey, A., {et~al.} 2017, The Astrophysical Journal, 837,
  172

\bibitem[{Yagi {et~al.}(2013)Yagi, Suzuki, Yamanoi, Furusawa, Nakata, \&
  Komiyama}]{Yagi2013}
Yagi, M., Suzuki, N., Yamanoi, H., {et~al.} 2013, PASJ, 65, 22

\bibitem[{Yajima {et~al.}(2013)Yajima, Li, \& Zhu}]{Yajima2013}
Yajima, H., Li, Y., \& Zhu, Q. 2013, The Astrophysical Journal, 773, 151

\bibitem[{Yajima {et~al.}(2012)Yajima, Li, Zhu, Abel, Gronwall, \&
  Ciardullo}]{Yajima2012b}
Yajima, H., Li, Y., Zhu, Q., {et~al.} 2012, Astrophysical Journal, 754, 118

\bibitem[{Zehavi {et~al.}(2004)Zehavi, Weinberg, Zheng, Berlind, Frieman,
  Scoccimarro, Sheth, Blanton, Tegmark, Mo, Bahcall, Brinkmann, Burles, Csabai,
  Fukugita, Gunn, Lamb, Loveday, Lupton, Meiksin, Munn, Nichol, Schlegel,
  Schneider, SubbaRao, Szalay, Uomoto, \& York}]{Zehavi2004}
Zehavi, I., Weinberg, D.~H., Zheng, Z., {et~al.} 2004, The Astrophysical
  Journal, 608, 16

\bibitem[{Zheng {et~al.}(2011)Zheng, Cen, Weinberg, Trac, \&
  Miralda-Escud{\'{e}}}]{Zheng2011}
Zheng, Z., Cen, R., Weinberg, D., Trac, H., \& Miralda-Escud{\'{e}}, J. 2011,
  The Astrophysical Journal, 739, 62

\end{thebibliography}

\end{document}